\documentclass[12pt,preprint]{aastex}

\usepackage{graphics,graphicx}
\usepackage{subfigure}
\usepackage{color}
\usepackage{lscape}
\usepackage{amsmath}

\begin{document}

\title{Delving Into X-ray Obscuration of Type 2 AGN, Near and Far}

\renewcommand{\thefootnote}{\fnsymbol{footnote}}

\author{Stephanie M. LaMassa$^1$, Tahir Yaqoob$^2$, Andrew F. Ptak$^2$, Jianjun Jia $^3$\footnote{J. Jia is now affiliated with ASML Brion Technologies}, Timothy M. Heckman$^3$, Poshak Gandhi$^4$, C. Meg Urry$^1$}

\affil{$^1$Yale University,$^2$NASA/Goddard Space Flight Center, $^3$The Johns Hopkins University, $^4$University of Durham}

\begin{abstract}
Using self-consistent, physically motivated models, we investigate the X-ray obscuration in 19 Type 2 [OIII] 5007 \AA\ selected AGN, 9 of which are local Seyfert 2 galaxies and 10 of which are Type 2 quasar candidates. We derive reliable line-of-sight and global column densities for these objects, which is the first time this has been reported for an AGN sample; 4 AGN have significantly different global and line-of-sight column densities. Five sources are heavily obscured to Compton-thick. We comment on interesting sources revealed by our spectral modeling, including a candidate ``naked'' Sy2. After correcting for absorption, we find that the ratio of the rest-frame, 2-10 keV luminosity (L$_{\rm 2-10keV,in}$) to L$_{\rm [OIII]}$ is 1.54 $\pm$ 0.49 dex which is essentially identical to the mean Type 1 AGN value. The Fe K$\alpha$ luminosity is significantly correlated with L$_{\rm [OIII]}$, but with substantial scatter. Finally, we do not find a trend between L$_{\rm 2-10keV,in}$ and global or line-of-sight column density, between column density and redshift, between column density and scattering fraction or between scattering fraction and redshift.

\end{abstract}

% Sy2s                    QSO2     total
% unabsorbed: 1           0        1
% mild: 1                 1        2
% moderate: 1             4        5
% heavy: 4                2        6
% heavy to C-thick: 2     3        5

% scattering fraction for 13 sources

\section{Introduction}

Supermassive black holes in the centers of galaxies grow by accretion in which they are observed as Active Galactic Nuclei (AGN). X-ray emission is thought to originate in a corona surrounding the accretion disk where optical and ultraviolet photons from the disk are inverse-Compton scattered to higher energies. According to conventional unified models \citep{antonucci,urry}, this central engine is enshrouded by a circumnuclear obscuring medium of dust and gas which has a toroidal geometry. The inclination of this system to the observer's line of sight therefore determines the observable properties. In this classification scheme, Type 1 AGN represent a face-on system, allowing an unobscured view of the central engine where broad and narrow optical emission lines are apparent, as well as the ultraviolet continuum, mid-infrared emission from the circumnuclear obscuring matter, and X-ray emission from the corona. Conversely, the Type 2 systems are oriented so that the accretion disk and corona are hidden from the observer's sight line, blocking the optical and ultraviolet continuum and optical broad lines, and possibly significantly impacting the observed X-ray emission.

The physical processes that X-ray photons undergo are imprinted in the observed spectrum. An X-ray photon that is injected from the corona into the obscuring medium can pass unimpeded directly into our line of sight, be absorbed, or undergo one or more Compton-scatterings, with multiple scatterings becoming more likely if the column density is Compton-thick (i.e., N$_H \geq (1.2 \times \sigma_T)^{-1} \simeq 1.25 \times 10^{24}$ cm$^{-2}$, where $\sigma_T$ is the Thompson cross-section for electron scattering). The probabilities of these outcomes depend on the column density of the reprocessor and the incident energy of the X-ray photon. Fluorescent line emission from continuum photons ejecting inner shell electrons from atoms or ions in the obscuring medium can potentially be a dominant component of the observed spectrum. AGN continuum photons can also not interact with the absorber by escaping through the opening of the torus and/or passing through gaps in a clumpy medium if the absorber is non-uniform and patchy, and then scatter off a more distant optically-thin medium, entering our line of sight. 

Simple {\it ad hoc} X-ray models that parametrize the emission with an absorbed power law model or partial covering model will fail to capture the detailed physics inherent in these systems, especially when the column density exceeds $10^{24}$ cm$^{-2}$, i.e., when Compton scattering becomes an important factor. Even models that treat relativistic reflection off an accretion disk are unsuitable for these systems as they assume a column density that is infinite and a disk geometry does not adequately describe the absorber that reprocesses the X-ray emission. Additionally, such modeling is one-dimensional, which does not allow a finite column density out of the line of sight to be measured, even though the Fe K fluorescence line can provide information about the global matter distribution in which it is produced. With recent advances in spectral modeling that self-consistently treat Compton scattering, fluorescent line emission, and scattering off a distant medium \citep{mytorus,spherical}, we can now delve into the details of the obscuration in AGN and obtain physically meaningful constraints on inclination angle, line-of-sight and global column density, and the percentage of scattered AGN light.

Here we use physically motivated X-ray models to study the circumnuclear obscuration of a sample of Seyfert 2 galaxies (Sy2s) and Type 2 quasar (QSO) candidates selected based on their [OIII] 5007 \AA\ line emission, where the Sy2s are local ($z<0.15$), lower luminosity obscured AGN compared with quasars, which have bolometric luminosities exceeding $10^{45}$ erg s$^{-1}$. The [OIII] 5007 \AA\ line forms in the AGN narrow line region, 100s of parsecs above and below the circumnuclear obscuring medium, and it is primarily ionized by the AGN continuum, making it a reliable indicator of intrinsic AGN power \citep{bassani,kauff,tim04,cappi,panessa,me10}. These samples are thus unbiased with respect to the circumnuclear obscuration while X-ray selected samples are inherently biased against the most obscured sources as the combined effects of photoelectric absorption and Compton scattering modify the observed X-ray emission. 

The Sy2 and QSO samples have previously been analyzed and published in \citet{me} and \citet{jj}, respectively, using simple power law or double power law models to fit the spectrum. The amount of implied obscuration was determined by normalizing the observed X-ray flux by the intrinsic luminosity ([OIII] for the Sy2 and QSO sample, and by the mid-infrared [OIV] 26 $\mu$m line and mid-infrared continuum for the Sy2 sample) and analyzing the equivalent width of the Fe K$\alpha$ line, which can exceed 1 keV in Compton-thick sources as the fluorescent line in the reflection spectrum is superimposed on a depressed continuum \citep[e.g.,][]{ghisellini,levenson}. \citet{jj} also estimate the absorber column density by comparing the observed ratio of X-ray to [OIII] flux to a simulated unabsorbed ratio, using the unabsorbed (Type 1 AGN) values from \citet{tim} as the basis for simulating 1000 random draws per source, and determining the column density that can account for the attenuation between the absorbed and simulated unabsorbed X-ray fluxes. In neither work is the fitted column density used as representative of the true obscuration as this is poorly determined with the simple models used. Instead, the observed 2-10 keV fluxes and Fe K$\alpha$ EWs, which are largely model-independent, were used as diagnostics of the column density. In both works, these obscuration proxies suggested that a majority of the Sy2s and Type 2 QSO candidates are heavily obscured, and possibly Compton-thick.

In this analysis, we use the models from \citet{spherical} and \citet[MYTorus]{mytorus} to unveil the physical properties of the obscuration responsible for reprocessing the emission from the central engine within a spherical and toroidal geometry, respectively. These models treat the transmitted continuum, Compton scattered emission and fluorescent line emission self-consistently, allowing for physically meaningful measurements of the global and line-of-sight column densities which are poorly determined with {\it ad hoc} models. We apply these model to the subset of [OIII]-selected Sy2s from \citet{me} and QSO2 candidates from \citet{jj} that have adequate signal-to-noise X-ray spectra to constrain the model parameters. 

We comment on what we can learn about the circumnuclear obscuration with the present data, including reliable measurements of the global as well as line-of-sight N$_{H}$, and highlight two interesting sources as revealed by our X-ray analysis. As the Fe K$\alpha$ emission is accommodated self-consistently within the global obscuring medium in these models, we overcome the shortcomings of previous analyses that used simple power-law modeling to measure the Fe K$\alpha$ emission either in front of or behind the line-of-sight absorbing screen. We then test whether the Fe K$\alpha$ luminosity indeed traces the intrinsic luminosity as reported in previous works \citep[e.g.,][]{ptak,me,jj}. We also investigate whether there are differences between these QSO2s and Sy2s in terms of their column densities and scattering fractions and if there is a relationship between the column density and AGN luminosity as posited by the ``receding torus'' model  \citep[e.g.,][]{lawrence,lawrence2,ueda,simpson}.

Throughout, we refer to sources as mildly obscured if the circumnuclear column density is below $<10^{22}$ cm$^{-2}$, moderately obscured for $10^{22}$ cm$^{-2}< N_H < 10^{23}$ cm$^{-2}$, heavily obscured for 10$^{23}$ cm$^{-2} < N_H < 10^{24}$ cm$^{-2}$, and Compton-thick if the column density exceeds $1.25\times10^{24}$ cm$^{-2}$. We adopt a cosmology of H$_{0}$=70 km Mpc $^{-1}$ s$^{-1}$, $\Omega_M$=0.27, and  $\Omega_\Lambda$=0.73.

\renewcommand{\thefootnote}{\arabic{footnote}}

\section{Sample Selection}
The Sy2 sample and Type 2 QSO candidate sample are both selected from SDSS based on their [OIII] 5007 \AA\ flux and luminosity, with full details given in \citet{me} and \citet{jj}. Sy2s are identified using the BPT diagram \citep{bpt}, a line ratio diagnostic plot of [NII]/H$\alpha$ vs. [OIII]/H$\beta$ that separates star-forming from active galaxies \citep{kewley,kauff}. The 20 objects lying in the Sy2 locus of the BPT diagram with the highest [OIII] flux ($>4\times10^{-14}$ erg s$^{-1}$ cm$^{-2}$) with $z<0.15$ from SDSS Data Release 4 \citep{dr4} comprise the parent sample. Of these 20, we were awarded {\it XMM-Newton} observing time for 15, and another 2 objects were available in the archive.

The Type 2 quasar candidates are selected following the procedures of \citet{zakamska} who identified sources at redshifts $0.3 < z < 0.83$ lacking broad emission lines yet having strong narrow emission lines where the luminosity of the [OIII] line exceeded 10$^8$ L$_{\sun}$. More recently, \citet{reyes} used the same technique on Data Release 6 of the SDSS \citep{dr6}, which expanded the \citet{zakamska} sample four-fold and included objects at $z<0.3$, garnering 887 Type 2 QSO candidates. \citet{jj} found 71 of these sources in the {\it XMM-Newton} and {\it Chandra} archive, with 54 having enough counts for at least a crude spectral fitting. One of these objects, SDSS J123843.43+092736.6, is part of the Sy2 sample.

Though the parent Sy2 and QSO2 samples are unbiased with respect to their X-ray properties, the detailed modeling here requires relatively good signal-to-noise. We focus on objects that have enough detected photons to bin by at least 15 counts per bin. These sub-samples are therefore neither complete nor unbiased: we may be missing a significant fraction of the Compton-thick sources from the parent samples that have weak observed X-ray emission. We therefore refrain from extrapolating our results to the general AGN population.  We note, however, that of the Sy2s from \citep{me} not included in this analysis, all show evidence of heavy obscuration (high Fe K$\alpha$ EW and/or low L$_{\rm 2-10 keV}$/L$_{\rm [OIII]}$ compared with Sy1s). A little over half of the Type 2 QSOs not analyzed here are marked as likely Compton-thick in \citet{jj} since the 1$\sigma$ confidence interval on the simulated $N_{H}$ values exceed 1.6$\times10^{24}$ cm$^{-2}$.

Nine of the 17 [OIII]-selected Sy2s with X-ray coverage have an adequate number of counts for more detailed spectral fitting and are examined in this work (Table \ref{sy2_sample}). Twelve of the 54 Type 2 QSOs have enough counts to be useful for our analysis, though we exclude the Sy2 (SDSS J123843.43+092736.6) and one source (SDSS J141120.52+521210.0, a.k.a. 3C 295) which is at the center of a galaxy cluster and has prominent hot spots which complicates correctly modeling emission from the central engine. The 10 Type 2 QSO candidates used in this analysis are listed in Table \ref{qso2_sample}. We note that 3 of these sources have both {\it XMM-Newton} and {\it Chandra} observations, but as noted in \citet{jj}, there is extended X-ray emission due to radio jets (J083454.89+553411.1) or star formation (SDSS J090037.09+205340.2 and SDSS J091345.48+405628.2), so we utilize the {\it Chandra} data to isolate the quasar-only emission.

\begin{deluxetable}{lrlrrrl}

\tablewidth{0pt}
\tablecaption{\label{sy2_sample} [OIII]-selected Sy2 Sample}
\tablehead{ \colhead{Source} & \colhead{RA} & \colhead{Dec} & \colhead{$z$} & \colhead{{\it XMM-Newton} ObsID}  \\
& \colhead{J2000} & \colhead{J2000}}

\startdata

Mrk 0609 & 03 25 25.4 & -06 08 37 & 0.034 & 0402110201 \\

IC 0486  & 08 00 21.0 & 26 36 49 & 0.027 & 0504101201 \\

SDSS J082443.28+295923.5 & 08 24 43.3 & 29 59 24 & 0.025 & 0504102001 \\

CGCG 064-017 & 09 59 14.8 & 12 59 16 & 0.034 & 0504100201 \\

SBS 1133+572 & 11 35 49.1 & 56 57 08 & 0.051 & 0504101001 \\

Mrk 1457     & 11 47 21.6 & 52 26 58 & 0.049 & 0504101401 \\

SDSS J115704.84+524903.6  & 11 57 04.8 & 52 49 04 & 0.036 & 0504100901 \\

SDSS J123843.43+092736.6 & 12 38 43.4 & 09 27 37 & 0.083 & 0504100601 \\

CGCG 218-007 & 13 23 48.5 & 43 18 04 & 0.027 & 0504101601 \\

\enddata
\tablenotetext{\dagger}{The 9 out of 17 [OIII]-selected Sy2s from \citet{me} with adequate signal-to-noise for the more complex modeling used in this work.}
\end{deluxetable}

\begin{deluxetable}{lrlrrrll}

\tablewidth{0pt}
\tablecaption{\label{qso2_sample} [OIII]-selected Type 2 Quasar Candidate Sample}
\tablehead{ \colhead{Source} & \colhead{$z$} & \colhead{Observatory} & \colhead{ObsID}}

\startdata

SDSS J083454.89+553411.1 & 0.241 & {\it Chandra}    & 04940      \\ % done - no XMM, extended X-ray (jets)

SDSS J083945.98+384319.0 & 0.425 & {\it XMM-Newton} & 0502060201 \\ % done

SDSS J090037.09+205340.2 & 0.236 & {\it Chandra}    & 10463      \\ % done - no XMM, extended X-ray (sf)

SDSS J091345.48+405628.2 & 0.441 & {\it Chandra}    & 10445      \\ % good - no XMM, extended X-ray (sf) 

SDSS J093952.74+355358.0 & 0.137 & {\it XMM-Newton} & 0021740101 \\ % done

SDSS J103408.59+600152.2 & 0.051 & {\it XMM-Newton} & 0306050701 \\ % done

SDSS J103456.40+393940.0 & 0.151 & {\it XMM-Newton} & 0506440101 \\ % done

SDSS J122656.40+013124.3 & 0.732 & {\it XMM-Newton} & 0110990201 \\ % done

SDSS J134733.36+121724.3 & 0.120 & {\it Chandra}    & 00836      \\ % done

SDSS J164131.73+385840.9 & 0.596 & {\it XMM-Newton} & 0204340101 \\ % done

\enddata
\tablenotetext{\dagger}{The 10 out of 71 [OIII]-selected Type 2 quasar candidates from \citet{jj} with adequate signal-to-noise for the more complex modeling used in this work.}
\end{deluxetable}

\section{Analysis}
The X-ray data are reduced with {\it xassist} version 0.9993 \citep{xassist}.\footnote{http://xassist.pha.jhu.edu} This task processes the raw events files, calling the relevant routines from XMMSAS and CIAO for the {\it XMM-Newton} and {\it Chandra} observations, respectively. The data are filtered to omit time periods of background flaring. Spectra and response files are extracted for user-inputted sources (i.e., at the coordinates of the [OIII]-selected Sy2s and Type 2 QSO candidates), with background spectra extracted from an annulus around the object of interest, or in the case of crowded fields, from a circular aperture within a source free region nearby. For the sources with {\it XMM-Newton} observations, the PN, MOS1, and MOS2 spectra are fit simultaneously.

\subsection{Modeling the Spectra}
The spectra are initially fit in XSpec with the models from \citet{spherical} which self-consistently describes Compton scattering and fluorescent line emission within a spherical obscuring medium. We include a model component for scattering of the AGN continuum off a distant optically thin medium if the data require it. Although the \citet{spherical} model assumes full global covering, leakage of the intrinsic continuum of a few percent or less does not significantly break the self-consistency of the model.

We also fit the MYTorus model to the spectra, which we run in both the coupled and decoupled modes as described in detail in \citet{yaqoob}; in the latter mode we are able to derive separate line-of-sight column densities (N$_{\rm H,Z})$ from the global average (N$_{\rm H,S}$). With both models, we test for the presence of thermal emission, which steepens the spectrum, by adding an {\it apec} model component \citep[see][for a more thorough discussion of the interplay between thermal and non-thermal emission for Sy2s in general]{me12}. An initial best fit photon index ($\Gamma$) exceeding 2 likely indicates that thermal emission impacts the observed soft X-ray spectrum, where adding an {\it apec} component better constrains our modeling of the AGN continuum.

\subsubsection{Spherical Distribution}
\citet{spherical} performed a series of Monte Carlo simulations to emulate Compton scattering of photons injected from a central X-ray point source into a spherical distribution of obscuring matter, with photoelectric cross sections from \citet{verner2} and cosmic abundances from \citet{abund}. The obscuring medium is assumed to be neutral. This model self-consistently treats fluorescent line emission, using the fluorescent yields of \citet{bambynek}. The total elemental abundances and iron abundances are free parameters, but we leave these frozen at solar. We also include model components for scattering of the intrinsic continuum off of distant material, as well as a Gaussian convolution model for velocity broadening of the Fe K complex ($\Sigma_E = \sigma_L$($E$/6 keV)$^{\alpha}$, with $\alpha$ fixed at 1, i.e., the velocity width is independent of energy), if required by the spectrum. The model is then
\begin{equation}
{\rm model = const} \times N_{\rm H,Gal} \times (\Sigma_E \times {\rm spherical} + f_{\rm scat}\times{\rm pow} + {\rm thermal}),
\end{equation}
where the first constant accounts for the cross-calibration among the {\it XMM-Newton} PN, MOS1, and MOS2 detectors, N$_{H\rm ,Gal}$ is the (fixed) Galactic absorption, and $f_{\rm scat}$ is the fraction of the AGN continuum that is scattered into our line of sight from a distant optically thin medium.
In XSpec ``speak,'' this translates into
\begin{equation}
\begin{split}
{\rm model = const * phabs * (gsmooth * atable\{sphere0708.fits\} + apec +}\\
{\rm const *zpow)}.
\end{split}
\end{equation}
If scattering off a distant medium is required by the data (i.e., the {\it const * zpow} components are needed to adequately model the spectra), the photon index ($1 \leq \Gamma \leq 3$) and powerlaw normalization are tied to the values from the spherical model. In this set-up, it is the same AGN continuum that is reprocessed by the absorber that leaks through the obscuration to be scattered off distant matter that ultimately enters our line of sight. The normalization of the thermal model is an independent parameter.

\subsubsection{MYTorus Model}
The MYTorus model \citep{mytorus} assumes that an isotropic emitting X-ray source is surrounded by a neutral, uniform absorbing toroidal medium with a half opening angle of 60$^{\circ}$ ($\Delta\Omega$/(4$\pi$) = 0.5). This medium has the cosmic abundances of \citet{abund} and photoelectric cross sections from \citet{verner1} and \citet{verner2}. The inclination angle of this system ($\theta_{obs}$) can range from 0$^{\circ}$ (face-on) to 90$^{\circ}$ (edge-on); angles below 60$^{\circ}$ do not intercept the torus.

When a photon is injected from the corona into this medium, it can pass through unimpeded, be absorbed, or undergo one or more Compton-scatterings, where the probabilities of these outcomes depend principally on the energy of the photon and the column density of the absorber. Within the circumnuclear material, absorption of photons can incite fluorescent line emission, which can escape the medium without scattering, forming the ``core'' of the observed emission line, or be scattered, resulting in a ``Compton-shoulder.'' MYTorus models the ``zeroth-order continuum'' (i.e., transmitted component), Compton scattered component and fluorescent line emission self-consistently by using pre-defined tables for a power law input spectrum realized from a suite of detailed Monte-Carlo simulations. These simulations include the K$\alpha$ transitions of Fe and Ni and the K$\beta$ transition of Fe, but the Ni K$\alpha$ line is not included in the spectral-fitting model.

The three model components are fit simultaneously in XSpec, where the power law index ($1.4 \leq \Gamma \leq 2.6$) and normalization, line of sight column density (N$_{\rm H,Z}$), inclination angle ($\theta_{obs}$) and redshift ($z$) are tied together. The relative normalizations among the zeroth-order continuum, Compton scattered component and fluorescent line emission may vary due to deviations from model assumptions (different elemental abundances and/or torus half opening angle, an ionized rather than neutral medium) and time delays between the transmitted and scattered component and the scattered component and line emission. Such differences may not be apparent in the spectrum since it may be observed on scales where the deviations are averaged over, especially between the scattered and fluorescent line emission. These unknowns are embodied in the relative normalizations $A_S$ and $A_L$, for the Compton scattered and line emission respectively, which are parametrized by a multiplicative constant in XSpec. We keep $A_{S}$ and $A_{L}$ tied together in order to preserve the self-consistency of the Compton-scattered continuum and the Fe K$\alpha$ line emission, within the context of the assumptions of the model (such as solar Fe abundance). Initially, $A_S$ (=$A_L$) is set to 1, i.e., the scattered and line emission have the same normalizations as the zeroth-order continuum. If the data require it ($f$-test probability $>3\sigma$), we untie $A_S$ (= $A_L$) from the zeroth-order continuum, but only accept the fit if $A_S$ is constrained.  As $A_S$ and $A_L$ parametrize our ignorance of different possible physical scenarios, and disentangling one possibility from another requires additional information that we lack, we only list the best-fit values for reference and refrain from drawing any physical interpretation from these normalizations.

In addition to the MYTorus zeroth-continuum, Compton scattered, and fluorescent line emission models, we include model components for Galactic absorption (N$_{\rm H,Gal}$), a multiplicative factor for the {\it XMM-Newton} PN, MOS1, and MOS2 detector cross-normalizations, a Gaussian convolution to the fluorescent line spectrum for velocity broadening, and possibly scattering of the continuum from an optically thin medium and/or a thermal component (if the data require these components), similar to the spherical absorption model. Our model in coupled mode, i.e. where the line-of-sight and global average column densities are equal, is then
\begin{equation}
\begin{split}
{\rm model = const} \times N_{\rm H,Gal} \times ({\rm pow} \times{\rm line-of-sight\ extinction}  + \\ 
A_S \times {\rm scattered} + A_L \times \Sigma_E \times{\rm line} +  \\
f_{\rm scat}\times {\rm pow} + {\rm thermal}),
\end{split}
\end{equation}
where the first constant, N$_{\rm H,Gal}$, $\Sigma_E$, and $f_{\rm scat}$ are the same as defined for the spherical geometry obscuration model, and $A_S$ and $A_L$ are defined above. Again, the photon indices, power law normalizations, inclination angles, and circumnuclear column densities are tied among all model components, while the thermal component normalization, if present, is independent.
In XSpec, the relevant commands are
\begin{equation}
\begin{split}
{\rm model = const * phabs * (zpow * etable\{mytorus\_Ezero\_v00.fits\} + }\\
{\rm const*atable\{mytorus\_scatteredH200\_v00.fits\} + }\\
{\rm const* gsmooth * atable\{mytl\_V000010nEp000H200\_v00.fits\} } + \\
{\rm const *zpow + apec)}.
\end{split}
\end{equation}

Here, {\it mytorus\_Ezero\_v00.fits}, {\it mytorus\_scatteredH200\_v00.fits}, and {\it mytl\_V000010nEp000H200\_v00.fits} are tables for the line-of-sight extinction, Compton scattered component, and fluorescent line emission, respectively, with an exponential cutoff energy of 200 keV for the latter two components; since we work in an energy range far below the termination energy, the choice of cutoff energy has a negligible effect on our results.

We can model the Compton-scattered continuum from the global matter distribution, which may in general be unrelated to the line-of-sight column density, by decoupling the inclination angle and column density of the MYTorus Compton-scattered continuum component. Such a scenario (which is no longer restricted to a toroidal geometry) can be realized if a portion of the observed X-ray emission is viewed after reflection off the back side of the obscuring medium without further interaction with this matter, mimicking a patchy medium where this reflected emission is seen through a hole. The spectrum can in fact be dominated by this unobscured reflection component. In this decoupled case, the inclination angle for the zeroth-order continuum is fixed at 90$^{\circ}$ so that its associated column density represents the line of sight column density only (N$_{\rm H,Z}$). There is a Compton scattered component for the far-side reflection, with the inclination angle fixed at 0$^{\circ}$ as it mimics face-on reflection, and its associated column density (N$_{\rm H,S}$) represents the global average. Like the coupled case, the normalizations and photon indices among the zeroth order continuum, the Compton scattered component, and scattering off a distant medium are tied together.

\subsubsection{Line-of-sight And Global Column Densities}
The absorption measured with the spherical model, N$_{\rm H,sph}$ refers to the radial distribution of the obscuring material and so the line-of-sight column density from the spherical model is directly comparable to the line-of-sight column density, N$_{\rm H,Z}$, measured by MYTorus. When MYTorus is run in the default (i.e., coupled) mode, the global column density, N$_{\rm H,S}$, is the same as the line-of-sight column density. As noted above, MYTorus can disentangle the global average column density from the line-of-sight N$_{H}$ when run in decoupled mode. For a toroidal distribution, the global column density is the angle average of the line-of-sight column density from the top to the bottom edge of the torus.  When assuming an isotropic distribution of input photons, the equatorial column of the torus, N$_{\rm H,S}$ is then equivalent to ($4/\pi$)$\times$N$_{\rm H,sph}$ \citep[see][]{mytorus}.

\section{Results}
If we are able to achieve a good fit to the spectra using the spherical absorption model of \citet{spherical}, then we accept this model as the preferred model since the added complexity of a toroidal geometry will not allow meaningful constraints on the inclination angle. In some cases where a decent fit (reduced $\chi^2 <2$) is obtained with the spherical absorption model, MYTorus provides a better fit of the Fe K complex, in which case we accept the MYTorus model fits as a better representation of the circumnuclear obscuration; indeed, these fits are a statistically significantly improvement (i.e., $f$-test probability exceeds 3$\sigma$). Though the spherical model may be preferred, this does not necessarily rule out the possibility that the absorber has a more complex geometry, but rather indicates that this latter possibility is not required to accomodate the spectrum. For the objects where the MYTorus fit is preferred, we inspect the results of fitting in coupled and decoupled mode. We accept the decoupled fit only if we obtain constraints on the fit parameters (namely $A_{S}$, N$_{\rm H,Z}$, and N$_{\rm H,S}$ which can be poorly determined in decoupled mode if there is a weak or absent iron line) or if the Fe K complex is better modeled in decoupled mode. We also note that even with the addition of the {\it apec} model component, the upper limit on the best fit $\Gamma$ is beyond the allowable model range (3 for the spherical model, 2.6 for MYTorus) for a handful of objects; in these cases, we quote the 90\% confidence lower limit. 

For several objects (Mrk 0609, SDSS J115704.84+524903.6, SDSS J123843.43+092736.6, SDSS J083454.89+553411.1, SDSS J103408.59+600152.2), the reduced $\chi^2$ gives a high null hypothesis probability value ($>99\%$) which may indicate a poor fit. However, with the exception of Mrk 0609, inspection of the fit residuals shows no systematic offsets which suggests that the preferred model is adequate and that a significant model component is not missing. The nature of the residuals do not compromise our primary objective of measuring reliable column densities for the circumnuclear matter. As we discuss below, Mrk 0609 has a typical type 1 X-ray spectrum, lacking X-ray evidence of circumnuclear obscuration. As the spherical absorption and MYTorus models require the presence of an X-ray reprocessor with a minimum column density of 10$^{20}$ cm$^{-2}$ and 10$^{22}$ cm$^{-2}$, respectively, they are not appropriate for this object; a power-law model is sufficient for fitting this source \citep{me}.

The spectral fitting results are reported in Tables \ref{sy2_sph_fits}-\ref{sy2_myt_decoup_fits} and \ref{qso2_sph_fits}-\ref{qso2_myt_decoup_fits}, while the [OIII], Fe K$\alpha$ and intrinsic, AGN-only rest-frame 2-10 keV luminosities are listed in Tables \ref{sy2_lums} and \ref{qso2s_lums} for the Sy2 and QSO2 sample, respectively. All [OIII] luminosities, unless otherwise noted, are observed and not corrected for dust extinction as H$\alpha$ falls out of the optical bandpass at $z\sim0.5$ and we are therefore unable to calculate the Balmer decrement (H$\alpha$/H$\beta$) for all objects in our sample. The estimation of the Fe K$\alpha$ luminosity depends on the best-fit model, since the Fe K$\alpha$ line cannot be independently varied in the spherical model. For the MYTorus model, the luminosity is integrated in XSpec around the rest-frame Fe K feature after turning off all model components other than $A_L$, {\it gsmooth} (if included), and {\it MYTorusL}. If a source is best-fit with the spherical model, the rest-frame luminosity of the continuum around the Fe K feature is estimated and then subtracted from the total luminosity within the same energy range, giving the Fe K$\alpha$ luminosity.

\subsection{Sy2s}
\begin{figure}[ht]
\centering

\subfigure[Mrk 0609]{\includegraphics[scale=0.3,angle=270]{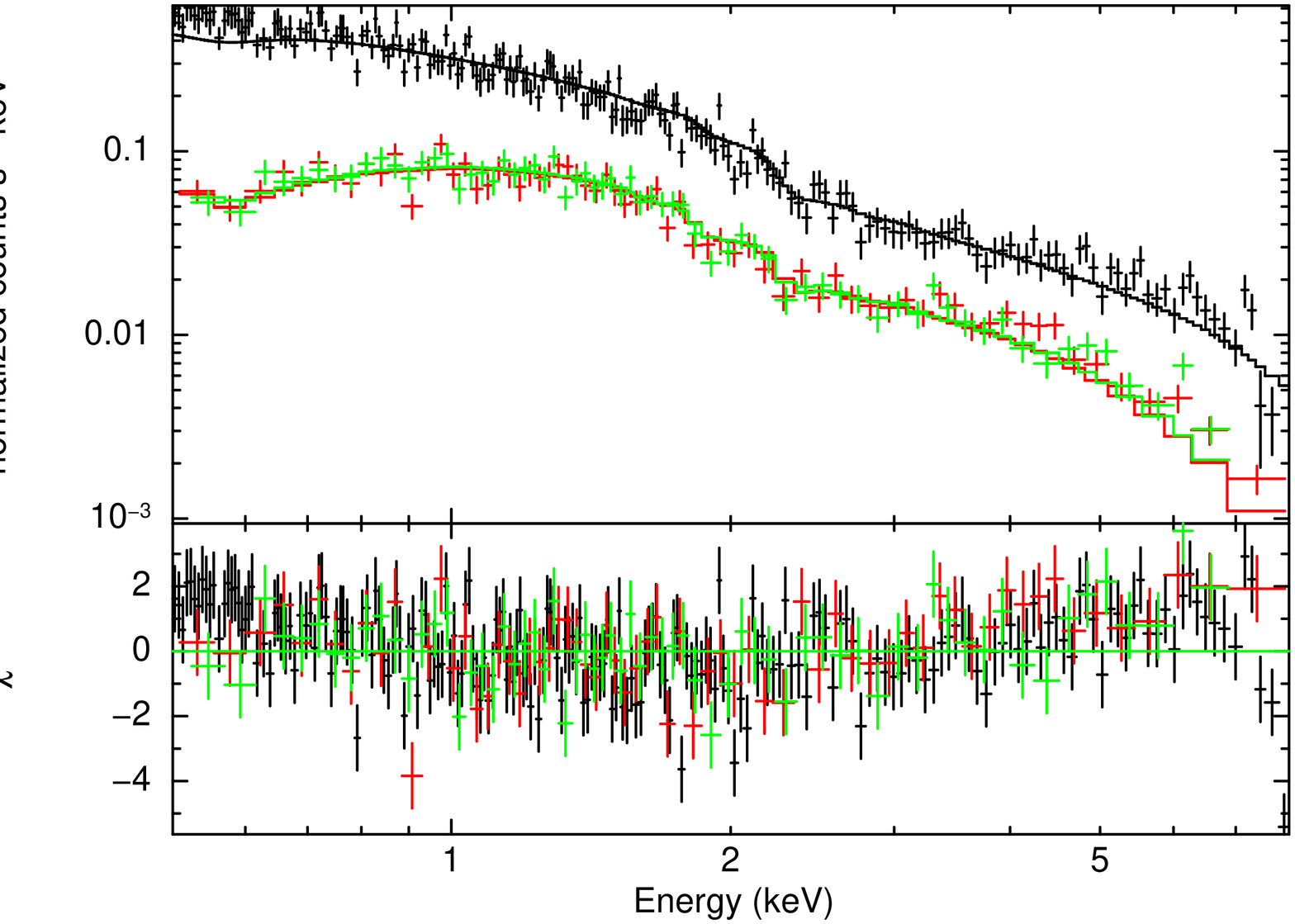}}
\hspace{0.2cm}
\subfigure[CGCG 064-017]{\includegraphics[scale=0.3,angle=270]{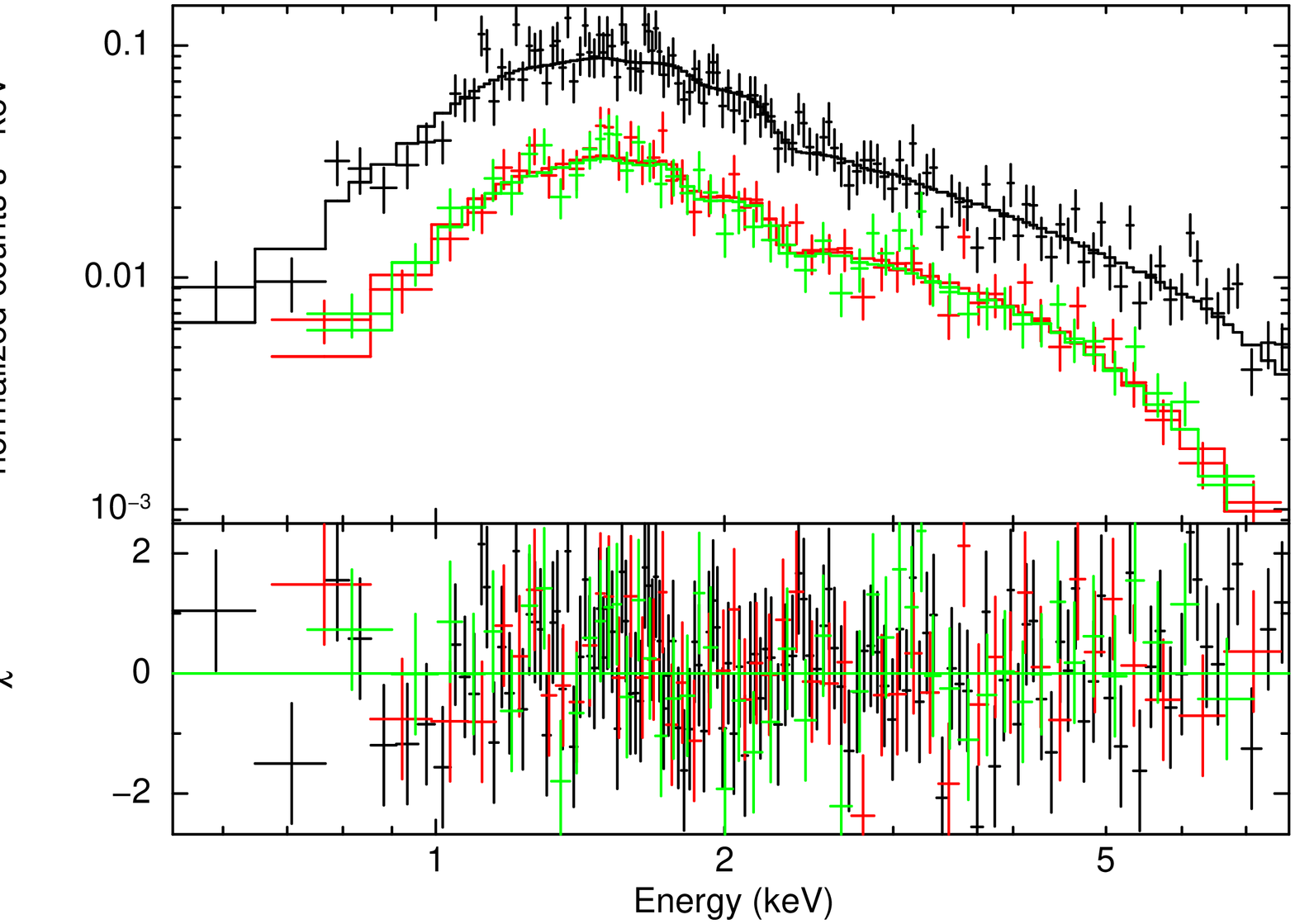}}

\subfigure[SBS 1133+572]{\includegraphics[scale=0.3,angle=270]{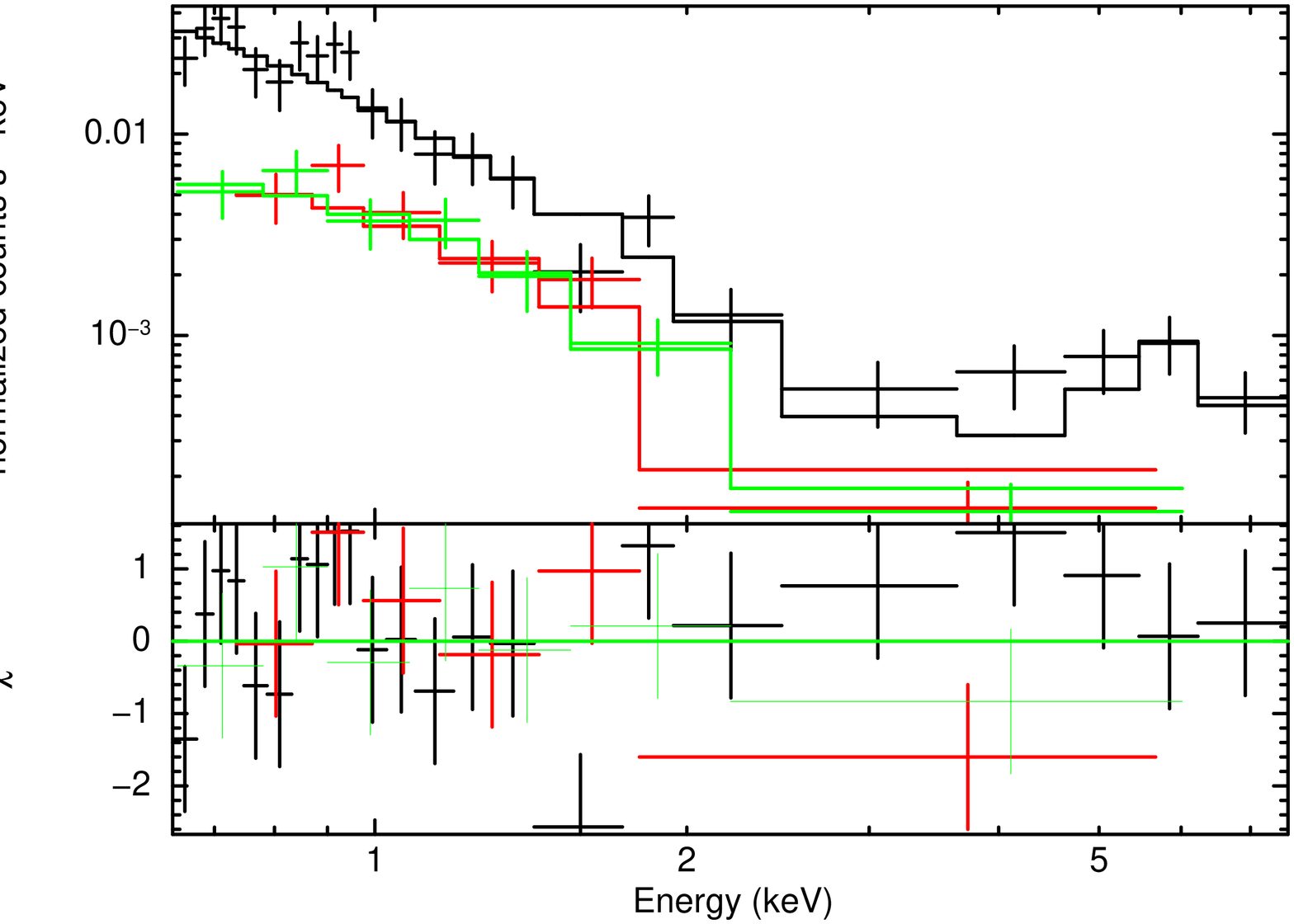}}
\hspace{0.2cm}
\subfigure[Mrk 1457]{\includegraphics[scale=0.3,angle=270]{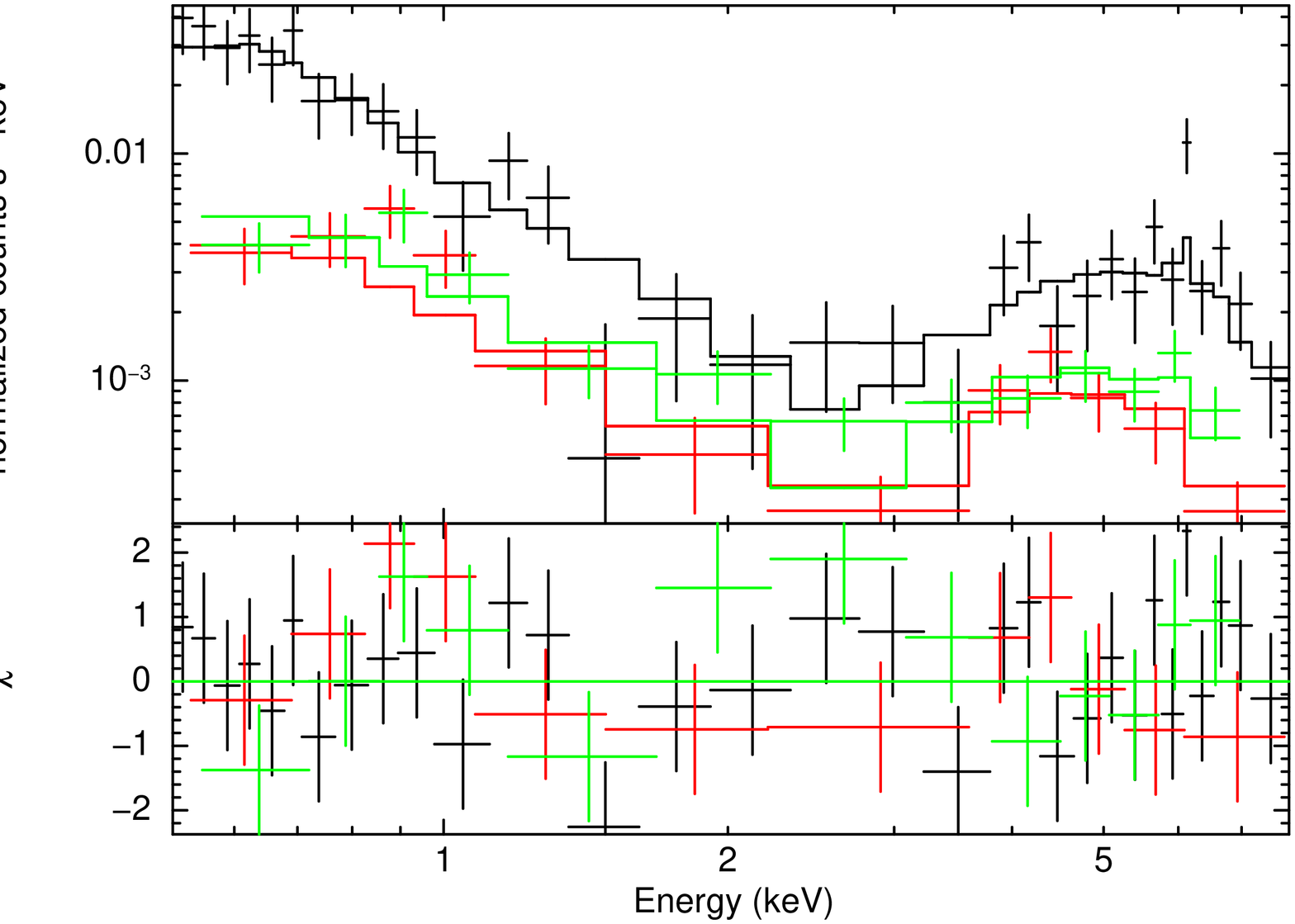}}

\subfigure[SDSS J115704.84+524903.6]{\includegraphics[scale=0.3,angle=270]{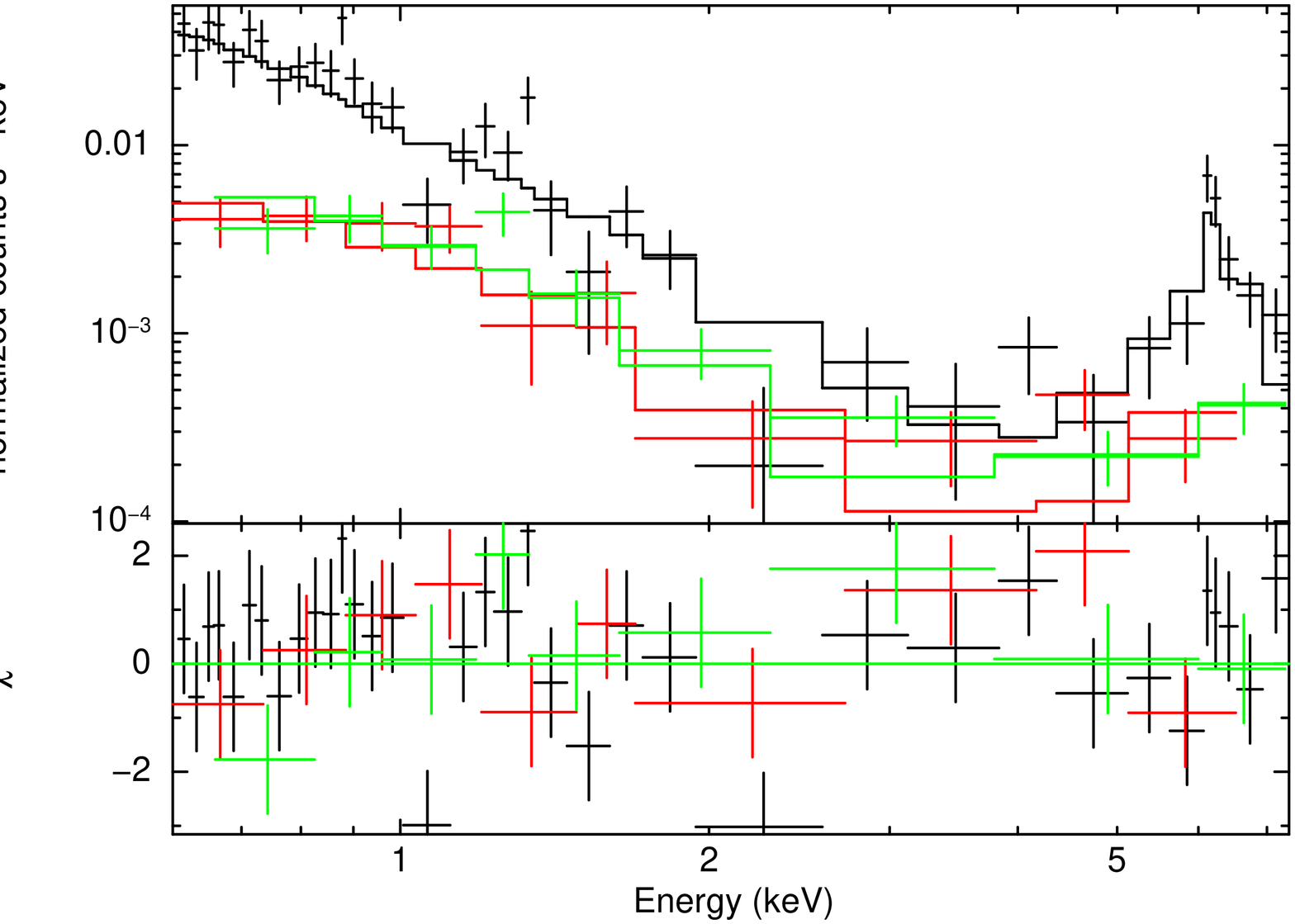}}
\hspace{0.2cm}
\subfigure[SDSS J123843.43+092736.6]{\includegraphics[scale=0.3,angle=270]{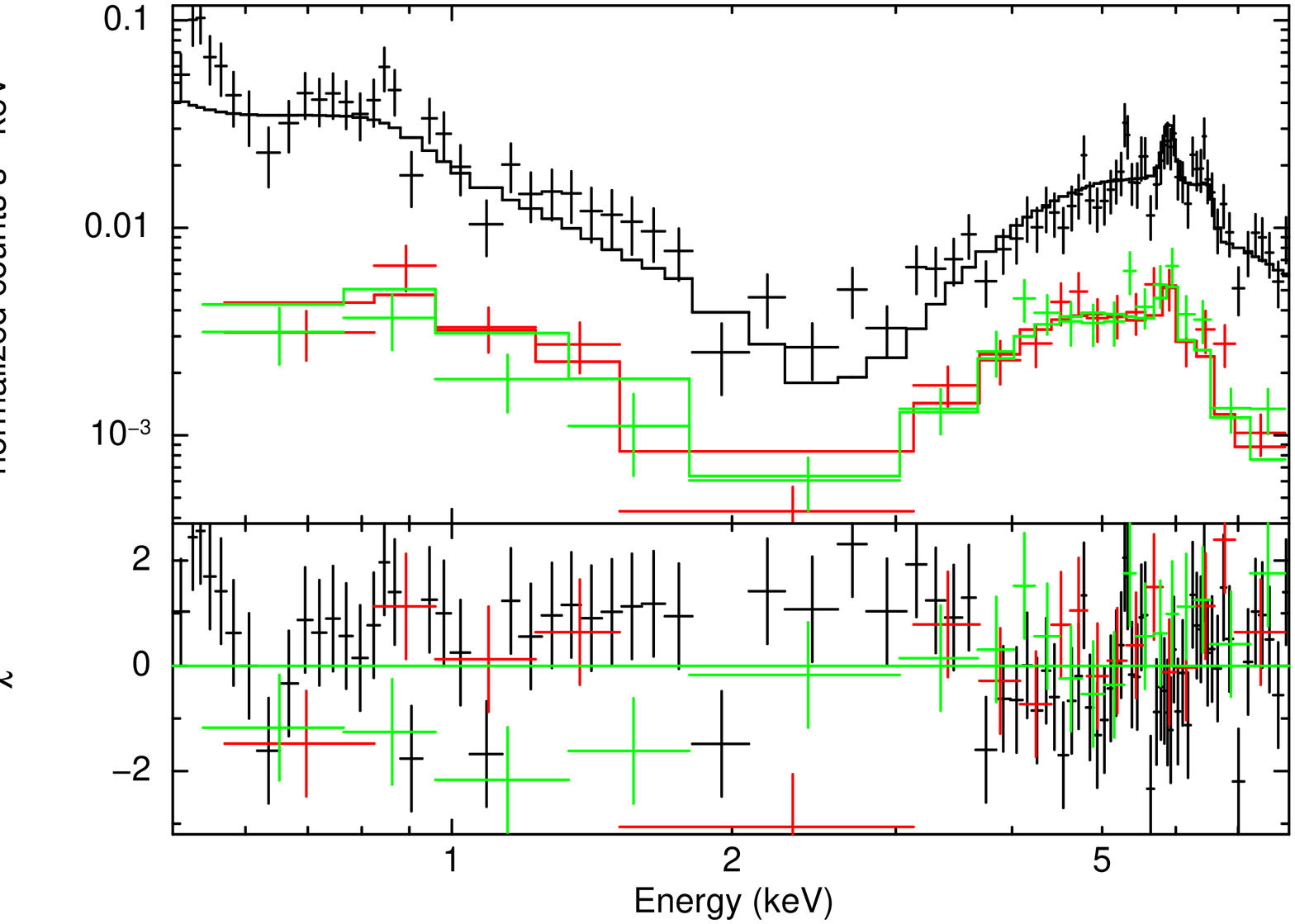}}

\subfigure[CGCG 218-007]{\includegraphics[scale=0.3,angle=270]{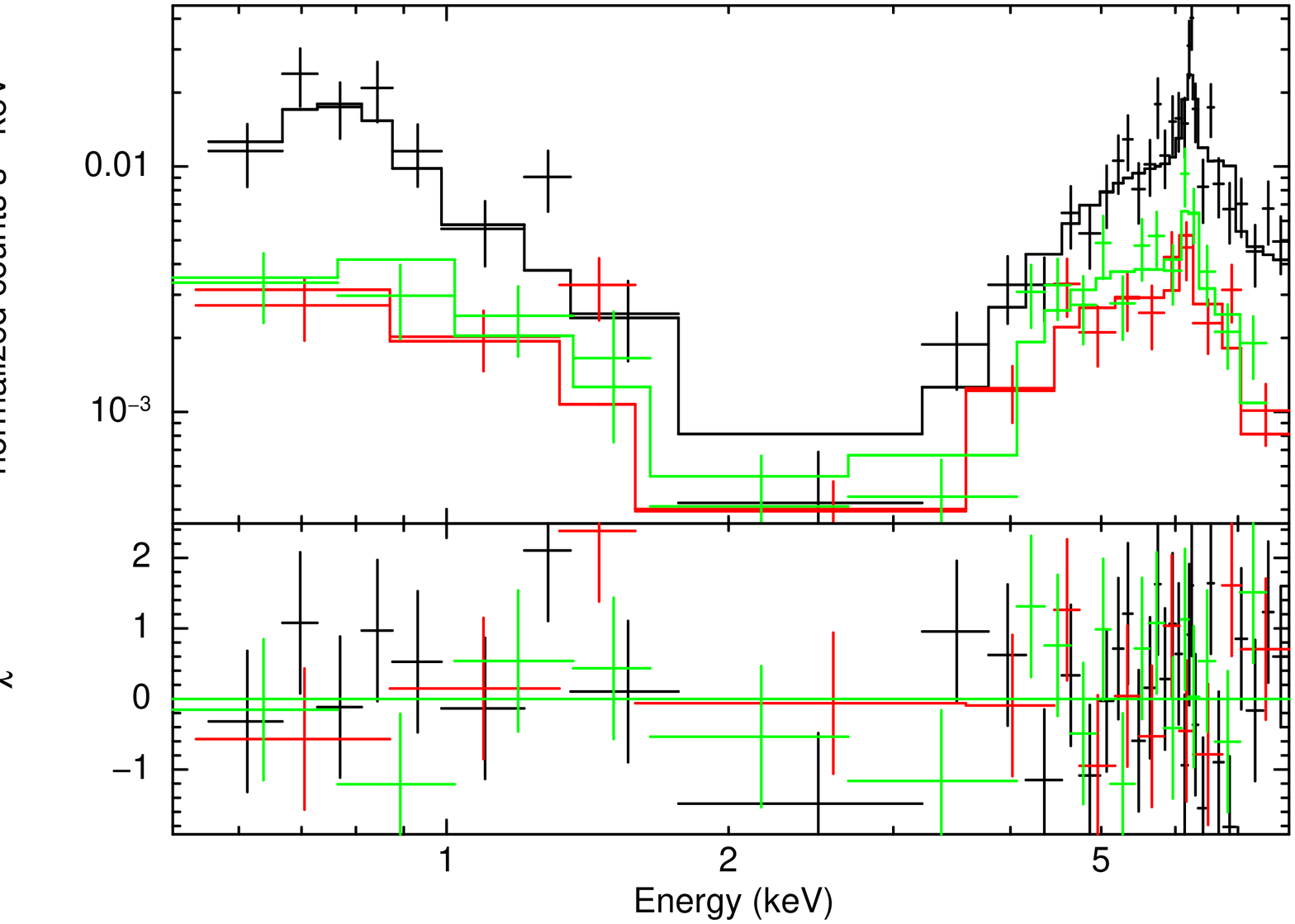}}

\caption[]{\label{sy2_sph_spec}\citet{spherical} spherical absorption model fits to {\it XMM-Newton} spectra of the [OIII]-selected Sy2s from \citet{me}. Black - PN spectrum, red - MOS1 spectrum, green - MOS2 spectrum.}
\end{figure}

Of the 9 [OIII]-selected Sy2s, 7 are best fit with the spherical absorption model, as summarized in Table \ref{sy2_sph_fits} and shown in Figure \ref{sy2_sph_spec}; for none of these sources was the width of {\it gsmooth} constrained and it is therefore not listed in the Table. The Fe K line was also unresolved for these AGN. Three of these had 90\% upper limits in $\Gamma$ beyond the maximum allowable limit of the model and are thus listed in Table \ref{sy2_sph_fits} with their 90\% lower limits. One object, SBS 1133+572, has a steep photon index that pegs at the maximum allowed value. Inspection of its spectra indicates that the soft emission dominates the spectrum while the hard emission ($>2$ keV) is heavily suppressed. However, the signal-to-noise of the spectrum precludes us from obtaining constraints on possible thermal emission by including an {\it apec} component. Conversely, IC 0486 has a shallow fitted photon index. Freezing $\Gamma$ to 1.8 leads to questionable best-fits for $A_S$ (11.1$^{+26.2}_{-2.9}$) and f$_{\rm scat}$ (16.7$^{+6.7}_{-6.9}$), and the line-of-sight and global column densities are nearly equivalent to those from the fit with $\Gamma$ free.

Only one source, CGCG 064-17, shows no evidence of scattering of the AGN continuum off a distant medium and a mildly obscured spectrum ($<10^{22}$ cm$^{-2}$). This source also does not require a thermal component to better constrain the photon index, consistent with \citet{me12} where we reported that adding a thermal model results in an unconstrained fit; we note that in modeling CGCG 218-007 in that work, using a double power law to model the AGN spectrum, a thermal model was unconstrained, yet with the spherical absorption model used here, thermal emission is detected. Three sources are heavily obscured and 2 have best fit N$_{\rm H,Z}$ ranges that are within the Compton-thick regime, SBS 1133+572 and 2MASX J11570483+5249036. As discussed in detail below, one source, Mrk 0609, is unabsorbed.

\begin{figure}[ht]
\centering
\subfigure[IC 0486]{\includegraphics[scale=0.3,angle=270]{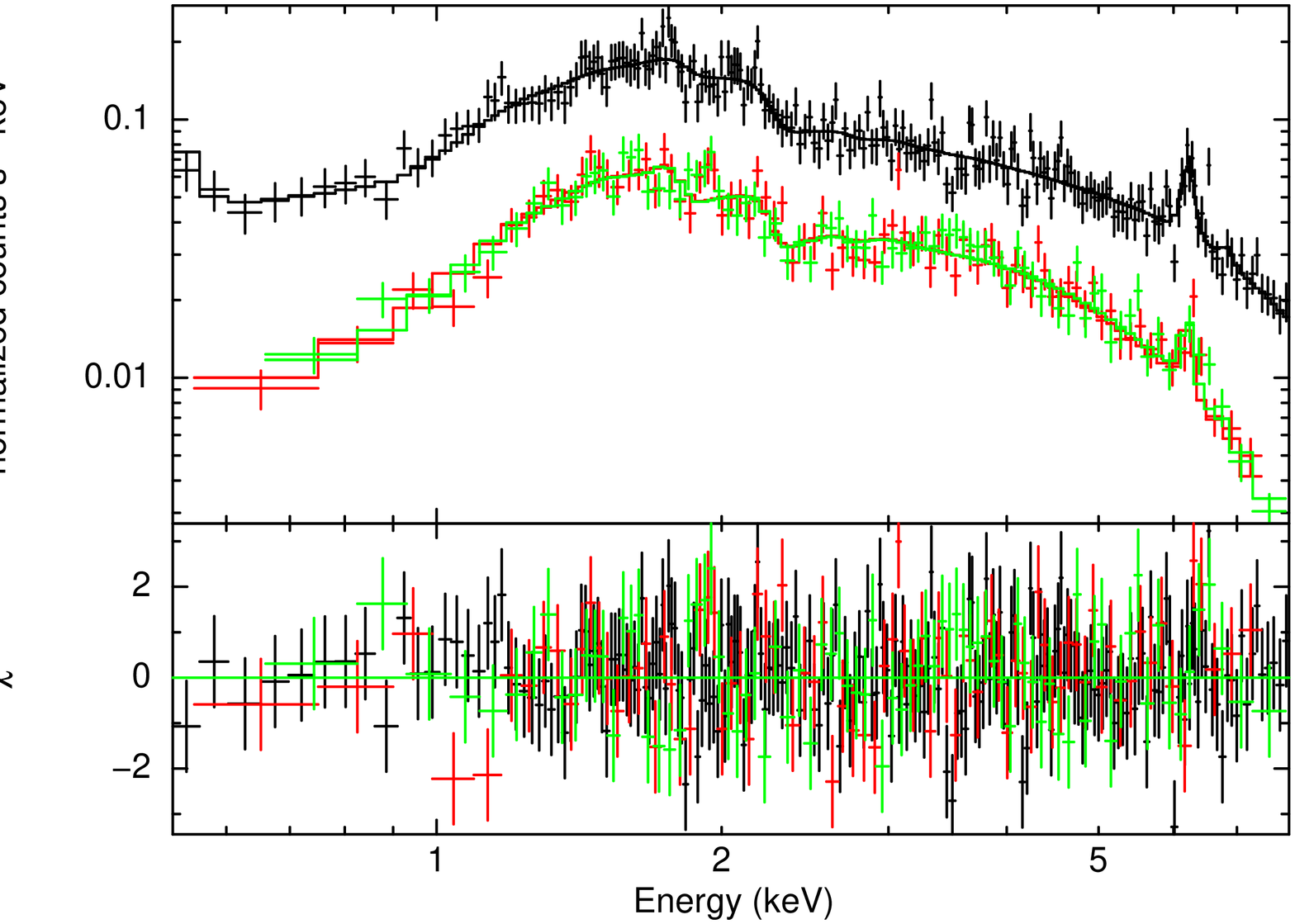}}
\hspace{0.2cm}
\subfigure[SDSS J082443.28+295923.5]{\includegraphics[scale=0.3,angle=270]{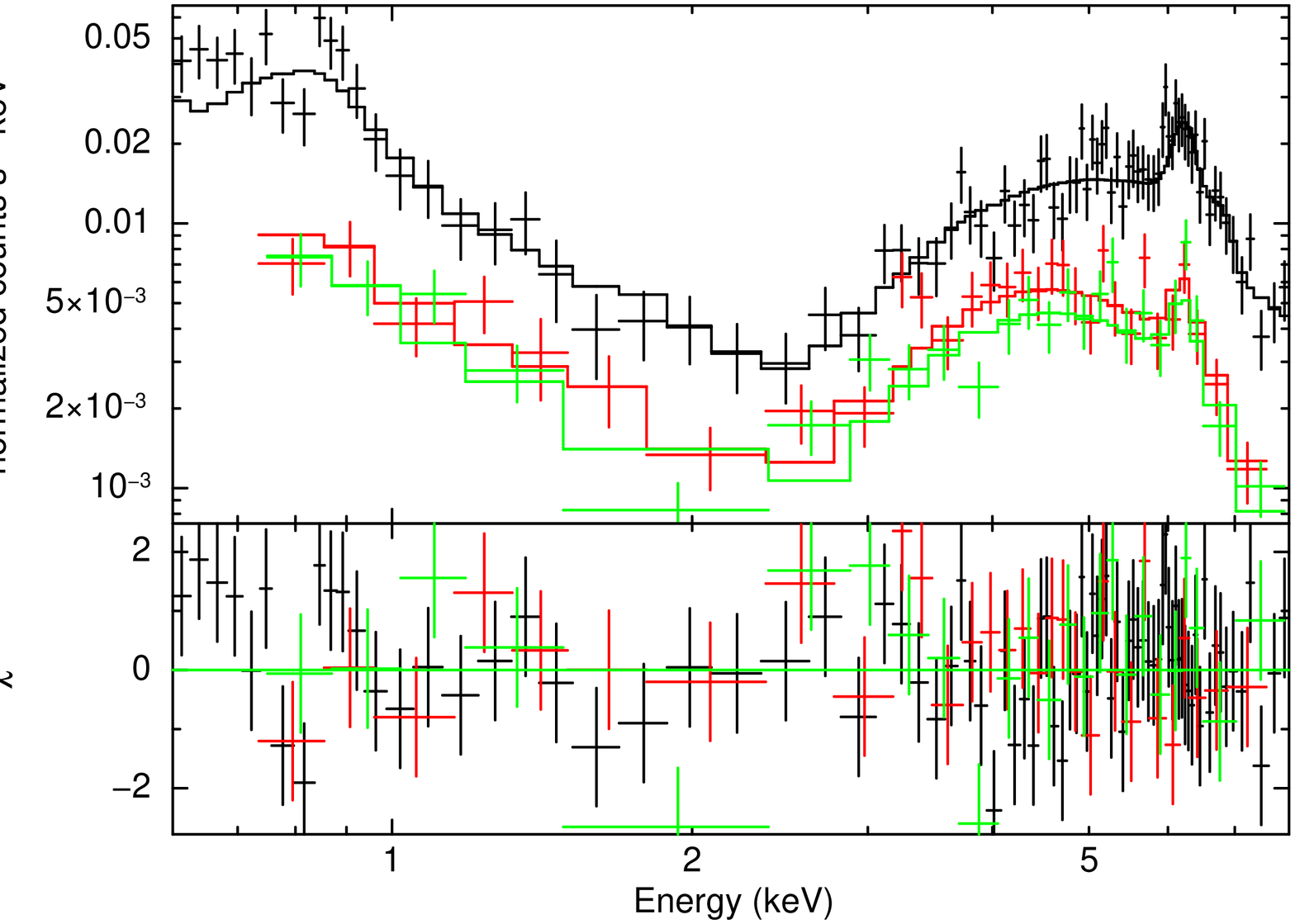}}

\caption[]{\label{sy2_myt_spec_decoup}Decoupled MYTorus fits to {\it XMM-Newton} spectra of the [OIII]-selected Sy2s from \citet{me}. Color coding same as Figure \ref{sy2_sph_spec}.}
\end{figure}

The remaining 2 objects are well fit with MYTorus in decoupled mode (see Figure \ref{sy2_myt_spec_decoup}), with line of sight column densities that differ from the global average by about an order of magnitude (Table \ref{sy2_myt_decoup_fits}).  We find the global obscuration in IC 0486 is high ($\sim10^{23}$ cm$^{-2}$), though not Compton-thick, while the global absorber in 2 MASX J08244333+2959238 permits a Compton-thick solution.

%------------------------
%% Sy2s:
%% Spherical fits
\begin{landscape}
\begin{deluxetable}{llllllr}

\tablewidth{0pt}
\tablecaption{\label{sy2_sph_fits} Spherical Absorption Fits to [OIII]-selected Sy2s}
\tablehead{ \colhead{Source} & \colhead{N$_{\rm H,Gal}$} & \colhead{$\Gamma$} & \colhead{N$_{\rm H,sph}$} & \colhead{$f_{\rm scat}$} & \colhead{kT} & \colhead{$\chi^2$(DOF)}  \\
& \colhead{$10^{22}$ cm$^{-2}$} & & \colhead{$10^{22}$ cm$^{-2}$}  & \colhead{\%} & \colhead{keV}}

\startdata
Mrk 0609 & 0.03 & 1.69$^{+0.03}_{-0.02}$ & $<$0.01 & ... & ... & 513.29(337)  \\

CGCG 064-017 & 0.03 & 1.89$^{+0.07}_{-0.09}$ & 0.76$^{+0.05}_{-0.06}$ & ... & ... & 225.38(227)  \\

SBS 1133+572\tablenotemark{1} & 0.01 & $>$2.84 & 68.7$^{+85.3}_{-33.0}$ & 1.52$^{+3.36}_{-1.33}$ & ... & 32.79(30) \\

Mrk 1457 & 0.02 & $>$1.56 & 26.3$^{+8.8}_{-5.6}$ & 1.86$^{+4.00}_{-1.22}$ & 0.25$^{+0.11}_{-0.08}$ & 57.34(49) \\

J1157+5249 & 0.02 & $>$2.11 & 106$^{+52}_{-32}$ & 0.40$^{+0.87}_{-0.30}$ & 0.24$^{+0.60}_{-0.16}$ & 77.4(47) \\

J1238+0927 & 0.01 & 2.32$^{+0.23}_{-0.22}$ & 36.1$^{+2.4}_{-2.5}$ & 0.57$^{+0.26}_{-0.20}$ & 0.86$^{+0.34}_{-0.21}$ & 170.21(120)  \\

CGCG 218-007 & 0.01 & 2.12$^{+0.50}_{-0.64}$ & 53.8$^{+6.1}_{-8.7}$ & 0.26$^{+0.45}_{-0.16}$ & 0.60$^{+0.28}_{-0.24}$ & 64.33(60) \\

\enddata
\tablenotetext{\dagger}{``...'' indicates that parameter was not included in the spectral fitting.}

\end{deluxetable}
\end{landscape}

%% MYTorus Fits: Decoupled
\begin{landscape}
\begin{deluxetable}{llllllllll}

\tablewidth{0pt}
\tablecaption{\label{sy2_myt_decoup_fits} Decoupled MYTorus Fits to [OIII]-selected Sy2s}
\tablehead{ \colhead{Source} & \colhead{N$_{\rm H,Gal}$} & \colhead{$\Gamma$} & \colhead{N$_{\rm H,Z}$\tablenotemark{1}} & \colhead{N$_{\rm H,S}$\tablenotemark{2}} & \colhead{$\Sigma_L$}
& \colhead{$A_{\rm S,0}$} & \colhead{$f_{\rm scat}$} &  \colhead{kT} & \colhead{$\chi^2$(DOF)}  \\
& \colhead{$10^{22}$ cm$^{-2}$} & & \colhead{$10^{22}$ cm$^{-2}$} & \colhead{$10^{22}$ cm$^{-2}$} & \colhead{keV} &  & \colhead{\%} & \colhead{keV}}

\startdata

IC 0486 & 0.03 & $<$1.43 & 1.21$^{+0.08}_{-0.06}$ & 10.2$^{+23.2}_{-5.34}$ & $<$0.09 & 6.84$^{+8.56}_{-4.43}$ & 7.09$^{+0.80}_{-0.81}$ & ... & 388.69(370)  \\

J0824+2959 & 0.03 & $>$2.08 & 24.6$^{+5.5}_{-4.0}$ & 90.2$^{+186}_{-43.7}$ & 0.12$^{+0.07}_{-0.06}$ & 3.65$^{+1.6}_{-1.29}$ & 0.43$^{+0.51}_{-0.25}$ & 0.74$^{+0.09}_{-0.13}$ & 142.91(129) \\

\enddata
\tablenotetext{\dagger}{``...'' indicates that parameter was not included in the spectral fitting.}
\tablenotetext{1}{\ Line of sight column density, associated with the zeroth order continuum.}
\tablenotetext{2}{Global column density, associated with Compton-scattering off the back wall of the X-ray reprocessor.}

\end{deluxetable}

\end{landscape}

\begin{deluxetable}{llll}

\tablewidth{0pt}
\tablecaption{\label{sy2_lums} Luminosities for [OIII]-selected Sy2s}
\tablehead{ \colhead{Source} & \colhead{L$_{\rm [OIII]}$\tablenotemark{1}} & \colhead{L$_{\rm Fe K\alpha}$\tablenotemark{2}} & \colhead{L$_{\rm 2-10keV,in}$\tablenotemark{3}} \\
& \colhead{L$_{\sun}$} & \colhead{erg s$^{-1}$} & \colhead{erg s$^{-1}$}}

\startdata
Mrk 0609                 & 7.72 & 40.77 & 42.58$^{+0.01}_{-0.01}$ \\

IC 0486                  & 7.30 & 41.21 & 42.77$^{+0.03}_{-0.02}$ \\

SDSS J082443.28+295923.5 & 7.51 & 40.95 & 42.62$^{+0.25}_{-0.31}$ \\

CGCG 064-017             & 7.77 & 40.22 & 42.48$^{+0.04}_{-0.05}$ \\

SBS 1133+572             & 7.88 & 40.27 & 42.72$^{+0.86}_{-0.51}$ \\

Mrk 1457                 & 7.80 & 40.56 & 42.61$^{+0.55}_{-0.65}$ \\

SDSS J115704.84+524903.6 & 7.59 & 40.49 & 42.99$^{+0.10}_{-0.58}$ \\

SDSS J123843.43+092736.6 & 8.39 & 41.84 & 43.97$^{+0.21}_{-0.20}$ \\

CGCG 218-007             & 7.25 & 41.16 & 42.97$^{+0.46}_{-0.59}$ \\

\enddata
\tablenotetext{\dagger}{\ Luminosities reported in log space.}
\tablenotetext{1}{Observed luminosity, not corrected for dust extinction.}
\tablenotetext{2}{Rest-frame luminosity.}
\tablenotetext{3}{Rest-frame, intrinsic (i.e., absorption corrected) luminosity.}
\end{deluxetable}

%%% - End of Sy2 spectra

\subsection{QSO2 Candidates}
\begin{figure}[ht]
\centering

\subfigure[]{\includegraphics[scale=0.3,angle=270]{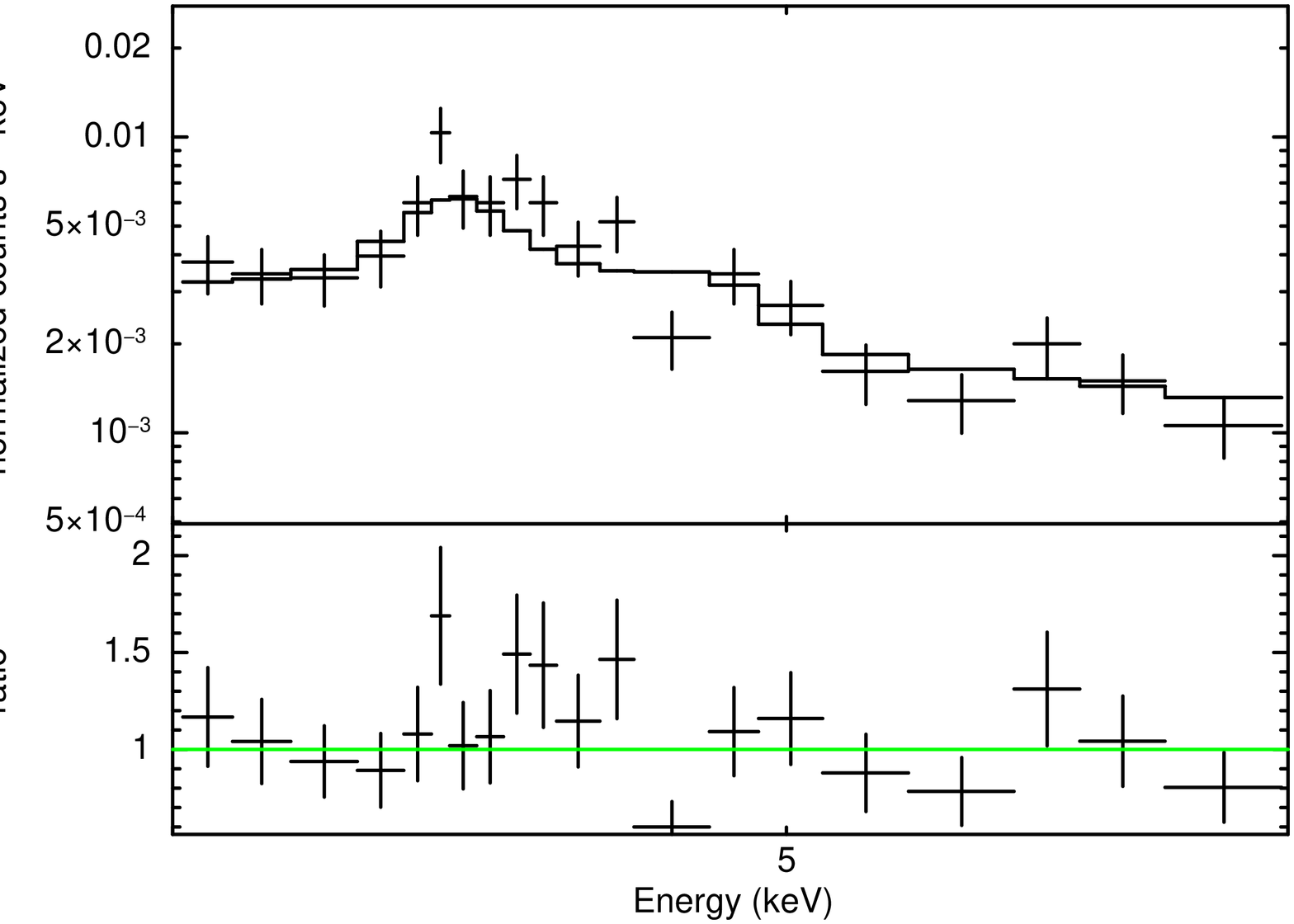}}~
\hspace{0.2cm}
\subfigure[]{\includegraphics[scale=0.3,angle=270]{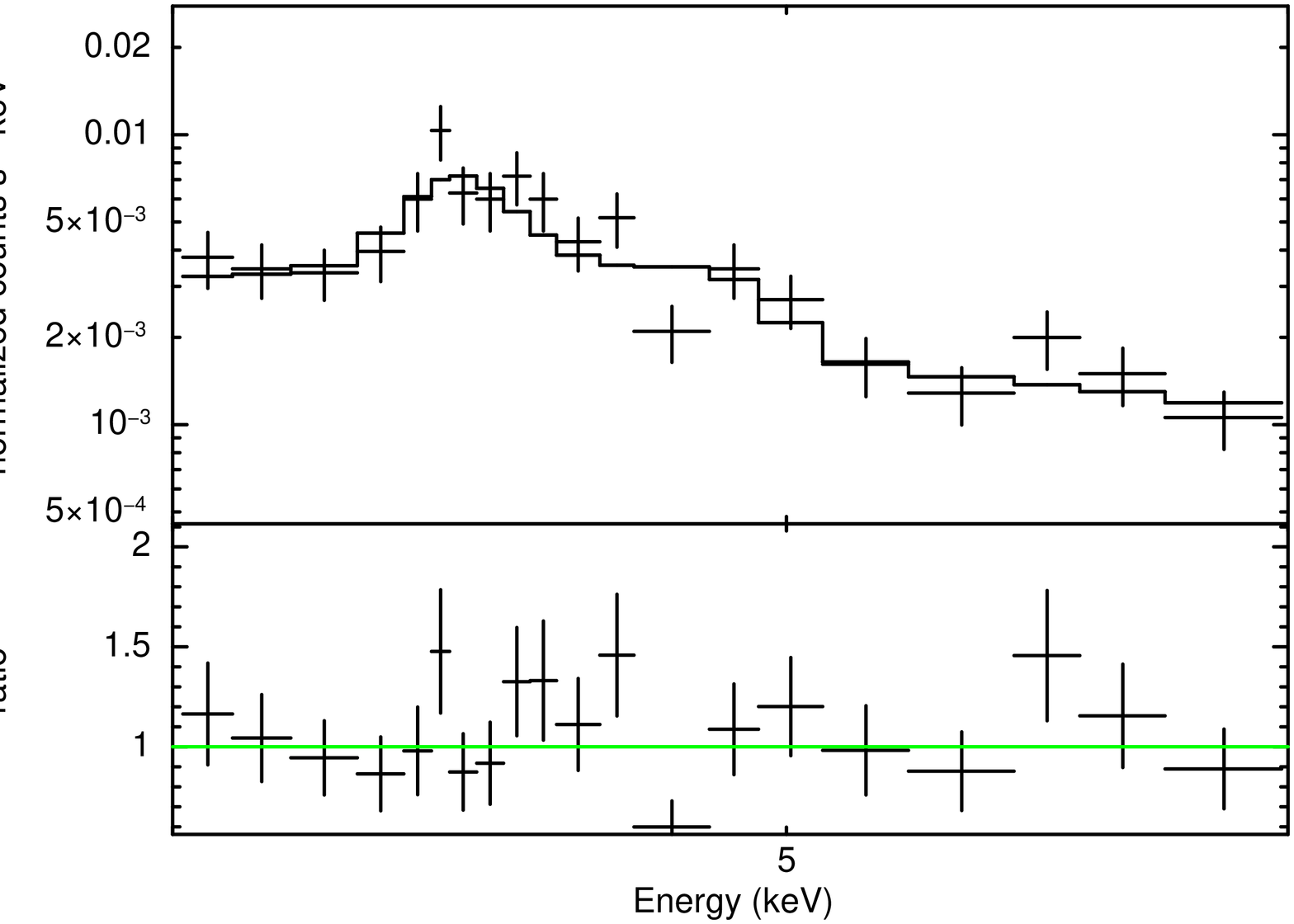}}

\caption[]{\label{fek_j0913} Close-up of the Fe K complex in SDSS J091345.48+405628.2 fit with the (a) spherical absorption model and (b) MYTorus model in decoupled mode, with the ratio between the model and data plotted in the lower panels. Though MYTorus provides a slightly better fit, the improvement is not significant.}
\end{figure}

For one of the 10 QSO2s, SDSS J091345.48+405628.2, both the spherical absorption and MYTorus model in decoupled mode provided an acceptable fit to the data. Though the Fe K complex is slightly better accommodated by the MYTorus model (Figure \ref{fek_j0913}), the improvement is marginal, so we present the results for both model fits. The predicted intrinsic 2-10 keV luminosity is comparable between both models (Table \ref{qso2s_lums}) and the line of sight column densities are consistent within the error bars to be heavily obscured to Compton-thick.  The decoupled MYTorus fit does not constrain the upper limit on the global column density. The combination of heavy obscuration with a relatively high scattering fraction precludes us from determining the geometry of the absorber, though clearly leakage of the intrinsic continuum indicates that the obscuration is non-uniform. In the analysis below, we use the global and line-of-sight column densities measured by the decoupled MYTorus model, noting that N$_{\rm H,Z}$ is consistent with the results from the spherical model. As the possible scattering fraction between the two models shows a relatively wide range, we exclude this source when examining the scattering fraction distributions.

\begin{figure}[ht]
\centering

\subfigure[SDSS J083454.89+553411.1]{\includegraphics[scale=0.3,angle=270]{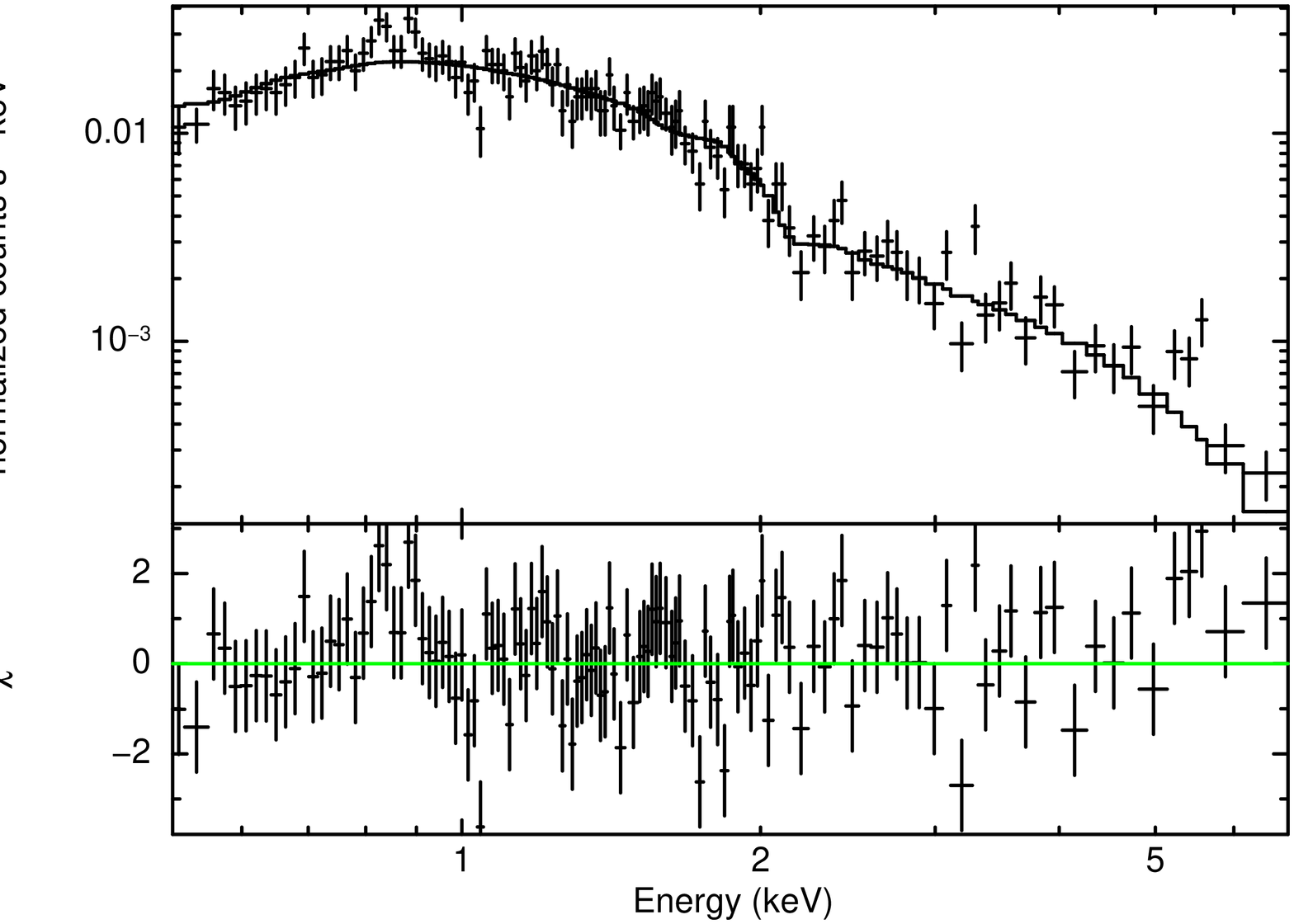}}~
\hspace{0.2cm}
\subfigure[SDSS J083945.98+384319.0]{\includegraphics[scale=0.3,angle=270]{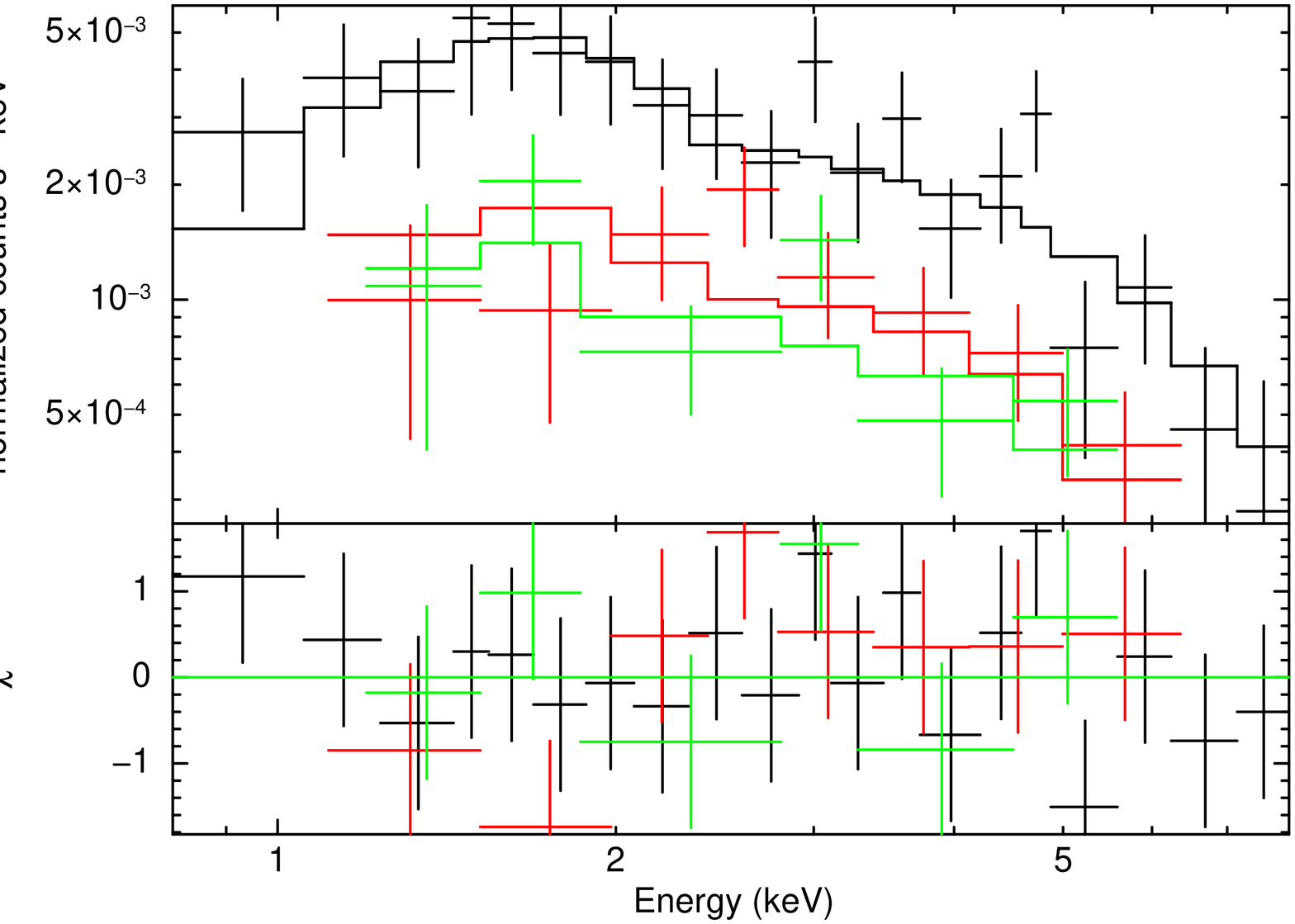}}

\subfigure[SDSS J090037.09+205340.2]{\includegraphics[scale=0.3,angle=270]{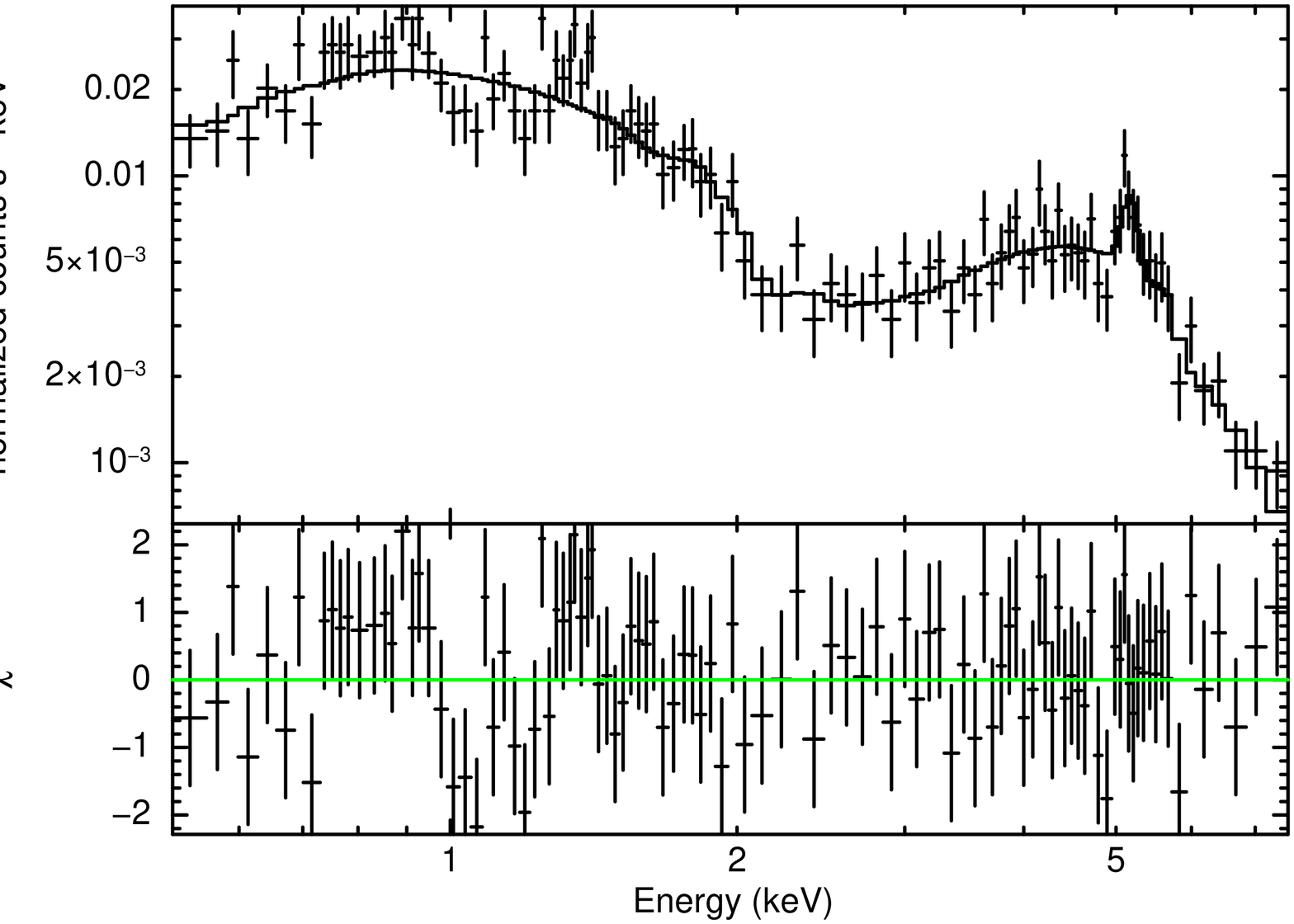}}~
\hspace{0.2cm}
\subfigure[SDSS J091345.48+405628.2]{\includegraphics[scale=0.3,angle=270]{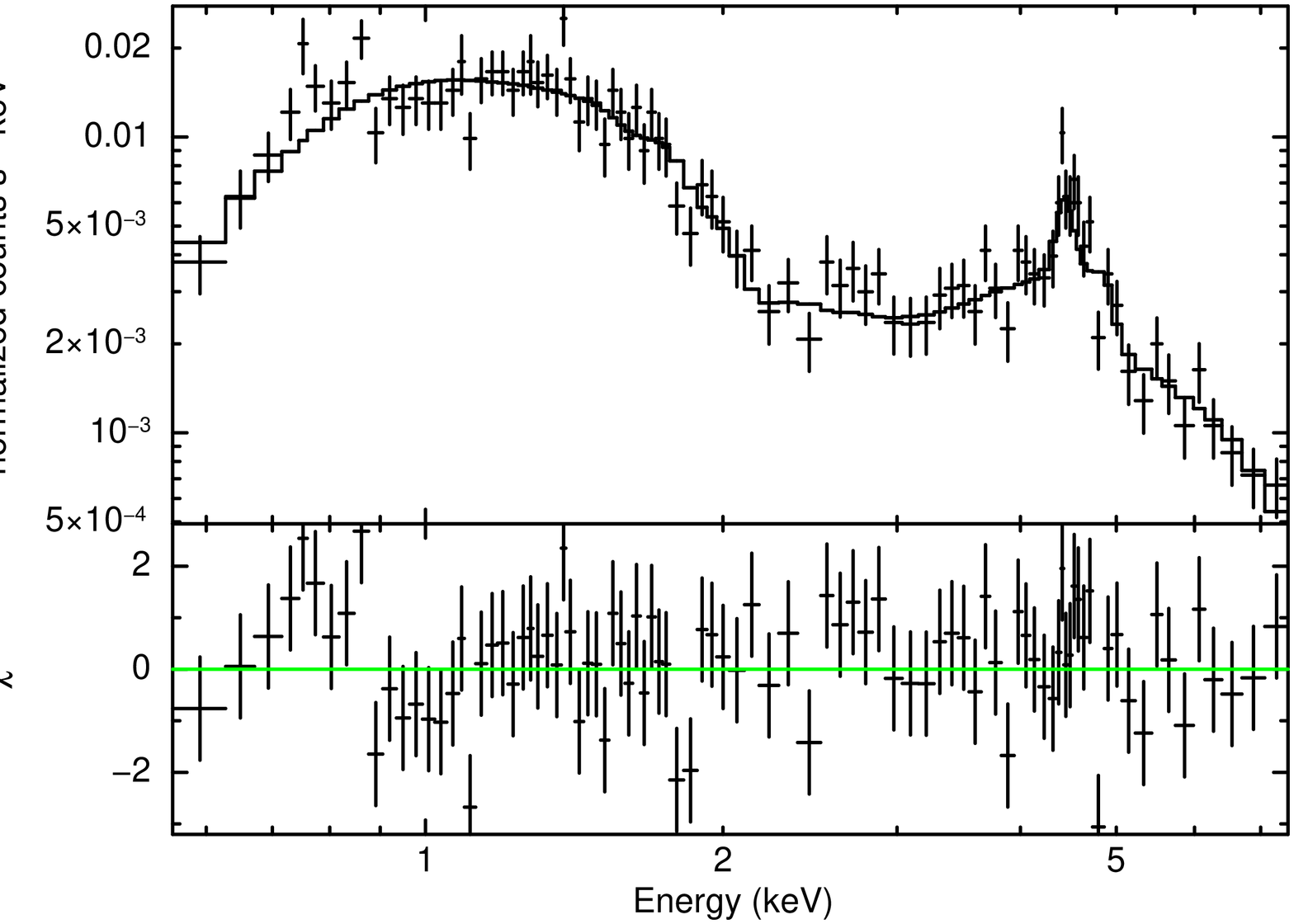}}

\subfigure[SDSS J103456.40+393940.0]{\includegraphics[scale=0.3,angle=270]{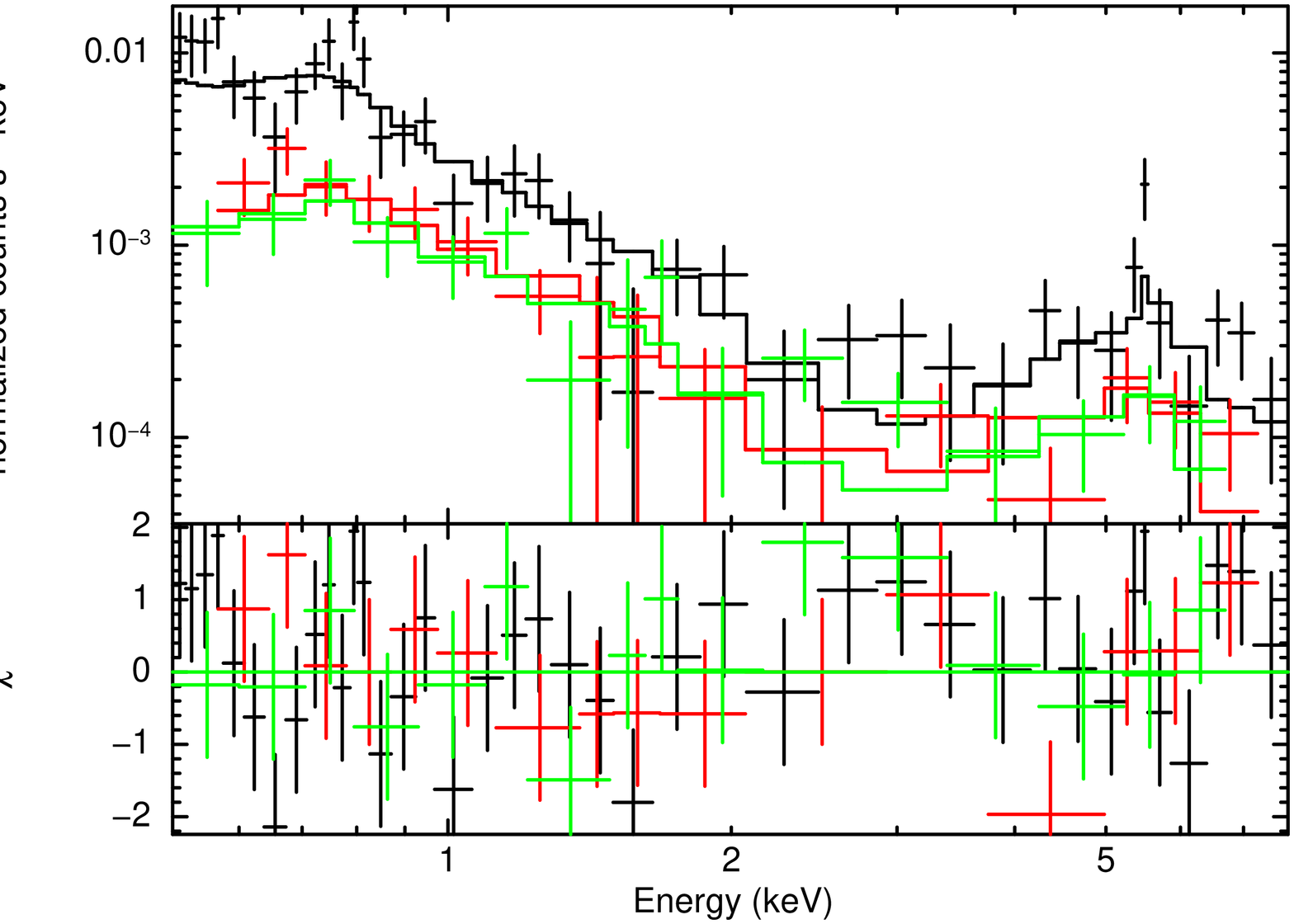}}~
\hspace{0.2cm}
\subfigure[SDSS J122656.40+013124.3]{\includegraphics[scale=0.3,angle=270]{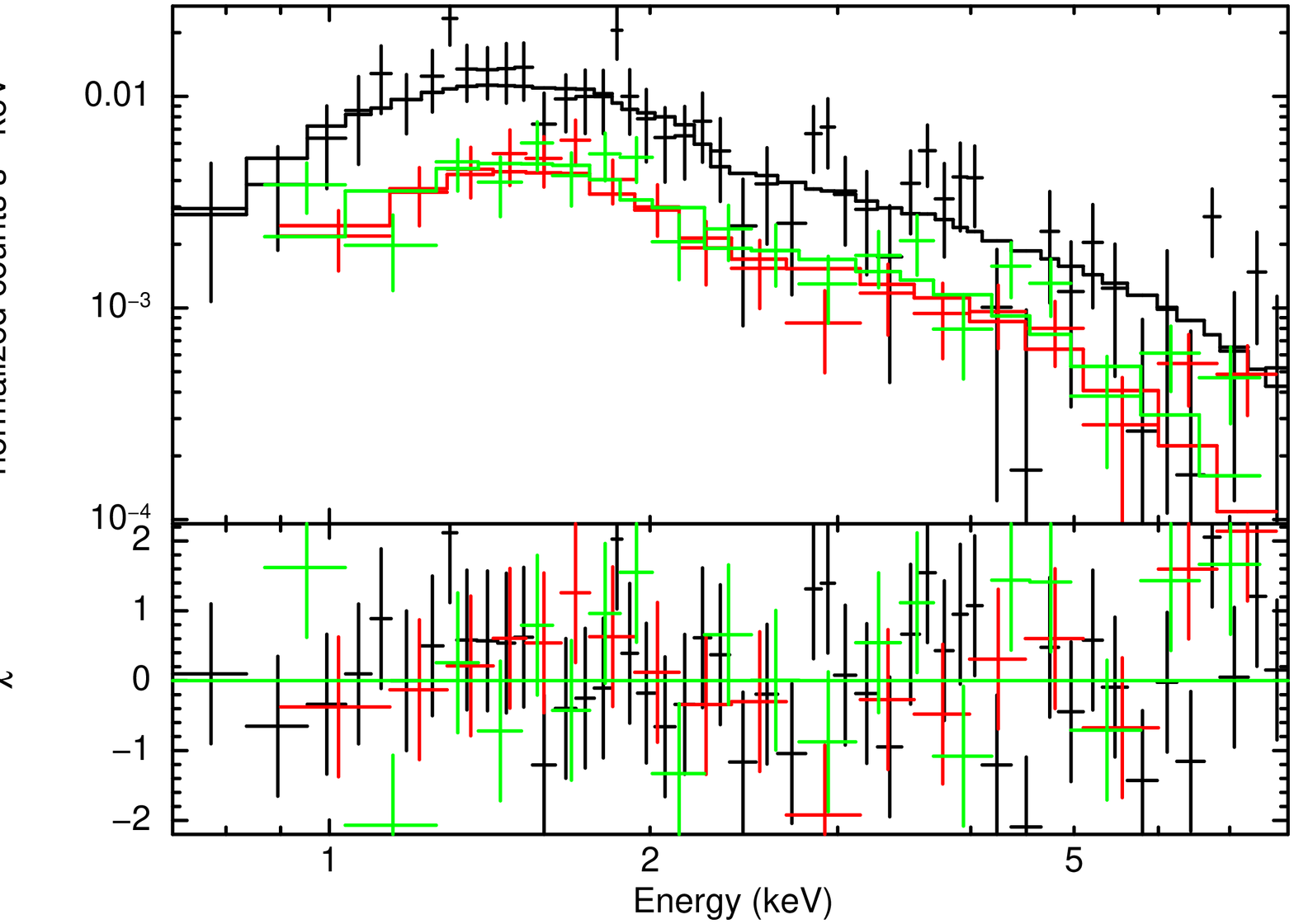}}

\subfigure[SDSS J134733.36+121724.3]{\includegraphics[scale=0.3,angle=270]{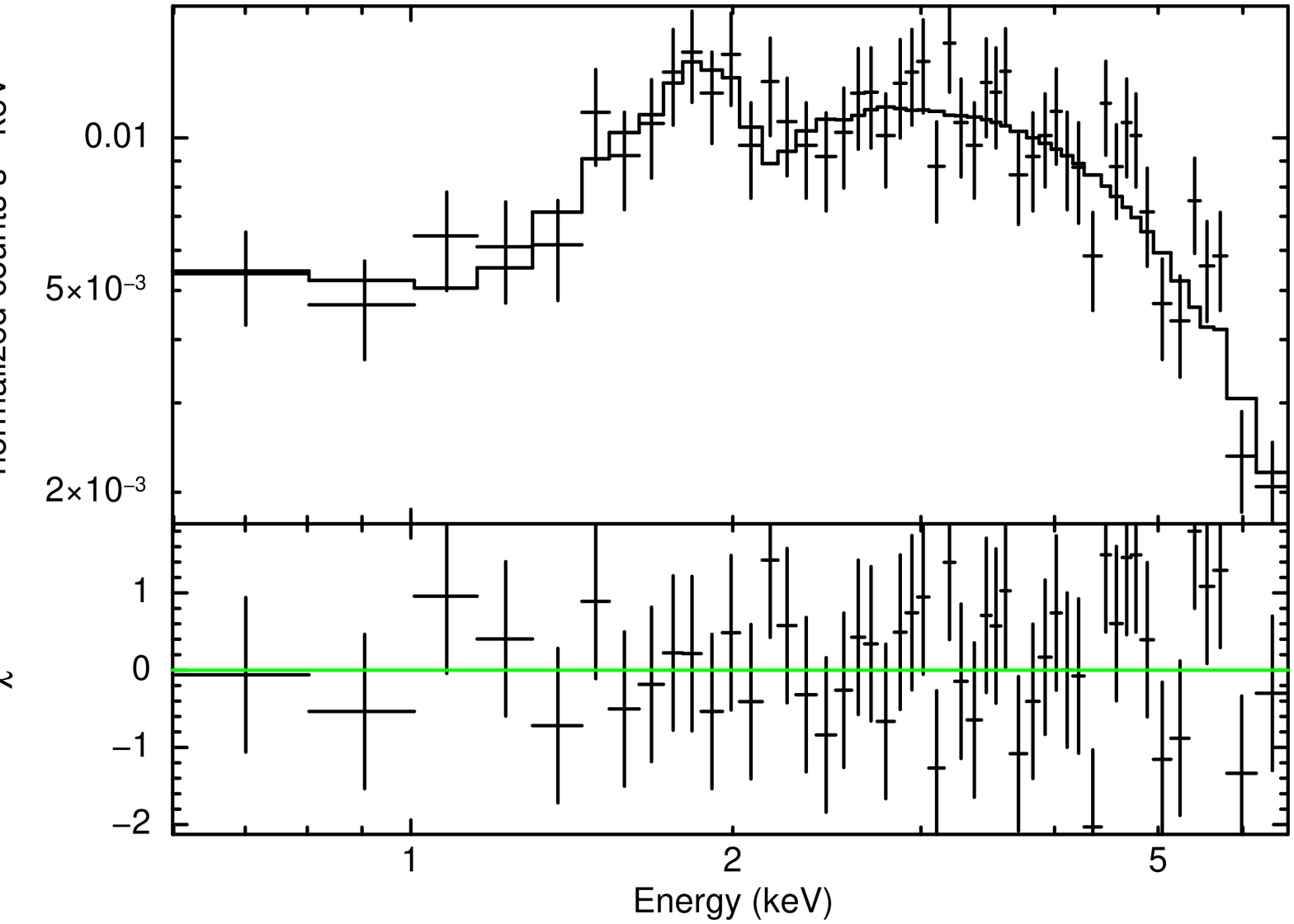}}~
\hspace{0.2cm}
\subfigure[SDSS J164131.73+385840.9]{\includegraphics[scale=0.3,angle=270]{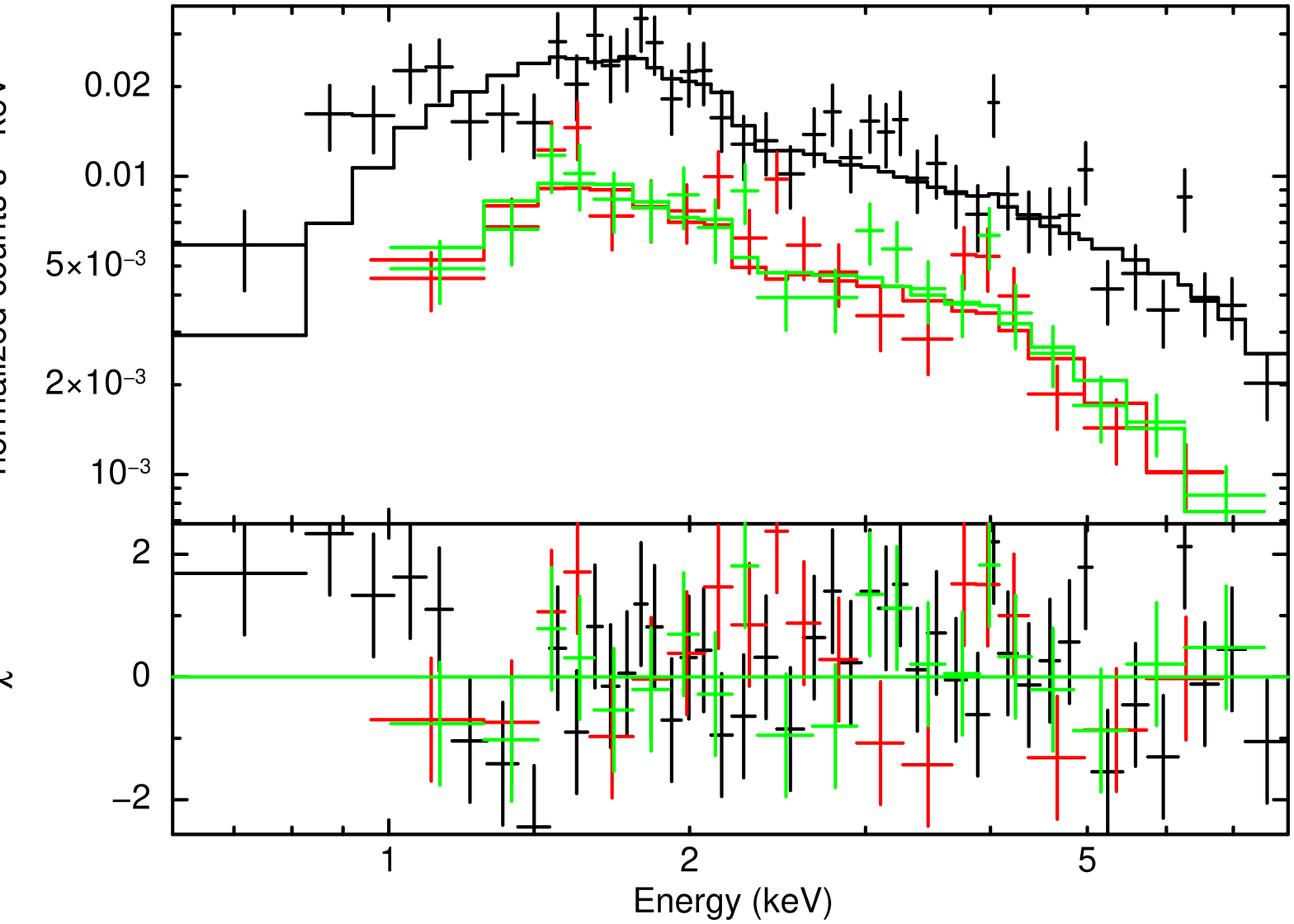}}

\caption[]{\label{qso2_sph_spec}\citet{spherical} spherical absorption model fits to X-ray spectra of the [OIII]-selected QSO2 candidates from \citet{jj}.  Black - PN or {\it Chandra} spectrum, red - MOS1 spectrum, green - MOS2 spectrum.}
\end{figure}

\begin{figure}[ht]
\centering

\subfigure[SDSS J103408.59+600152.2]{\includegraphics[scale=0.3,angle=270]{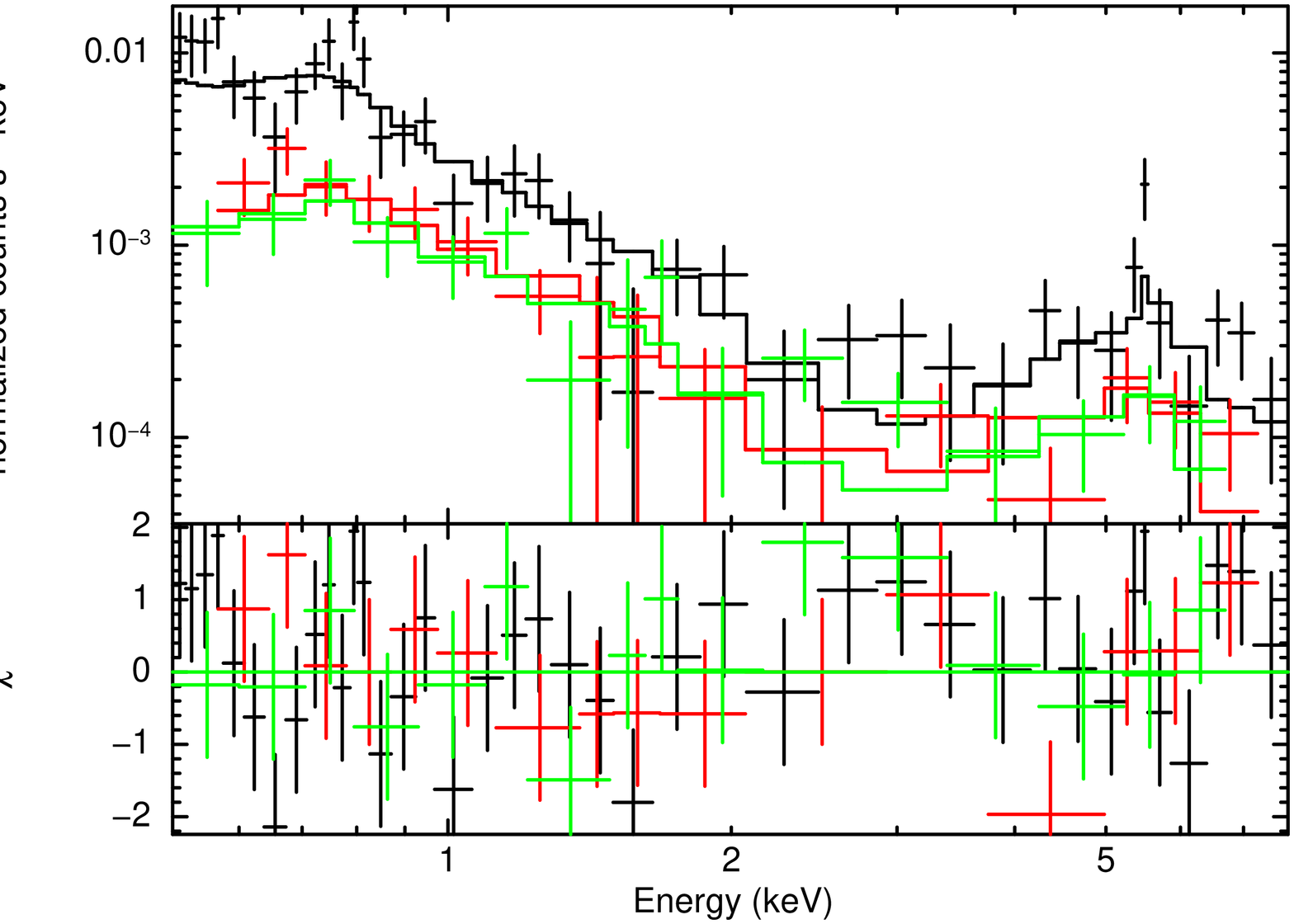}}~

\caption[]{\label{qso2_myt_coup} MYTorus coupled model fits to X-ray spectra of the [OIII]-selected QSO2 candidates from \citet{jj}. Color coding same as Figure \ref{qso2_sph_spec}.}
\end{figure}

\begin{figure}[ht]
\centering

\subfigure[SDSS J091345.48+405628.2]{\includegraphics[scale=0.3,angle=270]{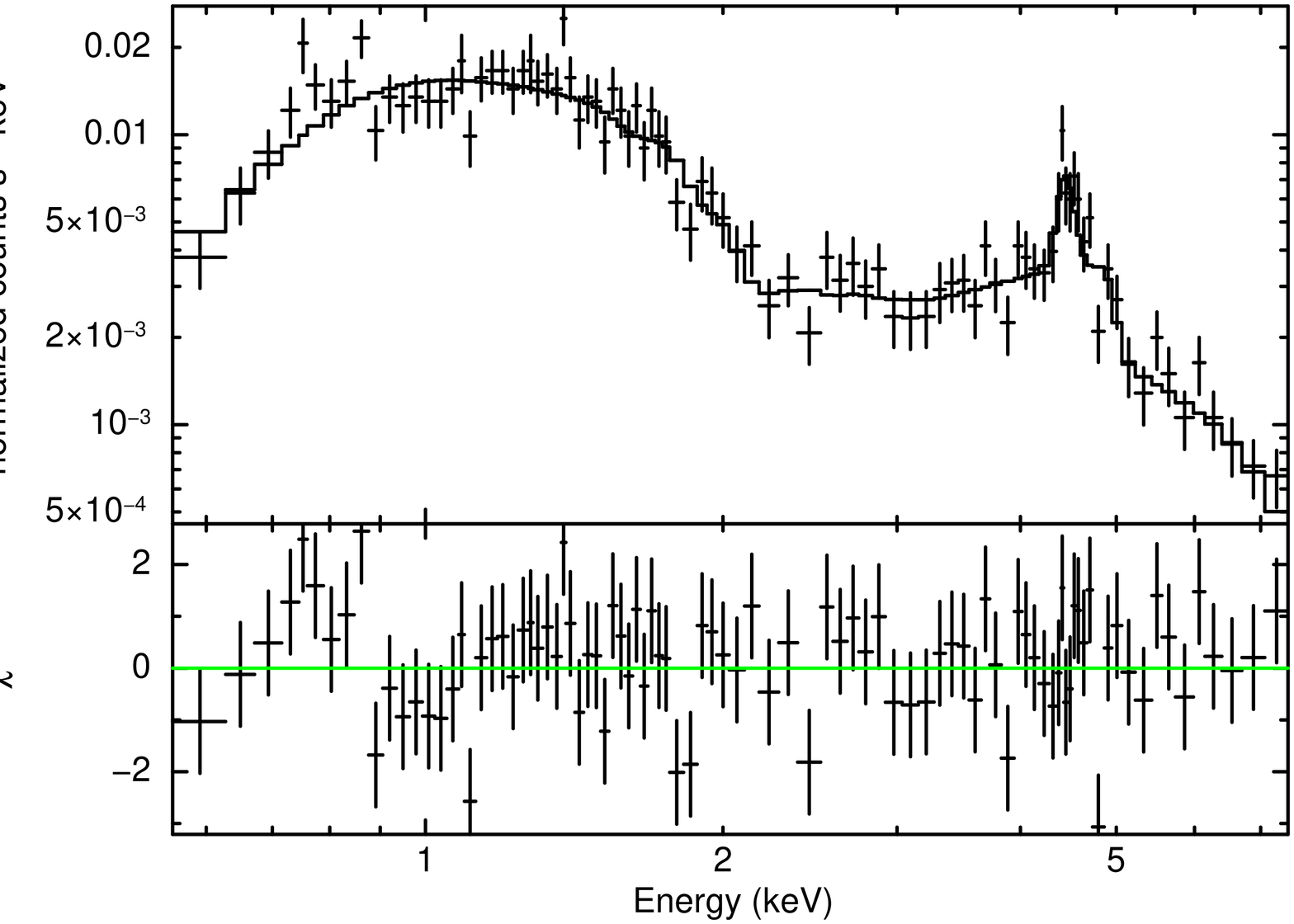}}~
\hspace{0.2cm}
\subfigure[SDSS J093952.74+355358.0]{\includegraphics[scale=0.3,angle=270]{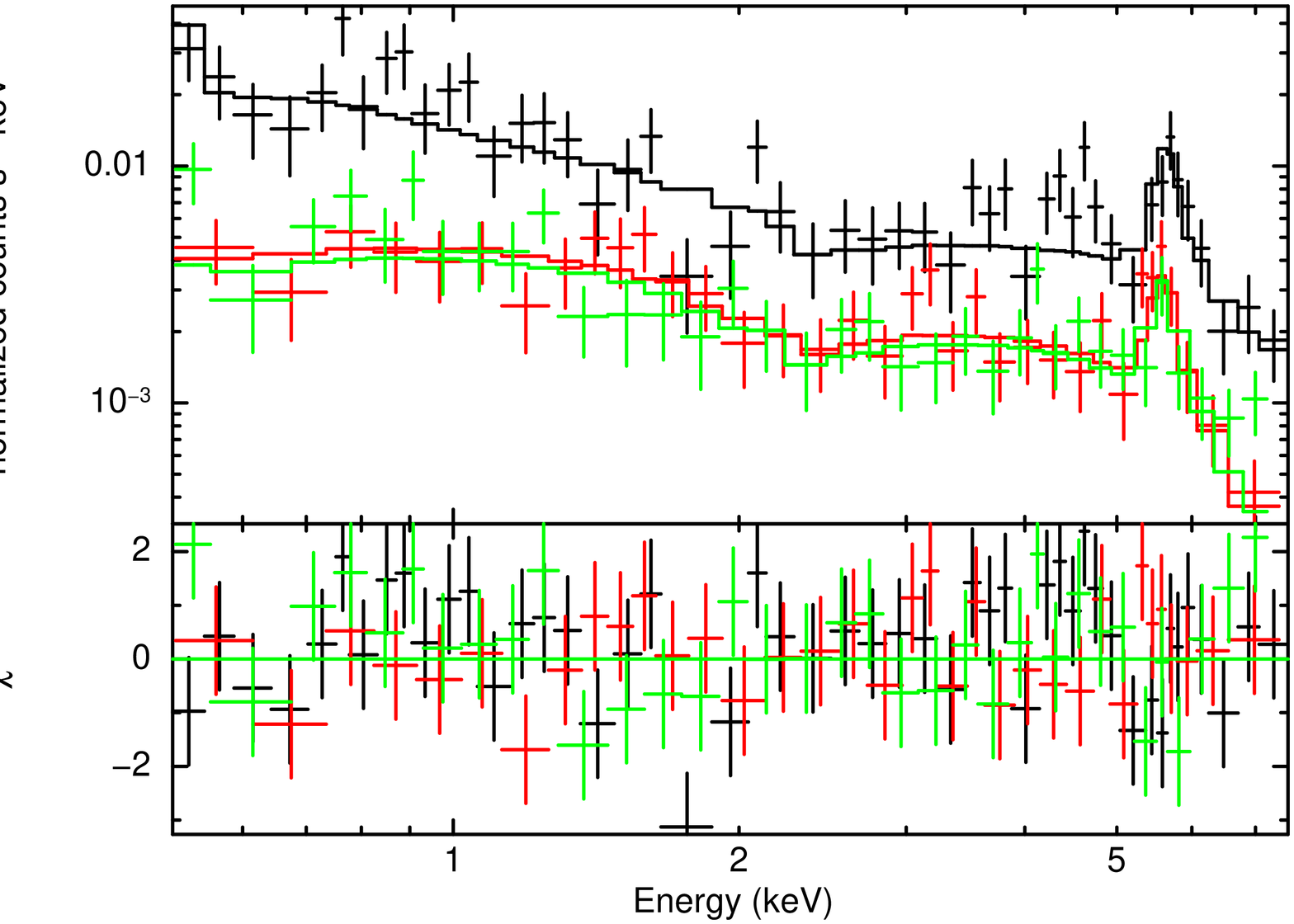}}

\caption[]{\label{qso2_myt_decoup} MYTorus decoupled model fits to X-ray spectra of the [OIII]-selected QSO2 candidates from \citet{jj}. Color coding same as Figure \ref{qso2_sph_spec}.}
\end{figure}

Of the remaining 9 QSO2s, 7 are best fit by the spherical absorption model (Table \ref{qso2_sph_fits}, Figure \ref{qso2_sph_spec}), 1 is well-accommodated by the coupled MYTorus model (Table \ref{qso2_myt_coup_fits}, Figure \ref{qso2_myt_coup}) and 1 requires the MYTorus model in decoupled mode (Table \ref{qso2_myt_decoup_fits}, Figure \ref{qso2_myt_decoup}), as discussed in greater detail below. We measured scattering fractions for 5 of these objects and obtain lower limits on the line of sight column density for 2 sources. For all but 2 objects, we derive constraints on $\Gamma$. One source is mildly obscured, 4 are moderately obscured, 2 are heavily obscured and 2 are heavily obscured to Compton-thick along the line of sight.

Though many of the Sy2s and QSO2s are best fit by the spherical absorption model, this does not necessarily mean that the X-ray reprocessor is a uniform sphere. Rather, limited statistics and model degeneracies preclude a more complex geometry from being required to explain the data. Additionally, many of the objects best fit by the spherical model show evidence of AGN continuum leakage, indicating that openings in the obscuring medium must exist for the photons to escape.

%%% QSO2 candidates
%% Spherical fits
\begin{landscape}
\begin{deluxetable}{lllllllr}

\tablewidth{0pt}
\tablecaption{\label{qso2_sph_fits} Spherical Absorption Fits to [OIII]-selected QSO2 Candidates}
\tablehead{ \colhead{Source} & \colhead{N$_{\rm H,Gal}$} & \colhead{$\Gamma$} & \colhead{N$_{\rm H,sph}$} & \colhead{$\Sigma_L$} & \colhead{$f_{\rm scat}$} & \colhead{kT} & \colhead{$\chi^2$(DOF)} \\
& \colhead{$10^{22}$ cm$^{-2}$} & & \colhead{$10^{22}$ cm$^{-2}$}  & \colhead{keV} & \colhead{\%} & \colhead{keV}}

\startdata

J0834+5534 & 0.04 & 2.08$^{+0.10}_{-0.12}$ & 0.08$^{+0.03}_{-0.04}$ & ... & ... & ... & 159.42(122) \\

J0839+3843 & 0.03 & $<$1.60 & 1.67$^{+1.56}_{-0.5}$ & ... &  ... & ... & 24.97(29)  \\

J0900+2053 & 0.03 & 1.77$^{+0.12}_{-0.12}$ & 34.5$^{+4.8}_{-5.4}$ & $<$0.14 & 9.17$^{+2.93}_{-2.06}$ & ... & 98.36(103) \\

J0913+4056 & 0.01 & 1.90$^{+0.14}_{-0.12}$ & 62.0$^{+15.7}_{-15.1}$ & ... & 12.7$^{+5.5}_{-3.8}$ & ... & 106.57(88) \\

J1034+3939 & 0.01 & $>$1.82 & 52.9$^{+38.4}_{-20.6}$ & ... & 1.82$^{+3.85}_{-1.28}$ & 0.77$^{+0.21}_{-0.17}$ & 70.97(64)  \\

J1226+0131 & 0.02 & 1.74$^{+0.28}_{-0.28}$ & 2.33$^{+0.72}_{-0.65}$ & ... & ... & ... & 84.12(82) \\

J1347+1217 & 0.02 & 1.41$^{+0.36}_{-0.26}$ & 3.50$^{+0.86}_{-0.55}$ & ... & 3.97$^{+2.33}_{-1.87}$ & ... & 38.53(46)  \\

J1641+3858 & 0.01 & 1.35$^{+0.13}_{-0.15}$ & 2.17$^{+0.49}_{-0.45}$ & ... & ... & ... & 99.75(81) \\

\enddata
\tablenotetext{\dagger}{``...'' indicates that parameter was not included in the spectral fitting.}
\end{deluxetable}
\end{landscape}

%% MYTorus Fits: Coupled
\begin{landscape}
\begin{deluxetable}{llllllllllr}

\tablewidth{0pt}
\tablecaption{\label{qso2_myt_coup_fits} Coupled MYTorus Fits to [OIII]-selected QSO2 Candidates}
\tablehead{ \colhead{Source} & \colhead{N$_{\rm H,Gal}$} & \colhead{$\Gamma$} & \colhead{N$_{\rm H,Z}$} & \colhead{$A_S$} & \colhead{Incl. Ang} & \colhead{$\Sigma_L$}
& \colhead{$f_{\rm scat}$} &  \colhead{kT} & \colhead{$\chi^2$(DOF)}  \\
& \colhead{$10^{22}$ cm$^{-2}$} & & \colhead{$10^{22}$ cm$^{-2}$} & & & \colhead{keV} &  \colhead{\%} & \colhead{keV}}

\startdata

J1034+6001\tablenotemark{1} & 0.01 & 1.87$^{+0.35}_{-0.42}$ & $>$41 & 1 & $>$60.2 & ... & 3.05$^{+2.91}_{-1.56}$ & 0.80$^{+0.20}_{-0.12}$ & 54.94(32) \\

\enddata
\tablenotetext{\dagger}{``...'' indicates that parameter was not included in the spectral fitting.}
\tablenotetext{1}{$A_S$ frozen to unity.}
\end{deluxetable}

\end{landscape}

% MYTorus Fits: Decoupled
\begin{landscape}
\begin{deluxetable}{lllllllllr}

\tablewidth{0pt}
\tablecaption{\label{qso2_myt_decoup_fits} Decoupled MYTorus Fits to [OIII]-selected QSO2 Candidates}
\tablehead{ \colhead{Source} & \colhead{N$_{\rm H,Gal}$} & \colhead{$\Gamma$} & \colhead{N$_{\rm H,Z}$\tablenotemark{1}} & \colhead{N$_{\rm H,S}$\tablenotemark{2}}\ & \colhead{$\Sigma_L$}
& \colhead{$A_{\rm S,0}$} & \colhead{$f_{\rm scat}$} &  \colhead{kT} & \colhead{$\chi^2$(DOF)}  \\
& \colhead{$10^{22}$ cm$^{-2}$} & & \colhead{$10^{22}$ cm$^{-2}$} & \colhead{$10^{22}$ cm$^{-2}$} & \colhead{keV} &  & \colhead{\%} & \colhead{keV}}

\startdata

J0913+4056\tablenotemark{3} & 0.01 & 2.03$^{+0.24}_{-0.18}$ & 97.8$^{+25.2}_{-31.2}$ & 34.5$^{+267}_{-21.4}$ & ... & 1 & 4.74$^{+5.76}_{-2.77}$ & ... & 101.61(87)   \\

J0939+3553\tablenotemark{3} & 0.01 & 1.71$^{+0.17}_{-0.20}$ & $>$167 & 14.4$^{+6.0}_{-4.2}$ & 0.15$^{+0.07}_{-0.05}$ & 1 & 0.56$^{+0.32}_{-0.18}$ & ... & 122.55(111) \\

\enddata
\tablenotetext{\dagger}{``...'' indicates that parameter was not included in the spectral fitting.}
\tablenotetext{1}{\ Line of sight column density, associated with the zeroth order continuum.}
\tablenotetext{2}{Global column density, associated with Compton-scattering off the back wall of the X-ray reprocessor.}
\tablenotetext{3}{$A_S$ frozen to unity.}
\end{deluxetable}

\end{landscape}

\begin{deluxetable}{llll}
\tablewidth{0pt}
\tablecaption{\label{qso2s_lums} Luminosities for [OIII]-selected QSO2 Candidates}
\tablehead{ \colhead{Source} & \colhead{L$_{\rm [OIII]}$\tablenotemark{1}} & \colhead{L$_{\rm Fe K\alpha}$\tablenotemark{2}} & \colhead{L$_{\rm 2-10keV,in}$\tablenotemark{3}} \\
& \colhead{L$_{\sun}$} & \colhead{erg s$^{-1}$} & \colhead{erg s$^{-1}$}}

\startdata
SDSS J083454.89+553411.1 & 8.69 & 41.61 & 43.52$^{+0.04}_{-0.05}$ \\ 

SDSS J083945.98+384319.0 & 8.61 & 42.28 & 44.36$^{+0.96}_{-0.07}$ \\ 

SDSS J090037.09+205340.2 & 8.98 & 42.92 & 44.67$^{+0.12}_{-0.13}$ \\ 

SDSS J091345.48+405628.2 (Spherical) & 10.34 & 43.20 & 45.23$^{+0.18}_{-0.17}$ \\ 

SDSS J091345.48+405628.2 (MYTorus) & 10.34 & 43.15 & 45.28$^{+0.40}_{-0.37}$ \\ 

SDSS J093952.74+355358.0 & 8.75 & 42.39 & 44.74$^{+0.14}_{-0.19}$ \\ 

SDSS J103408.59+600152.2 & 8.81 & 40.72 & 42.96$^{+0.22}_{-0.29}$ \\ 

SDSS J103456.40+393940.0 & 8.91 & 41.00 & 43.27$^{+0.59}_{-0.66}$ \\ 

SDSS J122656.40+013124.3 & 9.81 & 42.44 & 45.02$^{+0.21}_{-0.21}$ \\ 

SDSS J134733.36+121724.3 & 8.65 & 41.04 & 43.69$^{+0.01}_{-0.01}$ \\ 

SDSS J164131.73+385840.9 & 10.05 & 42.17 & 45.08$^{+0.11}_{-0.13}$ \\

\enddata
\tablenotetext{\dagger}{\ Luminosities reported in log space.}
\tablenotetext{1}{Observed luminosity, not corrected for dust extinction.}
\tablenotetext{2}{Rest-frame luminosity.}
\tablenotetext{3}{Rest-frame, intrinsic (i.e., absorption corrected) luminosity.}
\end{deluxetable}

\section{Discussion}

\subsection{Interesting Sources}

\subsubsection{Mrk 0609 - A ``Changing-Look'' AGN or True Sy2?}  
Though the SDSS spectrum from 2001 and the CTIO spectrum from 2007 \citep{trippe} shows an absence of broad optical emission lines, there is no X-ray evidence to support that the view to the central engine is obscured. Not only do we find no absorption above that from our Galaxy, the Fe K$\alpha$ feature is not significant above the 2$\sigma$ level \citep{me}, suggesting a lack of global obscuration. BeppoSAX observations from 2000 also detected no signatures of absorption. Additionally, the observed 2-10 keV X-ray luminosity of 4$\times10^{42}$ erg s$^{-1}$ is consistent with that of an unabsorbed nucleus, given the observed [OIII] 5007 \AA\ luminosity of 2$\times10^{41}$ erg s$^{-1}$ and [OIV] 26$\mu$m luminosity of 2.14 $\times 10^{41}$ erg s$^{-1}$ and the relationships found between 2-10 keV X-ray and [OIII] and [OIV] luminosities for Sy1s (log(L$_{2-10keV,\rm unabsorbed}$/L$_{\rm [OIII]}$) = 1.59 $\pm$ 0.48 dex, Heckman et al. 2005; log(L$_{2-10keV,\rm unabsorbed}$/L$_{[OIV]}$) = 1.92 $\pm$ 0.60 dex, Diamond-Stanic et al. 2009). A proposed scenario where the source is heavily obscured while a small percentage of scattered light dominates the spectrum is not supported by our data as the observed luminosity would have to be over an order of magnitude lower for this situation to be applicable. 

Furthermore, the {\it Spitzer} observation of this source suggests that circumnuclear dust is not a dominant component of the mid-infrared spectrum. In AGN with strong MIR emission from the circumnuclear obscuring medium, the polycyclic aromatic hydrocarbon (PAH) features are diluted by the MIR continuum and they consequently have weak PAH EWs. In Mrk 0609, the PAH EWs at 11.3 $\mu$m and 17 $\mu$m are the highest among the 20 Sy2s in the parent [OIII] sample \citep{me10}. 

As no {\it Chandra} data exist for this source, it is possible that a background unabsorbed AGN contaminates the relatively larger {\it XMM-Newton} and BeppoSAX PSFs. However, given that the extraction radius for {\it XMM-Newton} is $\sim30^{\prime\prime}$ at the center of the host galaxy, a chance superposition of an unassociated source is improbable. Mrk 0609 can possibly be a ``changing-look'' AGN \citep[e.g.,][]{ren,elvis04,puccetti,risaliti09,bianchi09} where the optical observations in 2001 and 2007 caught the source in an obscured state while the 2000 and 2007 X-ray observations detected the object while unobscured. Alternatively, Mrk 0609 may belong to the class of ``true'' Sy2s that lack circumnuclear obscuration: broad emission lines are not observed because they do not exist. 

Such ``naked'' Sy2s can be explained via an extension of the disk-wind model, where the broad line region forms in the wind launched by the accretion disk \citep{elvis}. Theoretical work suggests that the outflow can not be maintained at low luminosities ($<$10$^{42}$ erg s$^{-1}$) or low Eddington rates (L$_{\rm bol}$/L$_{\rm Edd}$), causing the broad line region to disappear \citep[e.g.,][]{elitzur}. The critical Eddington rate below which the outflow fails has been reported as 2.4$\times 10^{-3} M_{8}^{-1/8}$ \citep{nicastro} and 1.3 $\times 10^{-2} M_{8}^{-1/8}$ \citep{trump}, where $M_{8}$ is the black hole mass in units of 10$^{8}$ M$_{\sun}$ \citep{bianchi12}. Based on the observed velocity dispersion, Mrk 0609 has a black hole mass of 0.63 $M_{8}$ \citep{tremaine,me10}, constraining the upper limit on the Eddington rate to be 0.0024 - 0.012 for this prediction to explain our observations. The reddening corrected [OIII] luminosity of Mrk 0609 is $8.5\times10^{41}$ erg s$^{-1}$, which translates to a bolometric luminosity of $4.3\times10^{44} - 6.8\times10^{44}$ \citep{kh} and associated Eddington rate between 0.06 - 0.09. 

Both the bolometric luminosity and Eddington rate for Mrk 0609 are inconsistent with predictions of the failed disk wind to explain naked Sy2s. However, simultaneous optical and X-ray observations, which we are currently pursuing, are necessary to determine whether the obscuration in this system varies or if the broad line region is truly absent, and if by extension, the luminosity and accretion rate of Mrk 0609 challenges current theories that explain these types of objects. 

We exclude Mrk 0609 from the analysis below since we do not detect any intrinsic X-ray absorption in this source.

\subsubsection{SDSS J093952.74+355358.0 - Compton-thick Ring of Obscuration}
The X-ray spectrum of SDSS J093952.74+355358.0 reveals interesting characteristics in its circumnuclear absorber that suggests different global and line-of-sight column densities. The Fe K$\alpha$ feature is quite prominent \citep[EW$\approx$0.5 keV,][]{jj}, precluding the global and line-of-sight absorbers from being equal and Compton-thin since the line of sight continuum would dilute the line compared with a pure reflection spectrum. If both column densities are Compton-thick, then the continuum between 2 - 6 keV would be curved downwards rather than flat. This continuum curvature can be avoided if the global column density is Compton-thin. To obtain the large Fe K$\alpha$ EW, however, the direct continuum has to be hidden behind a Compton-thick absorber. The solid angle of this line of sight absorber must also be relatively small since a larger covering factor would produce more pronounced signatures of a reflection spectrum, such as a curved continuum between 2 - 6 keV. Consequently, this absorber is likely in the form of a toroidal ring. We note that an alternate scenario where the Fe abundance is super-solar is not supported by current data: freeing the Fe abundance in the spherical absorption model does not adequately fit the Fe K complex, while the MYTorus decoupled model fit shows no significant residuals around the Fe K$\alpha$ line.

\begin{figure}[ht]
\centering
\subfigure[]{\includegraphics[scale=0.3,angle=270]{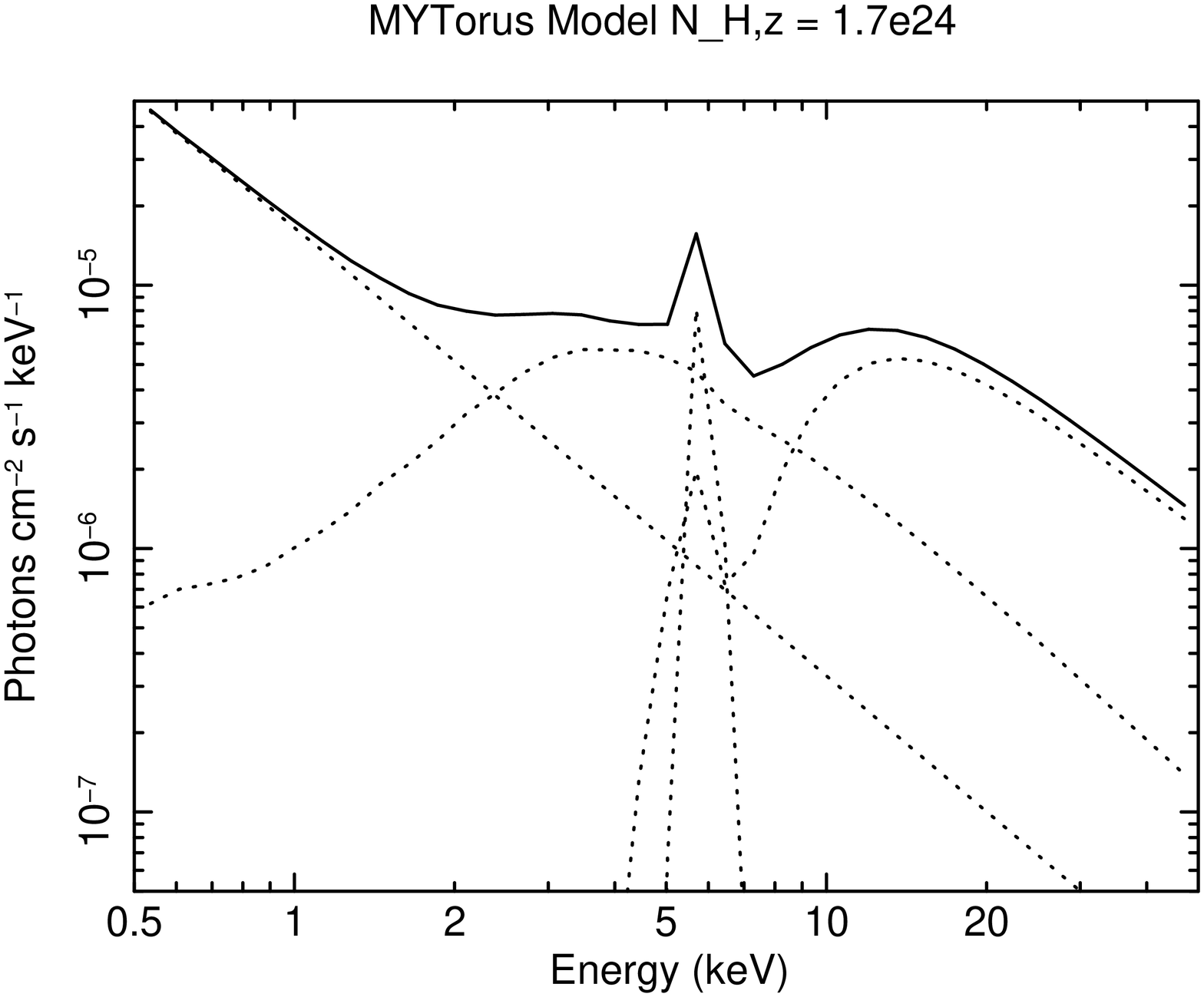}}
\subfigure[]{\includegraphics[scale=0.3,angle=270]{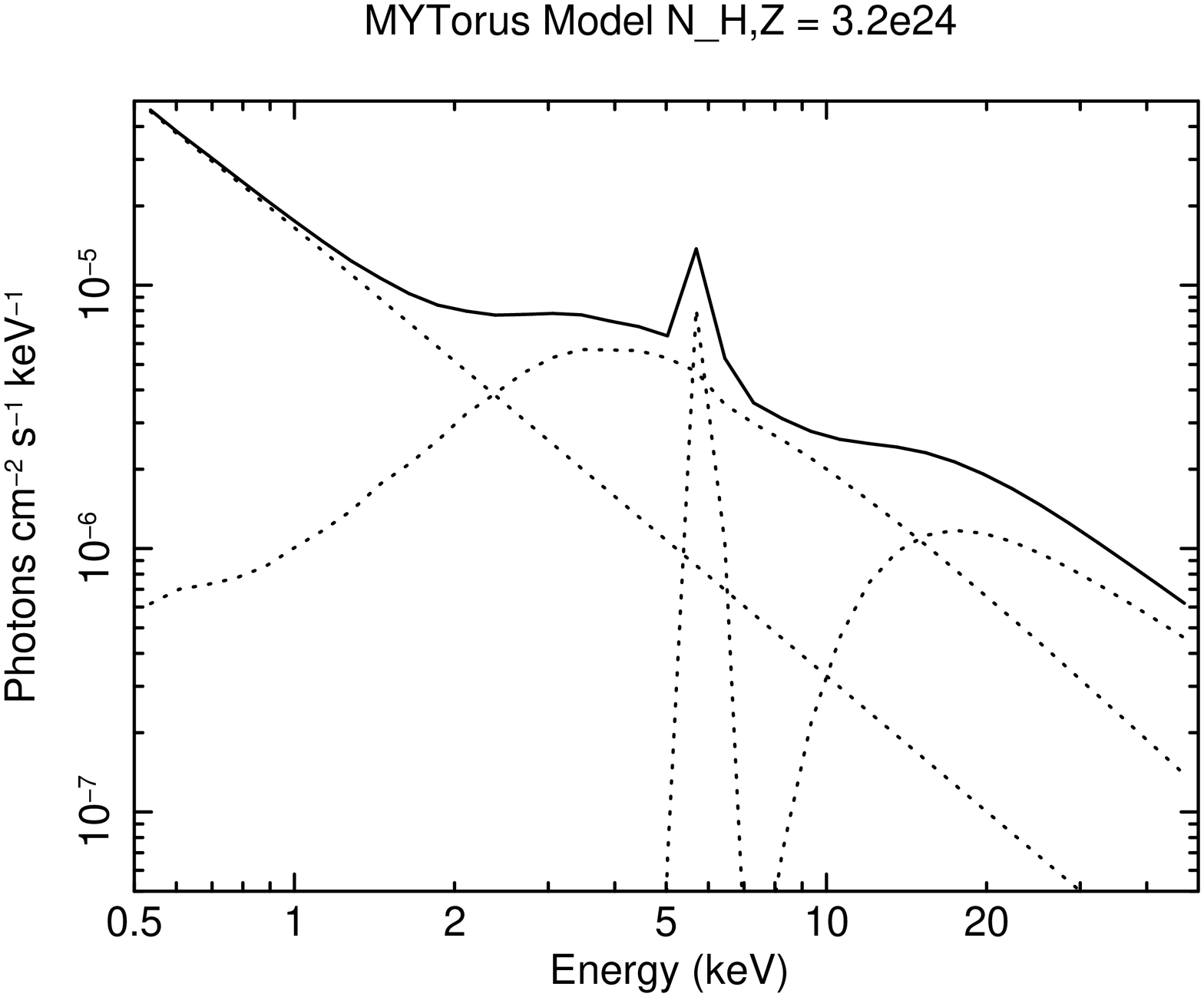}}
\subfigure[]{\includegraphics[scale=0.3,angle=270]{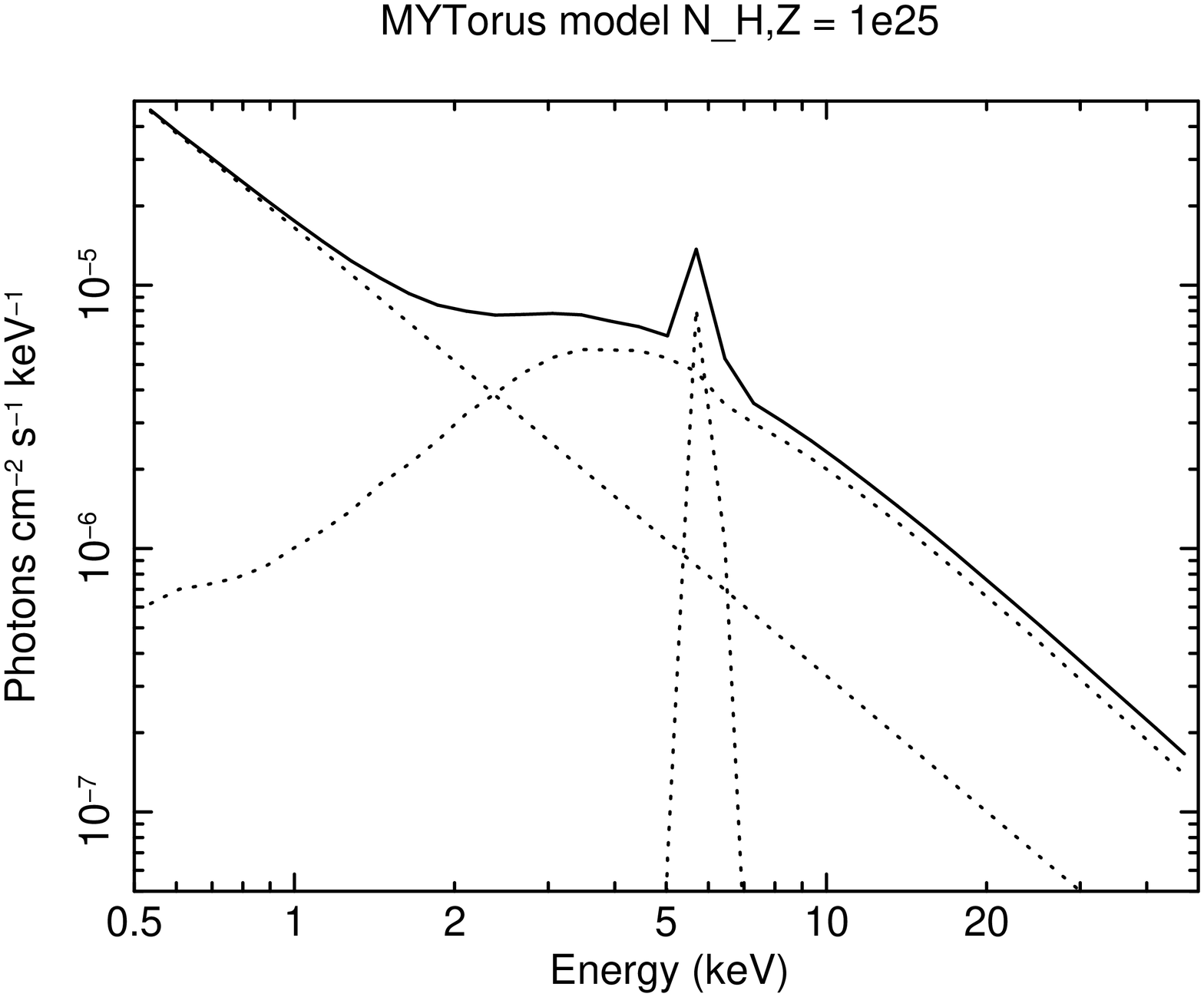}}
\caption[]{\label{j0935_highe}Decoupled MYTorus model describing the {\it XMM-Newton} data for permissible N$_{\rm H,Z}$ values for SDSS J093952.74+355358: (a) $1.7\times10^{24}$ cm$^{-2}$, (b) $3.2\times10^{24}$ cm$^{-2}$ (best-fit value), and (c) $10^{25}$ cm$^{-2}$. The model components shown are scattering of the intrinsic AGN continuum off a distant an optically thin medium (dominates $<$2 keV), the line of sight Compton scattered component (dominates above 20 keV when N$_{\rm H}<10^{25}$ cm$^{-2}$), and global Compton-scattered component (dominates between 2-10 keV); the Fe K complex is a combination of both Compton scattering components. The 3 possible N$_{\rm H,Z}$ values predict different spectral shapes above 10 keV, making hard X-ray observations, from, e.g., NuSTAR or Astro-H, necessary to accurately measure the column density.}
\end{figure}

As reported in Table \ref{qso2_myt_decoup_fits}, the global N$_{\rm H,S}$ is Compton-thin while N$_{\rm H,Z}$ is Compton-thick. However, the exact value of the N$_{\rm H,Z}$ is unknown, with possible values ranging from 1.67$\times10^{24}$ to above $10^{25}$ cm$^{-2}$. As shown in Figure \ref{j0935_highe}, the different permissible values of N$_{\rm H,Z}$, from the low end to the best-fit value of $3.2\times10^{24}$ cm$^{-2}$ to the high end of the range, predict very different spectral shapes at energies above 10 keV. Hard X-ray observations, from, e.g.,  NuSTAR (see, for instance, Lansbury et al., submitted) or Astro-H, are necessary to accurately constrain the line of sight column density in this source. No other Type 2 QSO is known to have a similar X-ray reprocessor geometry.

\subsection{Global vs. Line-of-sight N$_{H}$}
\begin{figure}[ht]
\centering
{\includegraphics[scale=0.5,angle=90]{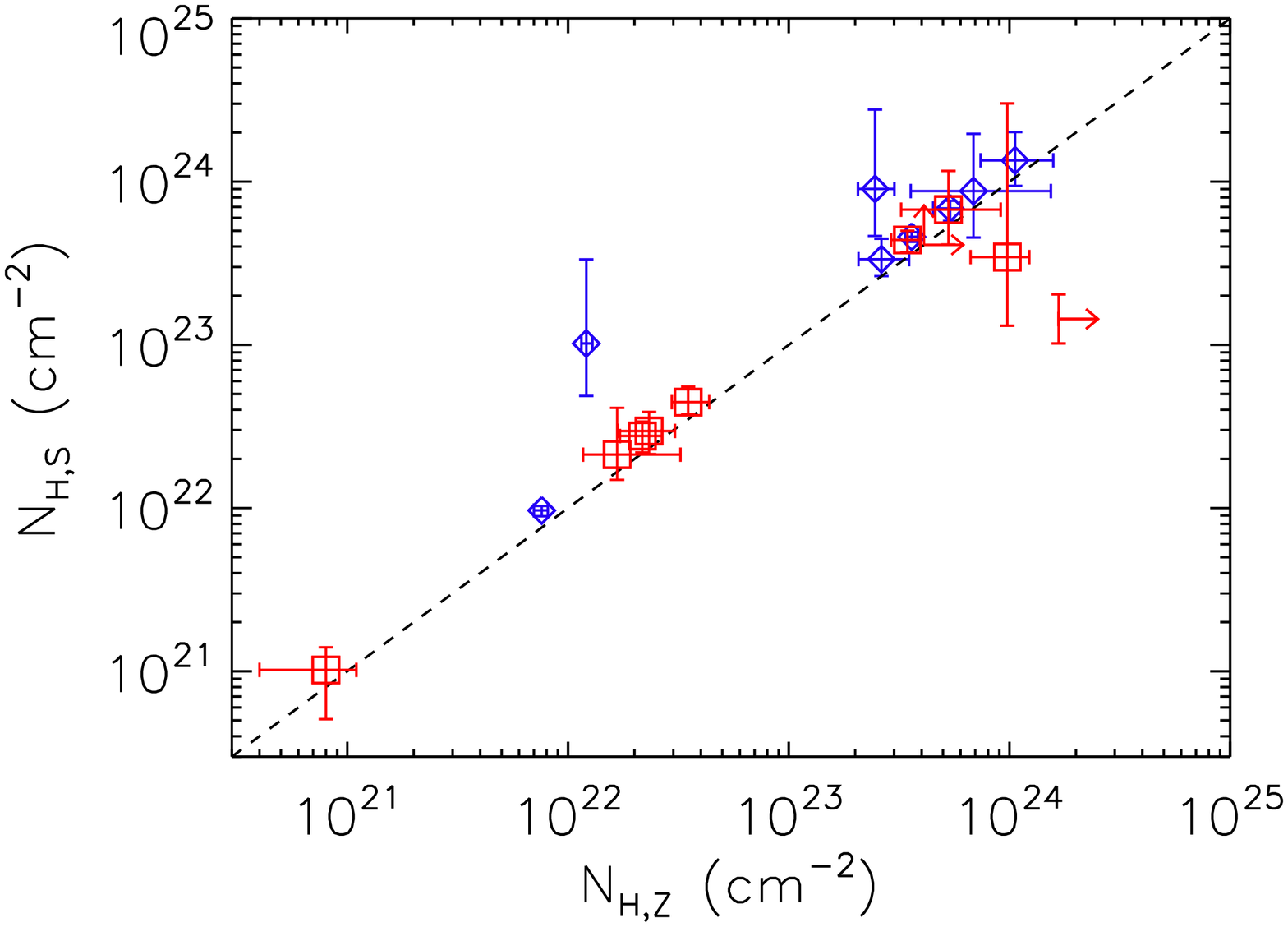}}
\caption[]{\label{nhz_v_nhs} Average global column density (N$_{\rm H,S}$) as a function of line-of-sight column density (N$_{\rm H,Z}$). For AGN best fit with the spherical absorption model, N$_{\rm H,S}$ = 4/$\pi \times$ N$_{\rm H,Z}$.In a majority of sources, the global N$_{H}$ is consistent with line-of-sight N$_{H}$ (the dashed line denotes equality between the two parameters), but a handful of sources have global column densities markedly different from that along the line-of-sight. Type 2 Seyferts shown as blue diamonds and Type 2 QSO candidates marked by red squares.}
\end{figure}

For every source, we measure a global average (N$_{\rm H,S}$), as well as line-of-sight (N$_{\rm H,Z}$), column density. As shown in Figure \ref{nhz_v_nhs}, N$_{\rm H,S}$ is largely consistent with N$_{\rm H,Z}$ for most objects, but this is not always necessarily the case. Four objects in particular are well fit MYTorus in decoupled mode, where we find that the global distribution of matter is significantly different from that along the line-of-sight. 

\subsection{Revealing the Intrinsic X-ray Luminosity}
As simpler models do not self-consistently treat the physics of the transmitted, Compton-scattered, and fluorescent line emission, the fitted column densities are unreliable, preventing an accurate calculation of the inherent X-ray flux. We bypass this shortcoming in the present work and are able to derive realistic, absorption corrected 2-10 keV X-ray luminosities (L$_{\rm 2-10keV,in}$, Tables \ref{sy2_lums} and \ref{qso2s_lums}). The mean log (L$_{\rm 2-10keV,in}$/L$_{\rm [OIII]}$) for the sample is 1.54 $\pm$ 0.49 dex (1.57$\pm$0.38 dex for Sy2s and 1.51 $\pm$0.59 dex for QSO2s), which is remarkably consistent with the average value for Sy1s of 1.59 $\pm$ 0.48 dex \citep{tim}. This similarity in normalized X-ray luminosities is to be expected if Type 1 and Type 2 AGN have the same physical central engine, as posited by unification models, and affirms that the sophisticated models do accurately measure the circumnuclear column density.

\subsection{Fe K$\alpha$ Luminosity as a Proxy of Intrinsic Luminosity?}
\begin{figure}[ht]
\centering
{\includegraphics[scale=0.5,angle=90]{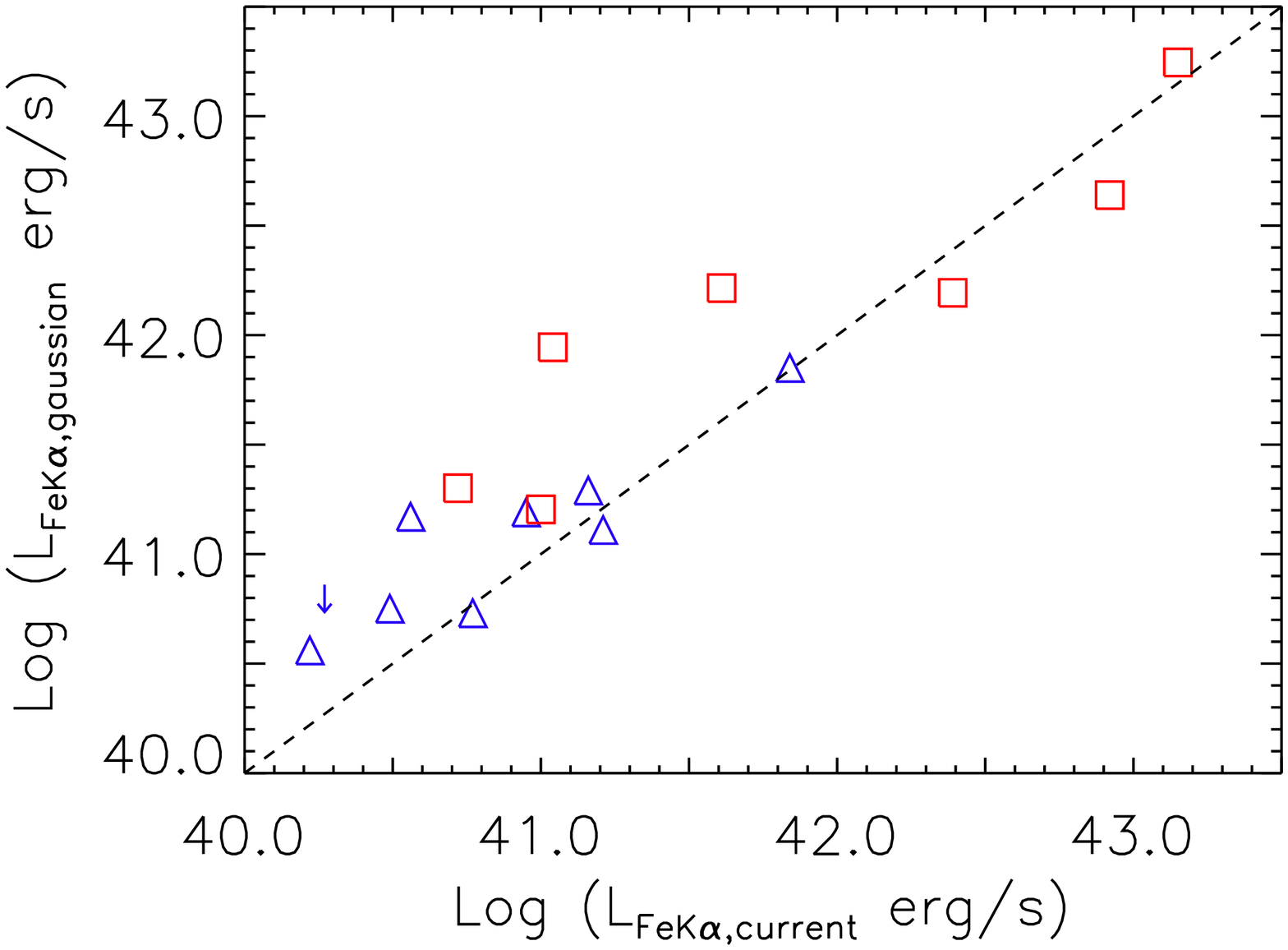}}~
\caption[]{\label{compare_fek} Fe K$\alpha$ luminosity measured using a Gaussian fit to the emission line as a function of the luminosity derived in the present analysis using self-consistent models. The dashed line indicates equality. Both methods are largely consistent, but in several instances, the simpler modeling over-predicts the Fe K$\alpha$ luminosity by around an order of magnitude with respect to the physically motivated models. Symbol and color coding same as Figure \ref{nhz_v_nhs}. }
\end{figure}

Past X-ray studies have reported a correlation between the Fe K$\alpha$ luminosity and intrinsic AGN luminosity \citep[e.g.,][]{ptak,me,jj}. However, such analyses relied on {\it ad hoc} X-ray modeling where the Fe K$\alpha$ feature was parametrized by a Gaussian component that is placed either in front of or behind the line-of-sight absorbing screen. As the models we use here self-consistently treat the fluorescent emission associated with the transmitted continuum and Compton-scattered emission within the global obscuring medium, our study better estimates the Fe K$\alpha$ luminosity. In Figure \ref{compare_fek}, we compare the Fe K$\alpha$ luminosity from modeling the feature with a Gaussian component in \citet{me} and \citet{jj} and with the values we derive using self-consistent models. There is general agreement between both approaches, but in several cases, disagreements on the order of one magnitude are apparent.

\begin{figure}[ht]
\centering
\subfigure[]{\includegraphics[scale=0.4,angle=90]{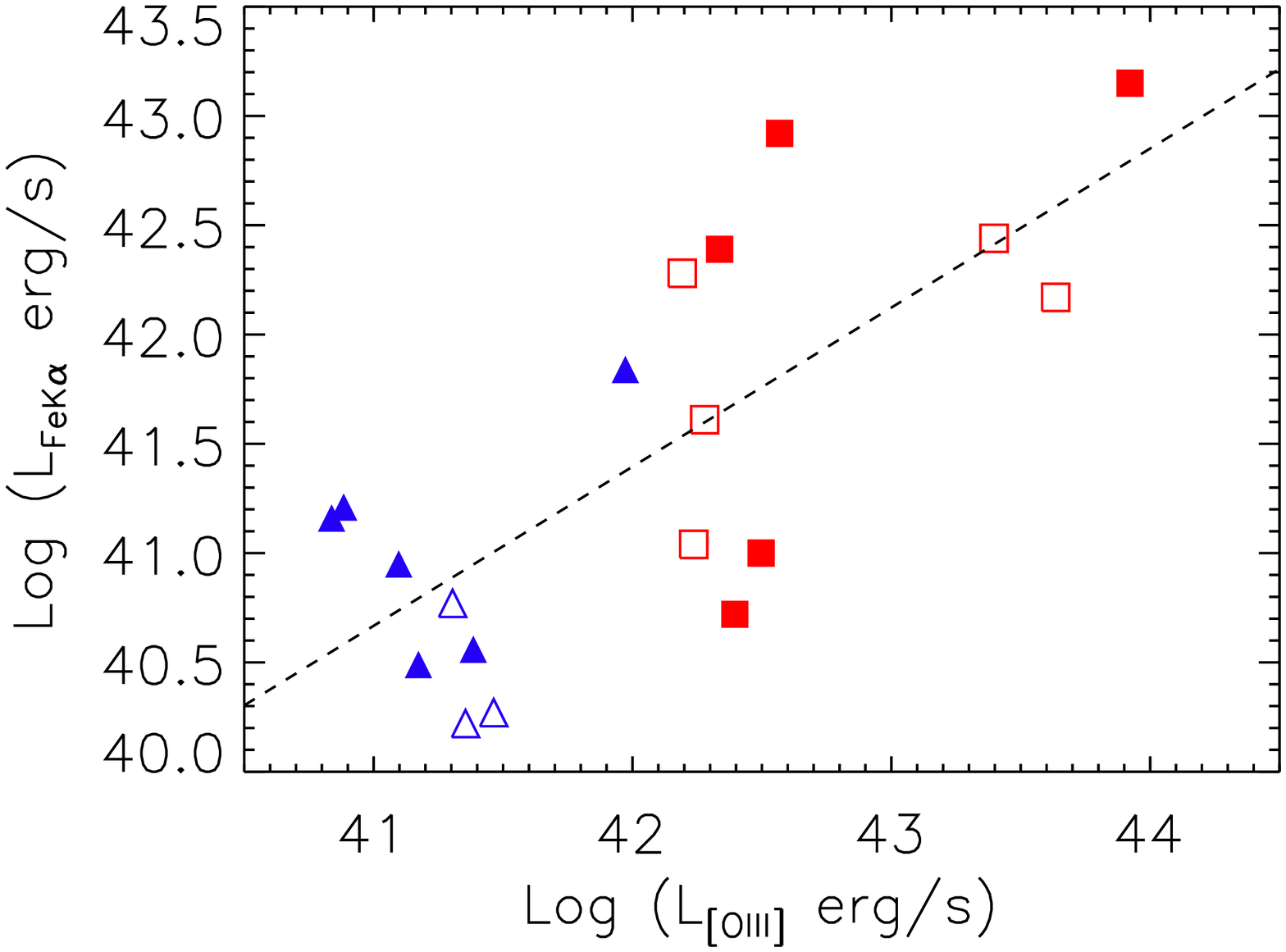}}~
\subfigure[]{\includegraphics[scale=0.4,angle=90]{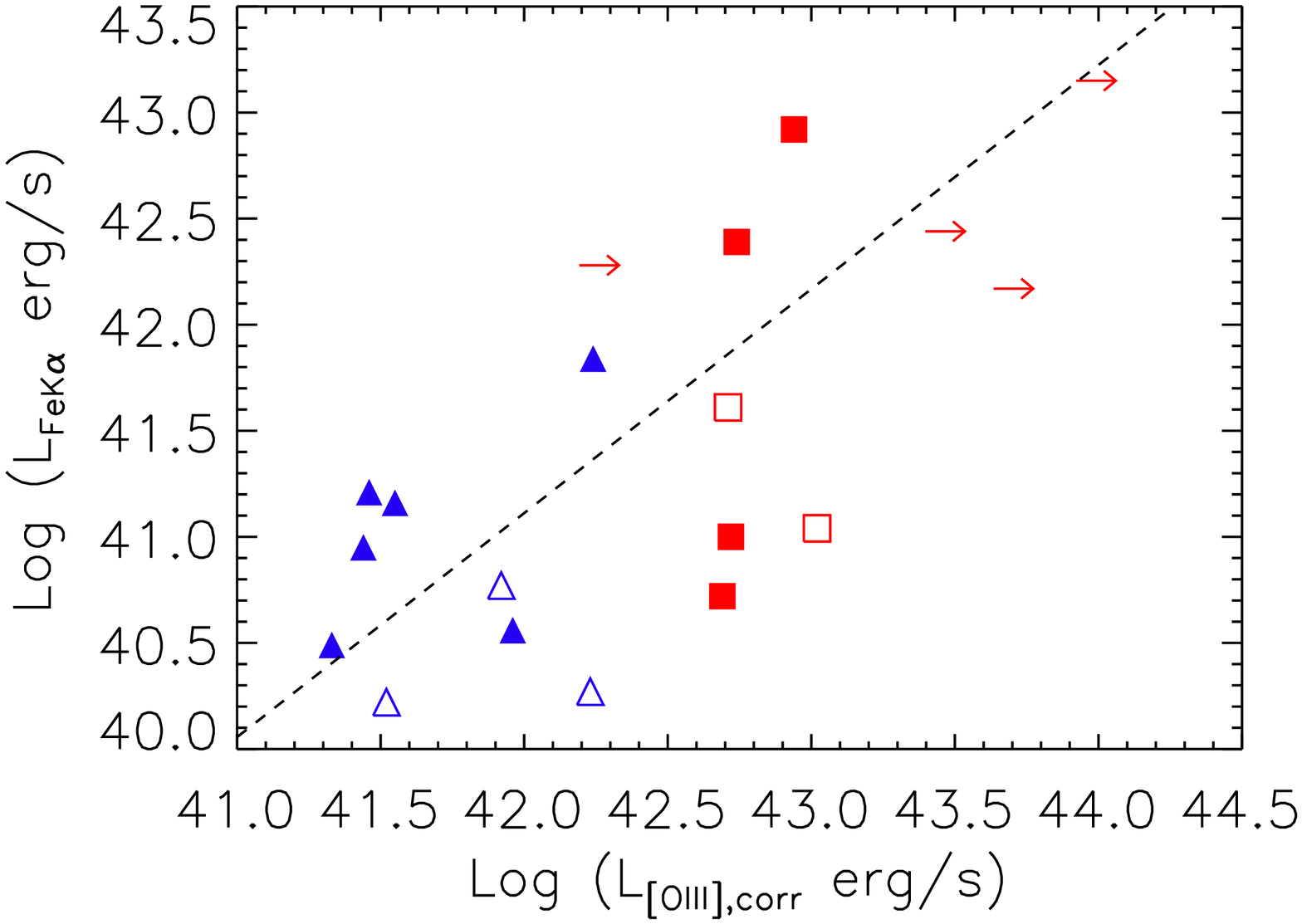}}
\subfigure[]{\includegraphics[scale=0.4,angle=90]{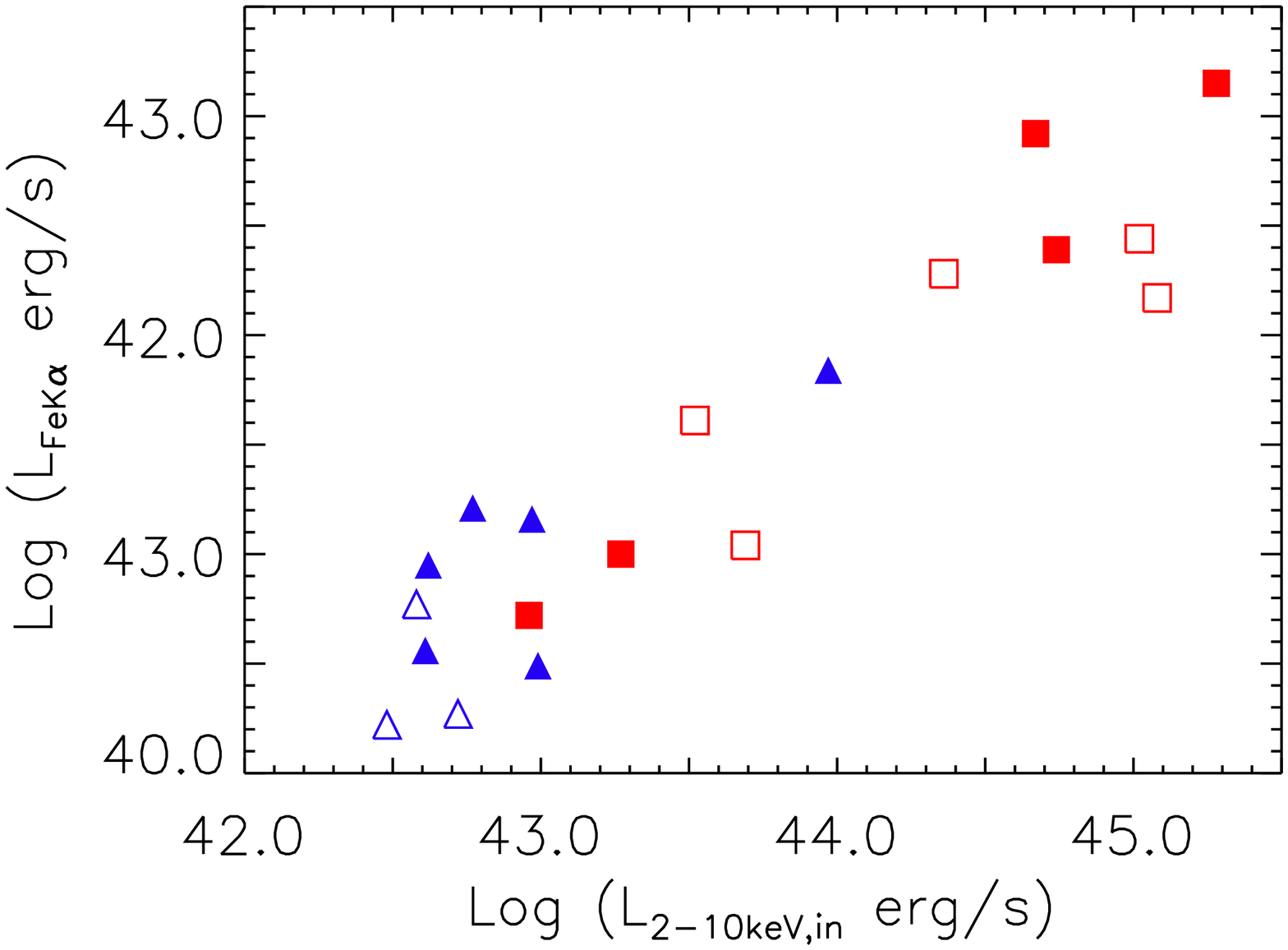}}
\caption[]{\label{fek_lum} (a) Fe K$\alpha$ luminosity as a function of [OIII] luminosity, with the best-fit (slope = 0.73$\pm$0.16) overplotted. Though the correlation is significant, there is wide scatter, with disagreements between the best-fit relation and data at an order of magnitude or greater for a substantial fraction of the sample. (b) Fe K$\alpha$ luminosity as a function of the redenning corrected [OIII] luminosity (L$_{\rm [OIII],corr}$) for the subset of 15 objects with H$\alpha$ coverage in their optical spectra; the observed [OIII] luminosity are lower limits for the remaining 4 sources. Using survival analysis, we find that the strength of the correlation does not improve and the scatter increases, indicating that the dispersion in (a) is not due to host galaxy dust affecting the [OIII] line emission. The best-fit relationship from the Buckley-James method (slope = 1.06$\pm$0.32) is overplotted. (c) Fe K$\alpha$ luminosity as a function of intrinsic 2-10 keV X-ray luminosity, where the slope, i.e., the difference in Fe K$\alpha$ line luminosity divided by the difference in X-ray luminosity, scales as the Fe K$\alpha$ EW. The small dispersion implies that the physical structure of the X-ray reprocessor, such as the global N$_{H}$ and covering factor, are similar among the objects in the sample. Filled symbols mark the sources with Fe K$\alpha$ emission visibly present in their spectra: trends and dispersions are independent of whether or not the Fe K$\alpha$ feature is prominent. Symbol and color coding same as Figure \ref{nhz_v_nhs}.}
\end{figure}

We show the Fe K$\alpha$ luminosity from the fits presented here as a function of the [OIII] luminosity in Figure \ref{fek_lum} (a). We highlight with filled symbols those Type 2 AGN that have the Fe K$\alpha$ feature visibly noticeable in their spectra to test whether the fluorescent emission has to be prominent for a trend to be apparent. The Fe K$\alpha$ and [OIII] luminosities are correlated significantly ($\rho$=0.735, which gives a correlation probability $>$99.9\%), and we find a best-fit slope of 0.73$\pm$0.16, which is largely consistent with the values of 1$\pm$0.5 from \citet{ptak}, 0.7$\pm$0.3 from \citet{me} and 1.13$\pm$0.15 from \citet{jj}. However, there is substantial scatter (dispersion of L$_{\rm FeK\alpha}$/L$_{\rm [OIII]}$ is 0.66), where the disagreement between the observed Fe K$\alpha$ luminosity and the best-fit relation is over an order of magnitude for half the sample, regardless of whether the Fe K$\alpha$ is visible or not. For the subset of 15 objects that have measured H$\alpha$ and H$\beta$ fluxes, we corrected [OIII] luminosity for redenning (L$_{\rm [OIII],corr}$), assuming the standard R(V)=3.1 extinction curve \citep{cardelli} and an intrinsic H$\alpha$/H$\beta$ ratio of 3.1, to test whether significant amounts of host galaxy dust affects the observed [OIII] emission, contributing to the scatter. With the remaining 4 objects, we used the observed [OIII] line luminosity as a lower limit and then employed survival analysis \citep[ASURV Rev 1.2,][]{asurv1,asurv2,asurv3} to test the strength of the correlation and find the trend between the two quantities. As depicted in Figure \ref{fek_lum} (b), the correlation does not become more significant (Spearman's $\rho$=0.713, giving correlation probability of $\sim$99.75\%), and the dispersion of L$_{\rm FeK\alpha}$/L$_{\rm [OIII],corr}$ increases somewhat (0.72). We can therefore conclude that dust attenuating the [OIII] line emission is not responsible for the dispersion in the Fe K$\alpha$-[OIII] luminsosity relationship. Such scatter reflects the complexities of radiative transfer effects when the obscuring medium is not optically-thin \citep{fek}.

The correlation in Figure \ref{fek_lum} (c) illustrates that the Fe K$\alpha$ line is excited by the X-ray continuum, where the slope is approximately proportional to the Fe K$\alpha$ EW (i.e., the difference in the Fe K$\alpha$ line luminosity divided by the difference in the X-ray luminosity provides an estimate of the line EW). The spread in this relation gives an indication of the range of physical properties, such as the global N$_H$ distribution and covering factor. The relatively tight correlation (dispersion of L$_{\rm FeK\alpha}$/L$_{\rm 2-10keV,in}$ is 0.36) implies that the differences in the structure of the X-ray processor are not large among these sources.

\clearpage

\subsection{Comparison with Previous Obscuration Diagnostics}
The ratio of the observed 2-10 keV luminosity to the intrinsic luminosity (e.g., L$_{\rm 2-10keV}$/L$_{\rm [OIII]}$) and EW of the Fe K$\alpha$ line are often used as model-independent proxies of circumnuclear obscuration \citep[e.g.,][]{bassani,cappi,panessa,ghisellini,levenson,me,me11,jj}. Here we test the relationship between these obscuration diagnostics and the reliable line-of-sight and global column densities we derive here. As our goal is to test the results of proxies calculated from the simple power-law modeling with the present analysis, we use the observed L$_{\rm 2-10keV}$ and Fe K$\alpha$ EWs reported in \citet{me} and \citet{jj}, rather than from the current modeling.

\begin{figure}[ht]
\centering
\subfigure[]{\includegraphics[scale=0.4,angle=90]{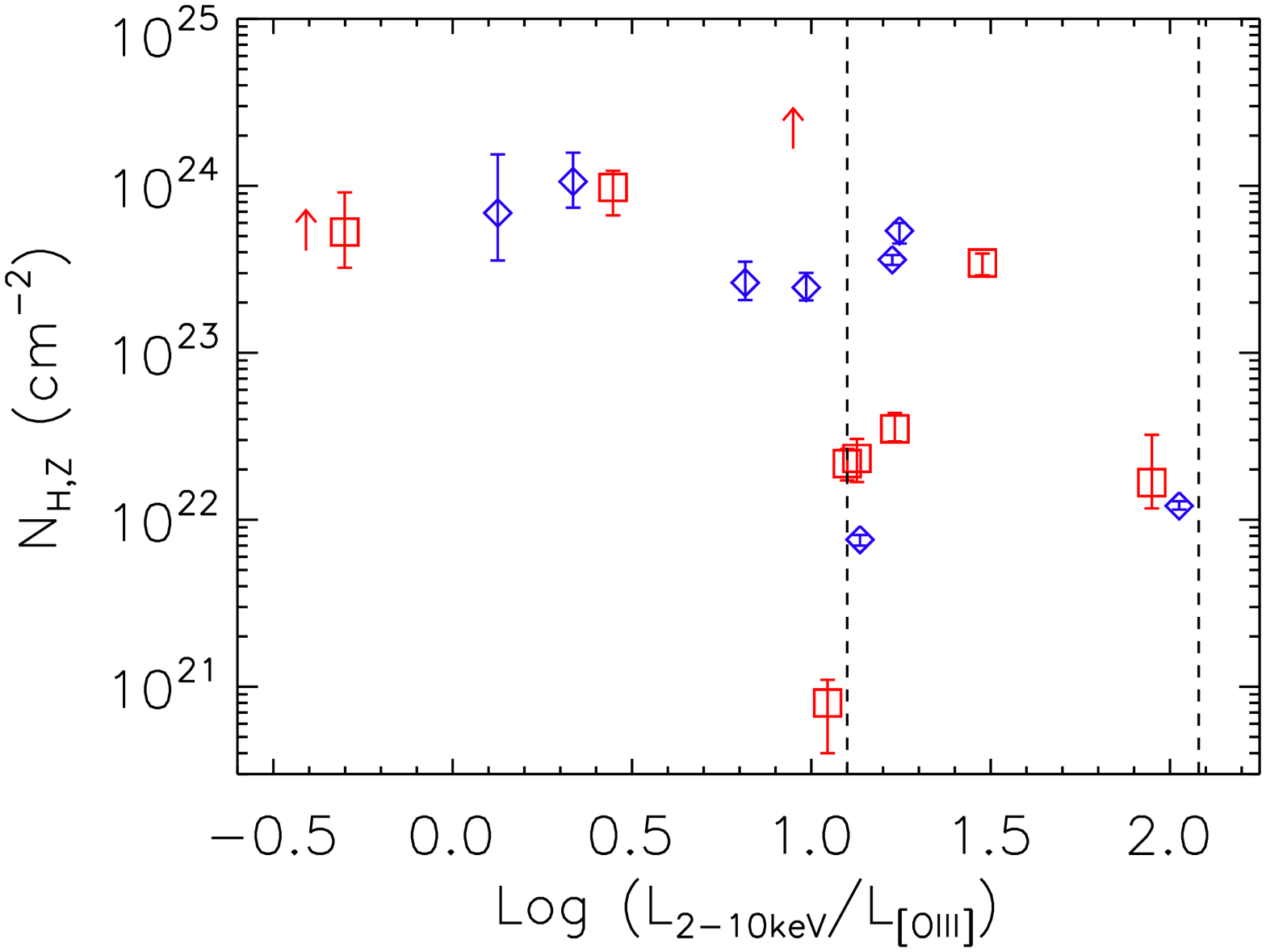}}~
\subfigure[]{\includegraphics[scale=0.4,angle=90]{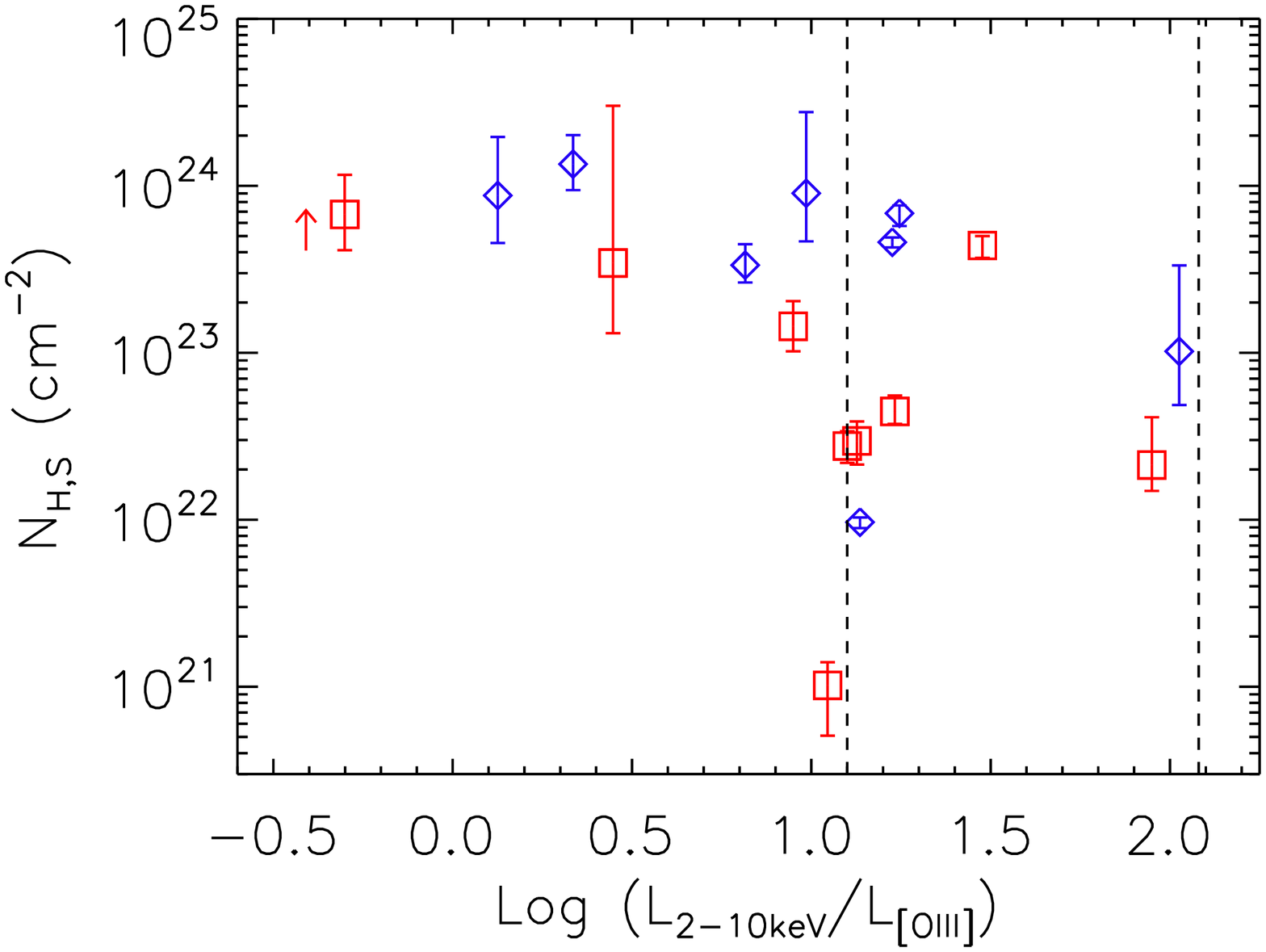}}
\subfigure[]{\includegraphics[scale=0.4,angle=90]{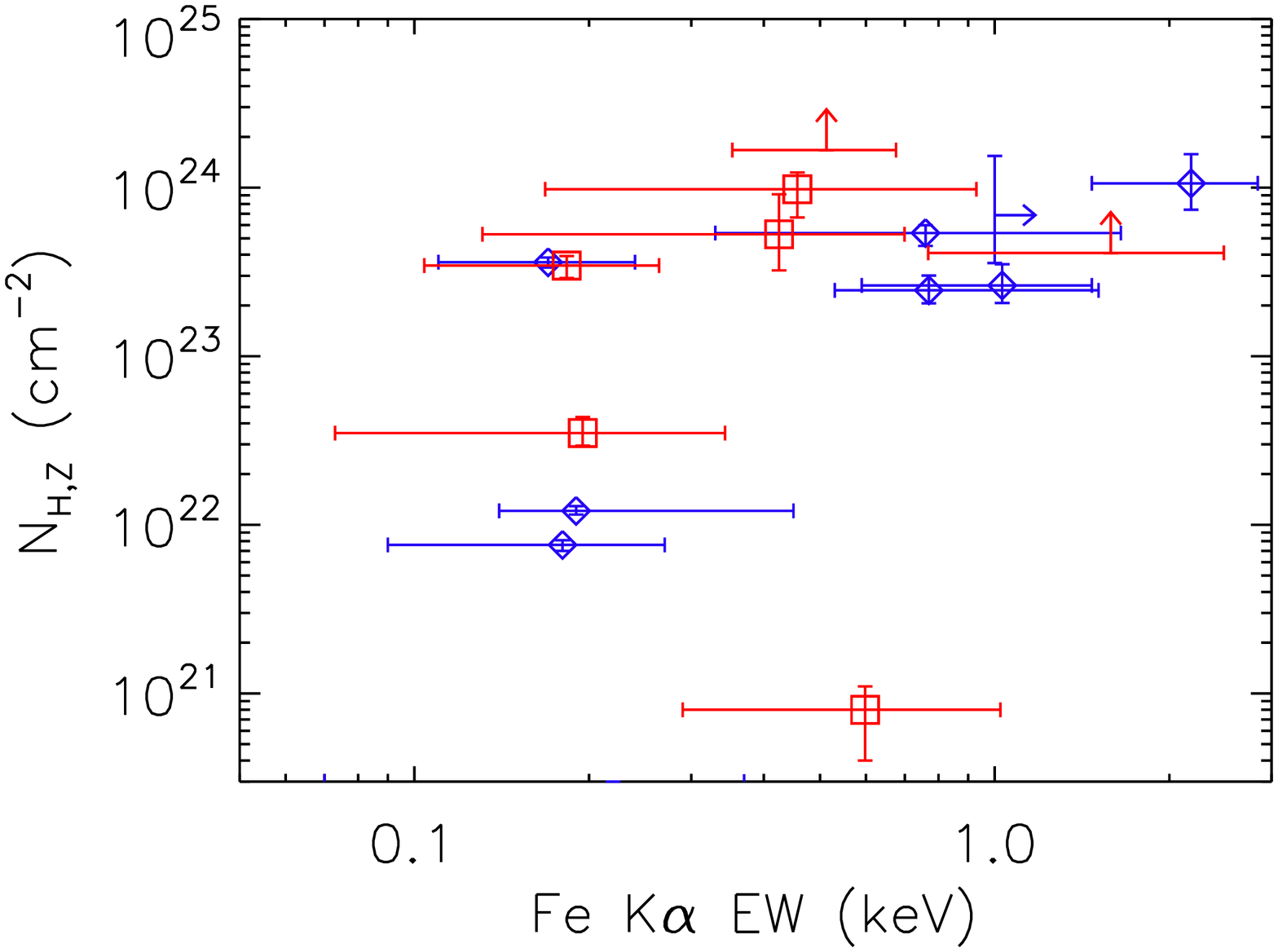}}~
\subfigure[]{\includegraphics[scale=0.4,angle=90]{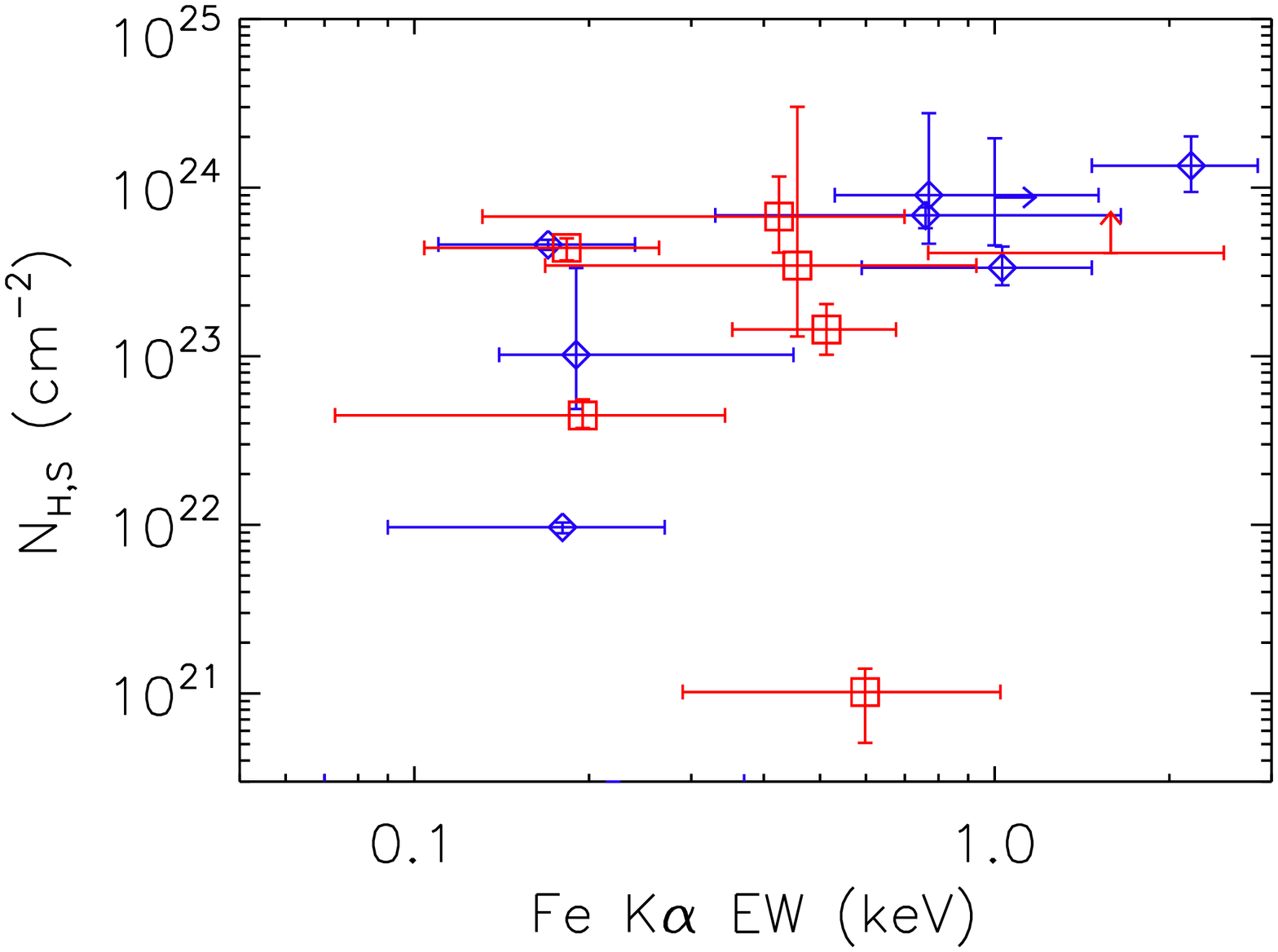}}
\caption[]{\label{nh-v-diag} (a,c) Line of sight column density and (b,d) global average column density as a function of often used, model-independent obscuration diagnostics: (top) L$_{\rm 2-10keV}$/L$_{\rm[OIII]}$, with the range of Type 1 Seyferts indicated by dashed lines, and (bottom) Fe K$\alpha$ EW. Low L$_{\rm 2-10keV}$/L$_{\rm[OIII]}$ values (i.e., $\lesssim1$) and high Fe K$\alpha$ EW ($\gtrsim$1 keV) values are consistent with heavily obscured/Compton-thick sources, but objects can be hidden behind high column densities (N$_{H} > 10^{23}$ cm$^{-2}$) with L$_{\rm 2-10keV}$/L$_{\rm[OIII]}$ values similar to Sy1s and EW values as low as $\sim$200 eV. Symbol and color coding same as Figure \ref{nhz_v_nhs}.}
\end{figure}

Figure \ref{nh-v-diag} (a,c) shows N$_{\rm H,Z}$ and N$_{\rm H,S}$ as a function of L$_{\rm 2-10keV}$/L$_{\rm [OIII]}$, where the vertical dashed lines indicate the range L$_{\rm 2-10keV}$/L$_{\rm [OIII]}$ values observed in Sy1s \citep{tim}. The sources with the lowest normalized X-ray flux are heavily obscured to Compton-thick. However, there are a handful of objects with normalized X-ray luminosities consistent with Sy1s yet are heavily obscured ($10^{23}$ cm$^{-2}<$ N$_{\rm H,Z} < 10^{24}$ cm$^{-2}$) while the rest are moderately obscured ($10^{22}$ cm$^{-2}< N_H < 10^{23}$ cm$^{-2}$) to mildly obscured (N$_H < 10^{22}$ cm$^{-2}$). These results suggest that a low L$_{\rm 2-10keV}$/L$_{\rm[OIII]}$ ratio is indicative of heavy obscuration, but the central engine can also be highly obscured when the normalized 2-10 keV luminosities are consistent with those of unabsorbed Seyferts. We note that 2 of these 4 anomalous sources (SDSS J082443.28+295923.5 and CGCG 218-007) have rather large Fe K$\alpha$ EWs ($\sim$0.75 keV), making it likely that attenuation of the observed [OIII] line by dust in the host galaxy or even the outer parts of the torus inflates the normalized X-ray luminosity.

Comparison of N$_{\rm H,Z}$ and N$_{\rm H,S}$ with the previously reported Fe K$\alpha$ EWs from \citet{me} and \citet{jj}, Figure \ref{nh-v-diag} (b,d), also shows mixed results. We note that the reliability of EWs calculated with previous modeling depends on how accurately the continuum around the line is measured. With the exception of SDSS J083454.89+553411.1 which has an EW of $\sim$600 eV,\footnote{We note that due to the higher binning we use in this work, the Fe line fit by \citet{jj} for SDSS J083454.89+553411.1 is marginal and is more consistent with a non-heavily obscured source in the present analysis.} AGN with an EW above 400 eV are heavily obscured or Compton-thick.  At an EW of $\sim$200 eV, however, about half of the sources are consistent with being heavily obscured while the other half are moderately to mildly obscured. Similar to the comparison with L$_{\rm 2-10keV}$/L$_{\rm [OIII]}$, a high Fe K$\alpha$ EW value ($>$400-500 eV) is a useful diagnostic of heavy absorption, but an AGN can be heavily obscured at lower EWs. We note that it is the same 2 sources with high L$_{\rm 2-10keV}$/L$_{\rm [OIII]}$ values and low Fe K$\alpha$ EWs that have fitted column densities exceeding 10$^{23}$ cm$^{-2}$ (SDSS J123843.43+092736.6 and SDSS J090037.09+205340.2).

\subsection{Sy2s vs. Type 2 QSOs}
The parent Sy2 and Type 2 QSO candidate samples from \citet{me} and \citet{jj}, respectively, are selected based on their optical emission, and are thus unbiased with respect to their X-ray properties. However, the detailed modeling used in this work requires relatively good signal-to-noise in X-rays, so the sub-samples analyzed here are unavoidably biased to be X-ray bright. We caution that the trends we examine between Sy2s and Type 2 QSOs pertain to the sources in this work and it would be inappropriate to extrapolate these results to the general AGN population.

\begin{figure}[ht]
\centering
\subfigure[]{\includegraphics[scale=0.4,angle=90]{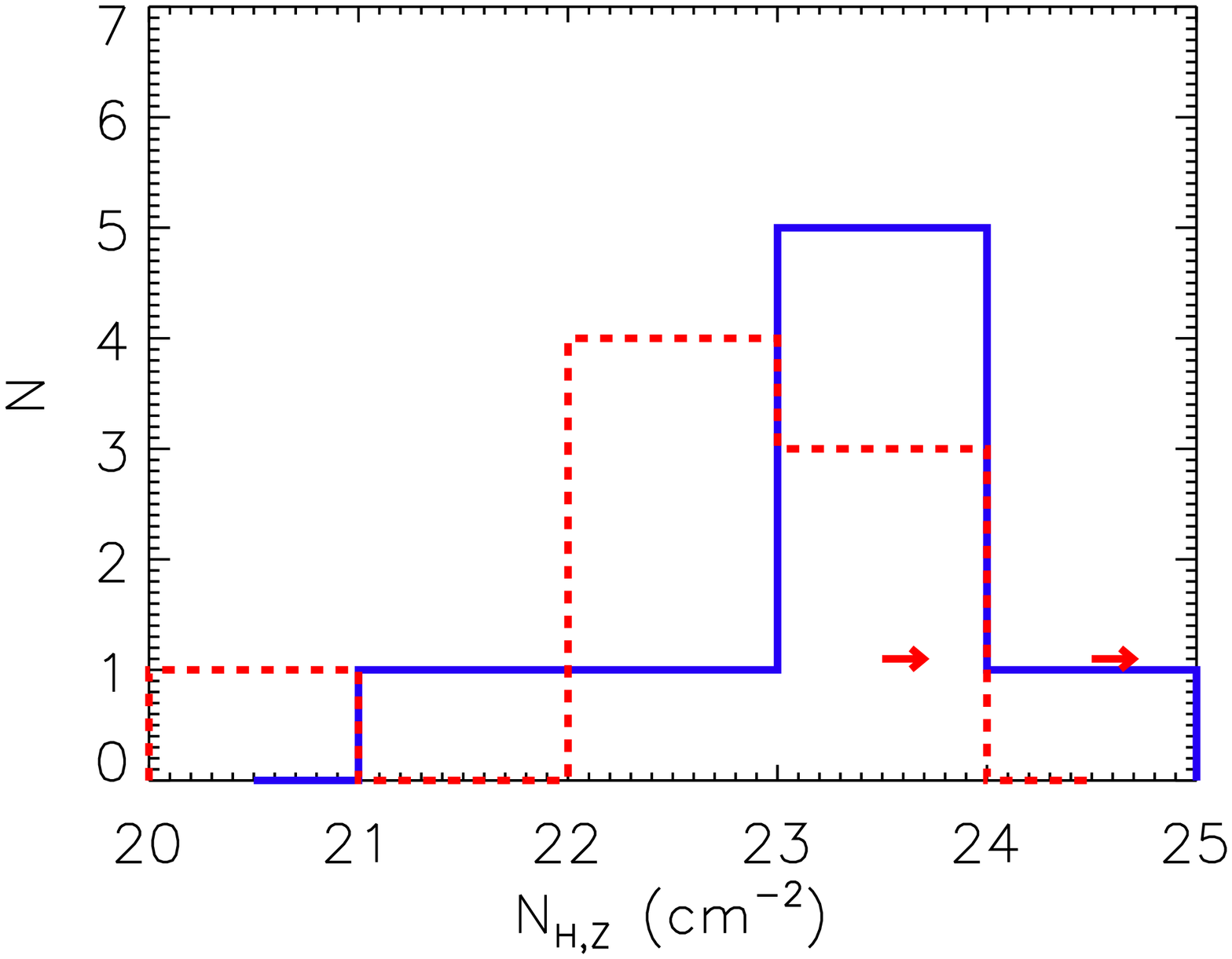}}~
\subfigure[]{\includegraphics[scale=0.4,angle=90]{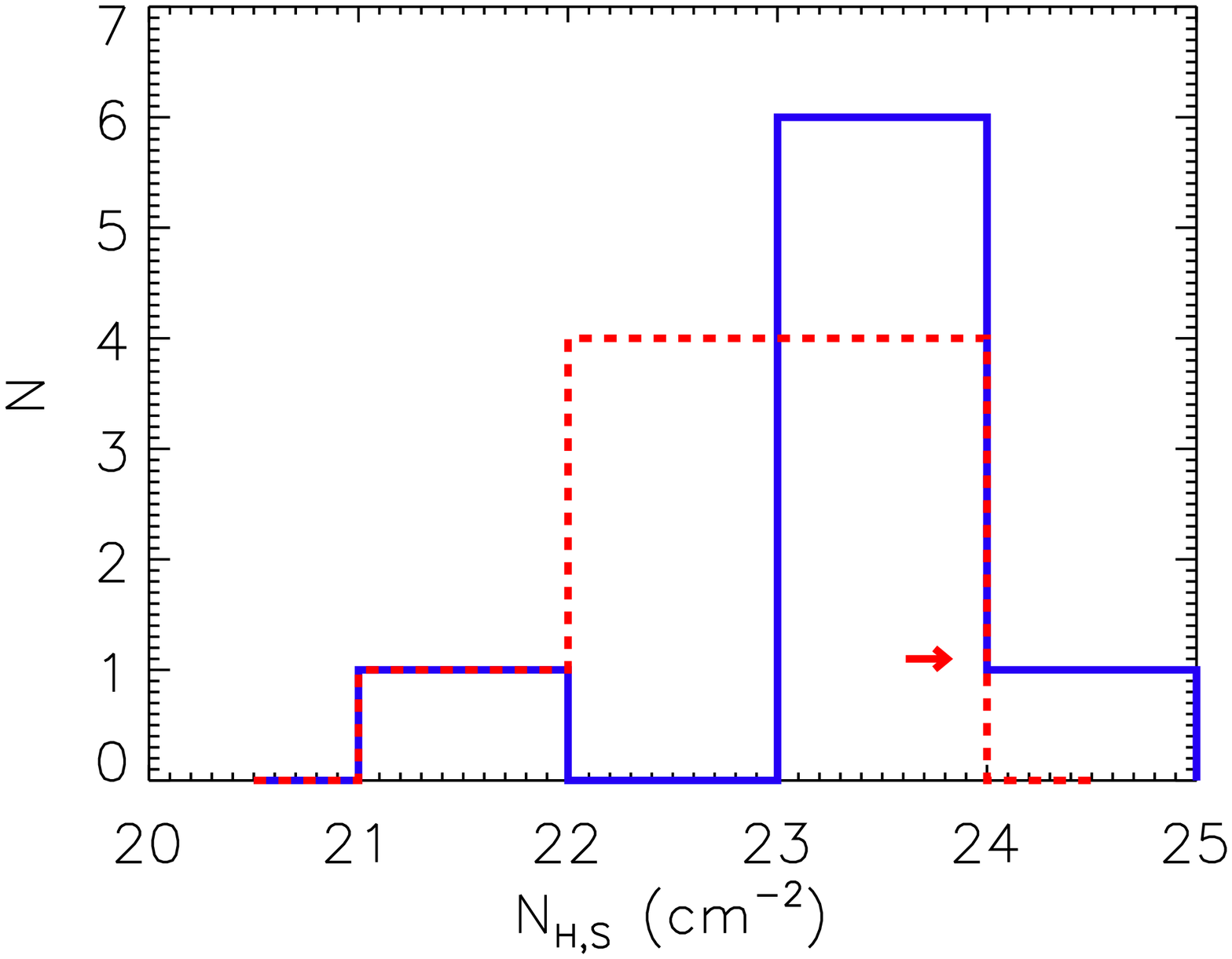}}
\subfigure[]{\includegraphics[scale=0.4,angle=90]{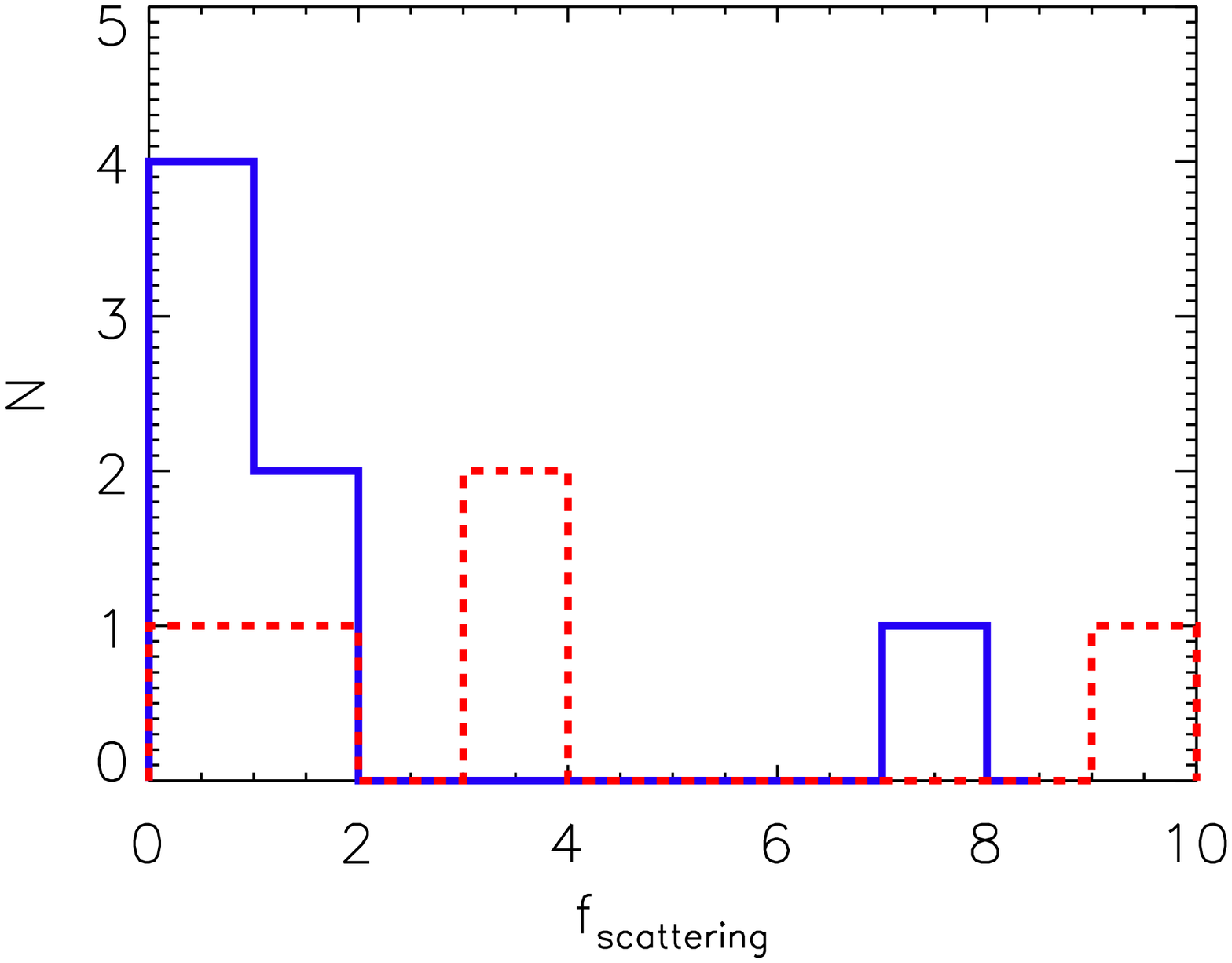}}
\caption[]{\label{hist} Distributions of (a) line of sight column densities (b) global average column densities and (c) scattering fractions for Sy2s (blue solid-line) and QSO2s (red dashed-line). A greater percentage of Sy2s are heavily obscured (N$_{H} > 10^{23}$ cm$^{-2}$) while a majority of Type 2 QSOs are moderately ($10^{22}$ cm$^{-2} <$ N$_{H} < 10^{23}$ cm$^{-2}$) to heavily obscured. For sources where a scattering fraction is measured, this fraction is largely under 2\% for both Sy2s and QSO2s.}
\end{figure}

In Figure \ref{hist}, we show the distributions of the line-of-sight (a) and global average (b) column densities for the Sy2s (solid blue) and Type 2 QSOs (dashed red). A larger percentage of Sy2s are heavily obscured while a majority of the quasars are almost evenly split between moderately and heavily obscured. Most of the Seyferts and quasars that have evidence for scattering of the AGN continuum have a leakage fraction under 2\%, though this fraction reaches between $\sim7-9$\% for individual sources (Figure \ref{hist} (c)).

\begin{figure}[ht]
\centering
{\includegraphics[scale=0.4,angle=90]{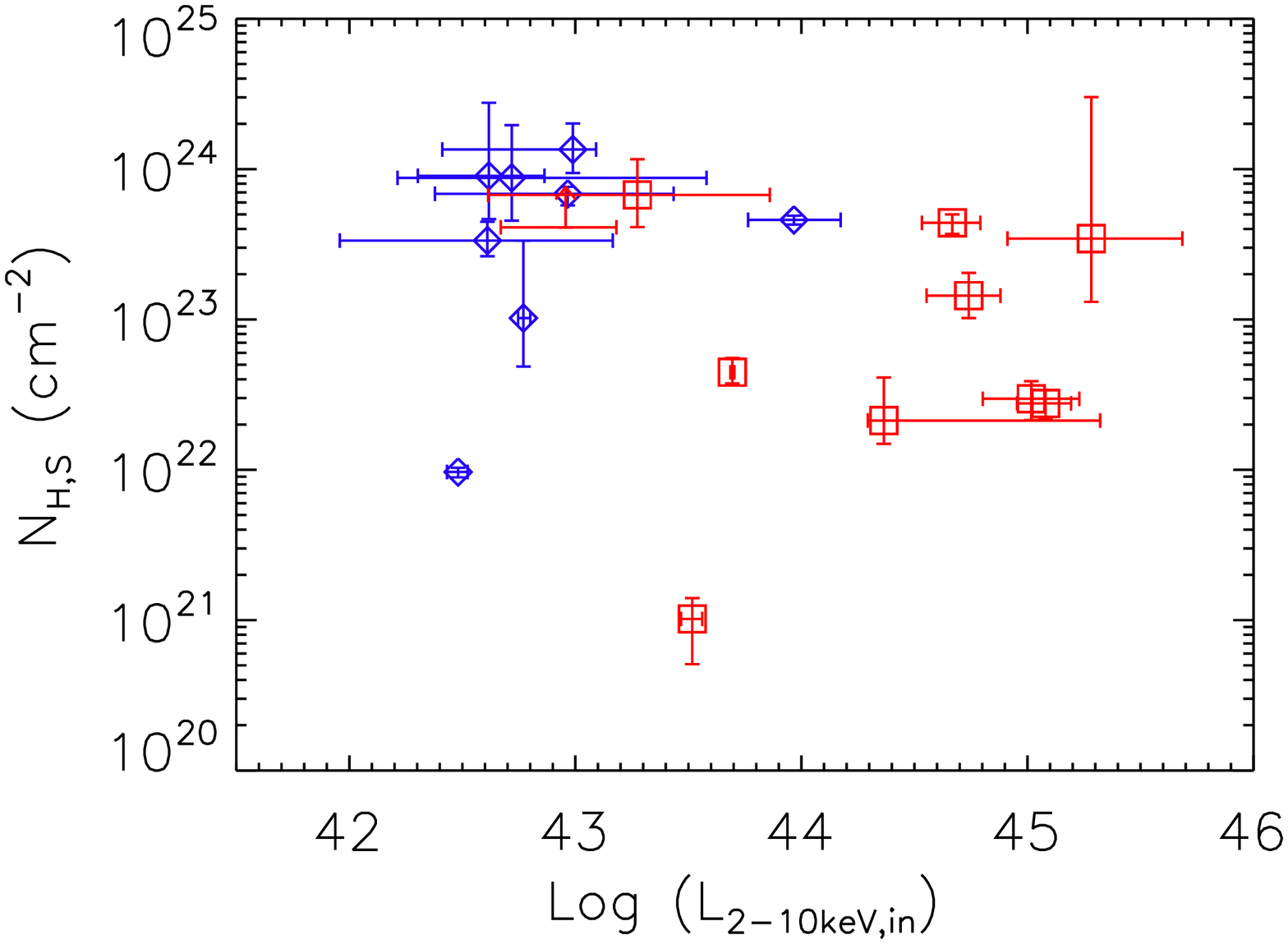}}~
{\includegraphics[scale=0.4,angle=90]{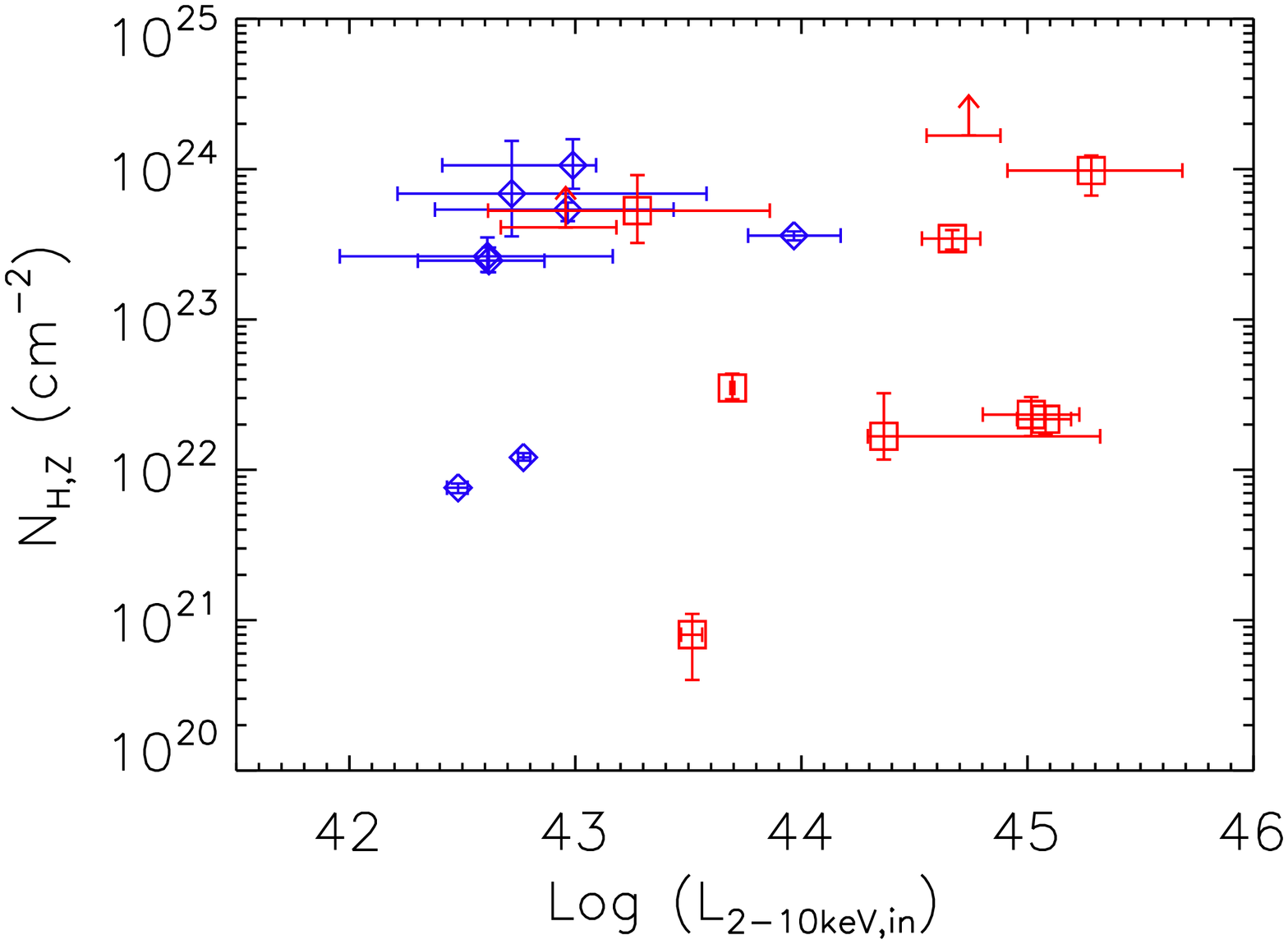}}
\caption[]{\label{lum_v_obs} (a) Line of sight column density and (b) global average column density as a function of intrinsic 2-10 keV luminosity. Similar to \citet{jj}, we do not find a dependence of obscuration as a function of AGN luminosity, inconsistent with the expectations of the ``receding torus model'' \citep{lawrence,lawrence2,ueda,simpson}, though the sources presented here represent neither an unbiased nor complete sample. Symbol and color coding same as Figure~\ref{nhz_v_nhs}.}
\end{figure}

Though Figure \ref{hist} (a,b) indicates that these Sy2s tend to be more heavily obscured than the QSO2s, we do not find a trend in general between N$_H$, line-of-sight or global, and the absorption corrected X-ray luminosity (Figure \ref{lum_v_obs}), as may be expected by the ``receding torus model'' \citep[e.g.,][]{lawrence,lawrence2,ueda,simpson}. This result is consistent with that of \citet{jj} where they used the [OIII] luminosity as a proxy of intrinsic AGN luminosity for the same parent sample from which this work represents a sub-set. However as the AGN we are studying are not representative of the general population, and in particular exclude objects with low column densities and unabsorbed Type 1 AGN, our results should not be construed as a challenge to  the receding torus model.

We also note that we find no trends between N$_{\rm H,Z}$ or N$_{\rm H,S}$ and redshift, N$_{\rm H,Z}$ or N$_{\rm H,S}$ and scattering fraction, scattering fraction and redshift, and scattering fraction and L$_{\rm 2-10keV}$/L$_{\rm[OIII]}$.

% Sy2s                    QSO2     total
% unabsorbed: 1           0        1
% mild: 1                 1        2
% moderate: 1             4        5
% heavy: 4                2        6
% heavy to C-thick: 2     3        5

% scattering fraction for 13 sources

\section{Conclusions}
We used the physically motivated, self-consistent spherical absorption model of \citet{spherical} and the MYTorus model \citep{mytorus} to unravel the complexities of the X-ray reprocessor in 19 Type 2 [OIII]-selected AGN, 9 of which are Seyfert 2 galaxies and 10 of which are obscured quasar candidates. We report, for the first time for an AGN sample, reliable measurements of the global, as well as line-of-sight, column densities.
\begin{itemize}
\item Along the line-of-sight, 1 Sy2 and 1 QSO2 are mildly obscured ($<10^{22}$ cm$^{-2}$), 1 Sy2 and 4 QSO2s are moderately obscured ($10^{22}$ cm$^{-2} <$ N$_{H} < 10^{23}$ cm$^{-2}$), 4 Sy2s and 2 QSO2s are heavily obscured ($10^{23}$ cm$^{-2} <$ N$_{H} < 10^{24}$ cm$^{-2}$), and 2 Sy2s and 3 QSO2s are heavily-obscured to Compton-thick ($> 1.25\times10^{24}$ cm$^{-2}$). One source is unabsorbed.

\item 13 objects have evidence of scattering of the intrinsic AGN continuum off a distant medium. The scattering fraction is largely under 2\%, but reaches between $\sim$ 7-9\% for individual sources.

\item In 4 AGN, we measure global column densities that are significantly different from the line-of-sight column density. In one of these objects, SDSS J082443.28+295923.5, the line-of-sight column density is Compton-thin while the global column density is Compton-thick, while in 1 other object, SDSS J093952.74+355358.0, the line-of-sight column density is heavily-obscured to Compton thick with a Compton-thin global column density.

\item We identified a candidate changing-look AGN or naked Sy2 galaxy, Mrk 0609. If Mrk 0609 is a true Sy2, both its bolometric luminosity and Eddington rate exceed the predicted critical values of disk-wind models that explain such sources \citep{nicastro,trump}. We are pursuing simultaneous X-ray and optical observations to test this possibility.

\item The spectral features of SDSS J093952.74+355358.0 indicate that the line-of-sight obscuration is in the form of Compton-thick toroidal ring which is embedded in a Compton-thin global matter distribution. This is the first Type 2 QSO where this geometry has been observed.

\item The ratio of the intrinsic 2-10 keV luminosity to the [OIII] luminosity, 1.54$\pm$0.49 dex, is basically equivalent to that for Type 1 AGN \citep[1.59$\pm$0.48]{tim}, affirming that these sophisticated models reliably measure the circumnuclear column density, allowing the inherent X-ray luminosity to be revealed.

\item We find a significant correlation between the Fe K$\alpha$ and [OIII] luminosities, though with wide scatter. Estimating the Fe K$\alpha$ luminosity from the [OIII] luminosity can lead to errors of over an order of magnitude.

\item We compared N$_{\rm H,Z}$ and N$_{\rm H,S}$ with obscuration diagnostics used often in the literature from {\it ad hoc} spectral modeling, i.e., L$_{\rm 2-10keV}$/L$_{\rm[OIII]}$ and the Fe K$\alpha$ EW \citep[e.g.,][]{ghisellini,bassani, cappi, levenson, panessa, me,me11,jj}, where lower values of the former (i.e., $\lesssim$1) and higher values of the latter (i.e., $\sim$1 keV) imply heavy-to-Compton-thick obscuration. Sources that are flagged as heavily obscured via these proxies do indeed have measured column densities consistent with heavy or Compton-thick obscuration, but a couple of objects with values similar to unabsorbed AGN also have measured N$_{\rm H,Z}$ indicative of heavy obscuration.

\item Though a larger percentage of the Sy2s are more heavily obscured than the QSO2s, no trend exists between line-of-sight or global obscuration and AGN luminosity, parametrized here as the intrinsic, rest-frame 2-10 keV luminosity. We also find no relationship between N$_{\rm H,Z}$ or N$_{\rm H,S}$ and scattering fraction, N$_{\rm H,Z}$ or N$_{\rm H,S}$ and redshift, scattering fraction and redshift, and scattering fraction and L$_{\rm 2-10keV}$/L$_{\rm[OIII]}$. We reiterate that these results hold for the objects we studied here which are not necessarily representative of Type 2 AGN in general.

\end{itemize}

\section{Acknowledgements}
We thank the referee for a careful reading of this manuscript and for helpful suggestions. P.G. acknowledges support from STFC (grant reference ST/J003697/1).

This work has made use of data from the SDSS. Funding for the SDSS and SDSS-II has been provided by the Alfred P. Sloan Foundation, the Participating Institutions, the National Science Foundation, the U. S. Department of Energy, the National Aeronautics and Space Administration, the Japanese Monbukagakusho, the Max Planck Society, and the Higher Education Funding Council for England. The SDSS Web site is http://www.sdss.org/. The SDSS is managed by the Astrophysical Research Consortium for the Participating Institutions. The Participating Institutions are the American Museum of Natural History, Astrophysical Institute Potsdam, University of Basel, University of Cambridge, Case Western Reserve University, University of Chicago, Drexel University, Fermilab, the Institute for Advanced Study, the Japan Participation Group, Johns Hopkins University, the Joint Institute for Nuclear Astrophysics, the Kavli Institute for Particle Astrophysics and Cosmology, the Korean Scientist Group, the Chinese Academy of Sciences (LAMOST), Los Alamos National Laboratory, the Max-Planck-Institute for Astronomy (MPIA), the Max-Planck-Institute for Astrophysics (MPA), New Mexico State University, Ohio State University, University of Pittsburgh, University of Portsmouth, Princeton University, the United States Naval Observatory, and the University of Washington.

\end{document}